\documentclass[11pt]{article}
\DeclareMathAlphabet{\scr}{U}{rsfs}{m}{n}
\usepackage[T1]{fontenc}
\usepackage{latexsym}
\usepackage{epsfig}
\usepackage[mathscr]{eucal}
\usepackage{amsfonts}
\usepackage{amscd}
\usepackage{array}
\usepackage{amssymb}
\usepackage{colordvi}
\usepackage{amsthm}
\usepackage[centertags]{amsmath}
\usepackage{enumerate}
\usepackage{graphicx}
\usepackage{booktabs}
\usepackage[footnotesize]{caption}

\usepackage{pdfpages}
\usepackage{hyperref}

\hypersetup{
    colorlinks,
    citecolor=blue,
    filecolor=black,
    linkcolor=red,
    urlcolor=magenta
}

\renewcommand{\d}{\mathrm{d}}

\newcommand{\N}{\mathbb{N}}
\newcommand{\newc}{\newcommand}
\newc{\be}{\begin{equation}}
\newc{\ee}{\end{equation}}
\newc{\bi}{\begin{itemize}}
\newc{\ei}{\end{itemize}}
\newc{\benu}{\begin{enumerate}}
\newc{\eenu}{\end{enumerate}}
\newc{\bc}{\begin{center}}
\newc{\ec}{\end{center}}
\newc{\bfig}{\begin{figure}}
\newc{\efig}{\end{figure}}
\newc{\qbar}{\bar{q}}
\newc{\go}{\tilde{g}}
\newc{\PB}{\textsc{Powheg-Box}}

\newcommand{\R}{\mathbb{R}}
\newcommand{\C}{\mathbb{C}}
\newcommand{\Z}{\mathbb{Z}}

\newcommand{\bI}{{\cal B}_{\mbox{\tiny I}}}
\newcommand{\bII}{{\cal B}_{\mbox{\tiny II}}}
\newcommand{\bIII}{{\cal B}_{\mbox{\tiny III}}}

\newtheorem{definition}{Definition}[section]
\newtheorem{theorem}{Theorem}
\newtheorem{proposition}{Proposition}[section]

\newtheorem{lemma}{Lemma}[section]
\newtheorem{remark}{Remark}[section]

\begin{document}

\title{
\hfill ~\\[-30mm]
\phantom{h} \hfill\mbox{\small Cavendish-HEP-19/04}
\\[1cm]
\textbf{A new type of charged black hole bomb}}

\date{}
\author{
Laurent DI MENZA\footnote{Laboratoire de Math\'ematiques de Reims, FR CNRS 3399, UFR SEN, Moulin de la Housse, BP 1039, 51687 Reims Cedex 2, France. E-mail: Laurent.Di-Menza@univ-reims.fr},
Jean-Philippe NICOLAS\footnote{LMBA, UMR CNRS n$^\mathit{o}$ 6205, Université de Brest, 6 avenue Le Gorgeu, 29238 Brest cedex 3, France. E-mail: Jean-Philippe.Nicolas@univ-brest.fr},
and Mathieu PELLEN\footnote{Cavendish Laboratory, University of Cambridge, Cambridge CB3 0HE, United Kingdom. E-mail: mpellen@hep.phy.cam.ac.uk}
}

\maketitle

\begin{abstract}
\noindent  
Black hole bombs are usually constructed by surrounding an ergoregion by a mirror.
The fields propagating between the event horizon and the mirror are prevented from escaping to infinity and reflected back to the ergoregion, thus undergoing repeated superradiant scattering which leads to a linear instability.
We introduce a new construction in which the field is outside the mirror and is therefore prevented from falling into the black hole but is free to escape to infinity.
Provided the mirror is inside the ergoregion, it turns out that this still causes linear instabilities. This behaviour is observed on Reissner-Nordström and de Sitter-Reissner-Nordström backgrounds using numerical simulations, based on a semi-implicit discretisation on a first-order system formulation of
the partial differential equations governing the evolution of the scalar field.
We also perform simulations for a standard black hole bomb and for another type of contraption: a sandwich black hole bomb.
\end{abstract}

\tableofcontents

\section{Introduction}
\label{ch:introduction}

In 1969, R. Penrose \cite{Pe1969} noticed the possibility of extracting rotational energy from a rotating black hole using test particles or material objects.
In a Gedankenexperiment now known as the \emph{Penrose process}, a test particle is sent from infinity into the ergoregion, where it splits in two particles. The first one is directed in such a manner that it has negative energy as measured by an observer static at infinity. It cannot escape, because its energy is conserved and negative energy is only possible inside the ergoregion. Eventually, it falls through the event horizon. The other particle is free to escape to infinity and since the total energy is conserved, its energy is positive and larger than that of the particle that was originally sent in.
The analogue of the Penrose process for fields is referred to as superradiance. It is a scattering process with a reflection coefficient that is larger than $1$.
In black hole spacetimes, superradiance can be caused by rotation, as first observed by Zeldovich in 1971 and 1972 \cite{Ze1971,Ze1972}, or by the interaction of the charges of the field and the black hole.
In the first case, the phenomenon is localised in a fixed geometrical region outside the black hole, called the ergoregion and determined uniquely in terms of the mass, charge, and angular momentum of the black hole as well as the cosmological constant. In the second case, the ergoregion depends on both the physical parameters of the black hole and the field and in some situations may cover the whole exterior of the black hole.

Rotational superradiance has been investigated more thoroughly than the charge induced one, probably because of its greater physical significance, since one expects stellar objects to be electrically neutral. The charged case is nevertheless interesting, albeit on a more abstract level, because of its very distinct analytical and geometrical features.
In particular, it may be a harder problem in terms of analysis of partial differential equations than the rotational case, which is solved at least partially for the Kerr and Kerr-de Sitter families\footnote{See M. Dafermos, I. Rodnianski and Y. Shlapentokh-Rothman \cite{DaRoShla2014} for the wave equation on Kerr backgrounds and V. Georgescu, C. Gérard and D. Häfner \cite{GeGeHa2017} for Klein-Gordon fields outside a slowly rotating Kerr-de Sitter black hole.
Note that in the slowly rotating Kerr-de Sitter, and Kerr-Newman-de Sitter cases, the global nonlinear stability has recently been established by P. Hintz and A. Vasy \cite{HiVa2016} and P. Hintz \cite{Hi2018}.}, whereas even the spherically symmetric situation is not well understood for charge induced superradiance.
For a long time, as far as we are aware, there was only one analytical study available in the literature, due to A. Bachelot \cite{Ba2004}. This remarkable work essentially gives a complete description of the spherically symmetric case.
It notably characterises the situations giving rise to the extraction of an infinite amount of energy by a simple analytic property of the radial speed of the radial null geodesics, but this property is difficult to verify in black hole spacetimes and the finiteness of the reflection coefficient is still an open question. Recently, N. Besset \cite{Be2018} proved the decay of the local energy of charged Klein-Gordon fields outside a subextremal de Sitter-Reissner-Nordström black hole, using a resonance expansion of the propagator.

The numerical investigations of the charged case are very scarce and were mostly done in the frequency domain\footnote{See S. Chandrasekhar's book \cite{ChaBook} and the more recent contribution by M. Richartz and A. Saa \cite{RiSa2011}.}.
A study in the time domain was published by two of the authors (LDM and JPN \cite{DiNi2015}) where a spontaneous behaviour, very similar to the Penrose process, was observed numerically for charged scalar fields outside a Reissner-Nordstr\o m black hole: a wave packet is sent from outside the ergoregion towards the horizon. As it enters the ergoregion, it splits spontaneously into an incoming wave packet with negative energy, which falls into the black hole and, an outgoing wave packet, with positive energy larger than that of the initial incoming one, which propagates to infinity. 
Another type of superradiant behaviour was also observed in Ref.~\cite{DiNi2015}, associated with so-called {\em flare} initial data. These are a class of initial data supported within the ergoregion but on which the energy is positive definite (the data for the field is zero and the data for its time derivative is a gaussian multiplied by an oscillatory exponential).
The energy gains obtained for flare data are much larger than for incoming wave packets and far exceed the theoretical limits for scattering experiments, given by reflection coefficients.
Similar initial data were studied by methods of geometric optics by G. Eskin in Ref.~\cite{E2016}. For a comprehensive list of references on superradiance and a thorough investigation of the phenomenon in its many aspects, see the book {\em Superradiance} by R. Brito, V. Cardoso, and P. Pani \cite{BCP}.

The concept of a black hole bomb was originally proposed in 1972 by W. H. Press and S. A. Teukolsky \cite{PreTe1972,Bardeen:1972fi}. A spherical mirror is constructed around a Kerr black hole, in the exterior region. A scalar field propagating outside the black hole and inside the mirror is thus prevented from escaping to infinity and undergoes repeated superradiant scattering inside the ergoregion. The result is a linear instability, \emph{i.e.}\ an exponential growth of the amplitude of the field.
Since it is usual to set reflecting boundary conditions at the timelike conformal boundary of anti-de Sitter spacetime\footnote{This is done for mathematical reasons, so as to ensure a unitary evolution, and does not seem to be guided by any physical motivation.}, asymptotically anti-de Sitter black holes provide a realisation of a black hole bomb without the need to construct a mirror, and has been the subject of numerous studies, see for example H. Witek, V. Cardoso, C. Herdeiro, A. Nerozzi, U. Sperhake and M. Zilhão \cite{WCHNSM2010}, S. Green, S. Hollands, A. Ishibashi and R. Wald \cite{GreHoIWa2016} and references therein and also the recent numerical study of the fully coupled Einstein-charged scalar field system by P.M. Chesler and D.A. Lowe \cite{CheLo2018}.
A similar construction to the one proposed in Ref.~\cite{PreTe1972} can be performed outside a charged black hole.
These charged black hole bombs have been somewhat less investigated than their rotating analogues but there is still a fair amount of studies, mostly numerical, available in the literature, for instance Ref.~\cite{BCP,DeHe2014,DiMa2018,HeDeRu2013,Ho2013,Ho2016}.

In this article, we propose a new type of black hole bomb in which the field is located outside the mirror.
Our numerical simulations show that provided the mirror is located inside the ergoregion, this still leads to a linear instability.
Although the field is free to escape the ergoregion, we observe that a part of the solution with negative energy creeps along the mirror and sends out repeated bursts of positive energy while its amplitude increases exponentially.
For completeness, we also study the more classic black hole bombs and another new type of bomb, the sandwich bomb, for which one naturally anticipates linear instability.
The simulations in these cases give the expected results and provide a validation of our numerical scheme. We use the following terminology for our different versions of the bombs:
\begin{itemize}
\item {\bf classic, or type I, charged black hole bomb}; it is analogous to the Press-Teukolsky bomb; the field propagates between the mirror and the horizon;
\item {\bf type II charged black hole bomb}; this time the mirror is set within the ergoregion and the field propagates outside of the mirror;
\item {\bf type III, or sandwich, charged black hole bomb}; the field is sandwiched between two mirrors, the innermost one being set inside the ergoregion.
\end{itemize}
The simulations are performed in the time domain for charged scalar fields in the exterior of a subextremal Reissner-Nordström or de Sitter-Reissner-Nordström black hole. The initial data are of flare-type and are supported inside the ergoregion.

The article is organised as follows. Section \ref{GAF} is devoted to a brief presentation of de Sitter-Reisser-Nordström and Reissner-Nordström metrics and of the charged Klein-Gordon equation on such backgrounds. The three types of black hole bombs are described in Section \ref{BHBombs}. 
We describe in Section \ref{SIM} the numerical
method that is used for the simulations, with particular attention paid on the energy conservation
at a discrete level. Finally, the numerical simulations of black hole bombs for types I, II, and III, are presented and discussed in Section \ref{NumExp}.


\section{Geometrical and analytical framework} \label{GAF}

\subsection{Reissner-Nordström and de Sitter-Reissner-Nordström metrics}

The de Sitter-Reissner-Nordström metric is a $3$-parameter family of solutions of the Einstein-Maxwell equations with positive cosmological constant. In Schwarzschild coordinates $(t,r,\omega) \in \R \times ]0,+\infty [ \times S^2$, it has the following expression:
\begin{equation} \label{dsrnMetric}
 g = F(r) \d t^2 - F(r)^{-1} \d r^2 - r^2 \d\omega^2\, ,
\end{equation}
with $\d \omega^2 $ the euclidean metric on $\mathbb{S}^2$,
\[ F(r) = 1 - \frac{2M}{r} + \frac{Q^2}{r^2} - \Lambda r^2 \, ,\]
$M$ being the mass of the black hole, $Q$ its charge, and $\Lambda >0$ the cosmological constant.

Setting $\Lambda = 0$, we recover the Reissner-Nordström solution while the de Sitter-Schwarzschild family corresponds to $Q =0$.
We also obtain the Schwarzschild metric for $Q=0$, $\Lambda =0$ and Minkowski spacetime when all three parameters $M,Q,\Lambda$ are zero.

The zeros of the function $F$ on the positive real axis correspond to event horizons. Note that
\[ F(r) = \frac{-\Lambda r^4 + r^2 - 2Mr + Q^2}{r^2} \]
and the coefficient of the term in $r^3$ in the numerator is zero. Hence, the sum of the roots of $F$ counted with their multiplicity is zero. It follows that there are at most three event horizons.
The subextremal case corresponds to the presence of exactly three event horizons. In this case, the function $F$ has one negative zero and three distinct positive ones: $r_n< 0 < r_- < r_0 < r_+$, with $r_n= - (r_-+r_0+r_+)$.
The innermost horizon $\{ r=r_-\}$ is the Cauchy horizon, $\{ r=r_0\}$ is referred to as the horizon of the black hole, and $\{ r=r_+\}$ is the cosmological horizon.
The spacetime is made of four distinct regions separated by the horizons:
\begin{itemize}
\item Block I is the region beyond the inner horizon ($0<r < r_-$); it is a static region containing a timelike curvature singularity at $ r=0 $;
\item Block II is the part of spacetime between the black hole and the Cauchy horizons ($r_- < r < r_0$); it is a dynamic region, $\partial_t$ is spacelike;
\item Block III is between the black hole and the cosmological horizons ($r_0 < r < r_+$); we refer to it as the exterior of the black hole; it is static;
\item Block IV is the dynamic region beyond the cosmological horizon ($r > r_+$).
\end{itemize}
Due to the rather large number of horizons which are all bifurcate, the construction of the maximal analytic extension of the subextremal de Sitter-Reissner-Nordström spacetime is intricate. For a clear and thorough description of it, we refer the reader to the recent detailed study by M. Mokdad \cite{M}. The extreme cases correspond to double or triple horizons.
There are three distinct extreme cases corresponding to $r_n< 0 < r_- = r_0 < r_+$, $r_n< 0 < r_- < r_0 = r_+$, and $r_n< 0 < r_- = r_0 = r_+$.
Precise conditions on the parameters $M$, $Q$, and $\Lambda$ characterising the subextremal, extreme, and superextremal (no horizon) cases will be derived in a subsequent work.

For $\Lambda =0$, the function $F$ simplifies to
\[ F(r) = 1 - \frac{2M}{r} + \frac{Q^2}{r^2} \]
and the conditions for extremality, subextremality, and superextremality are easily obtained.
\begin{enumerate}
\item Subextremal case: for $M>|Q|$, the function $F$ has two real roots
\begin{equation}
r_- = M - \sqrt{M^2 -Q^2} \, ,~ r_0 = M+\sqrt{M^2 -Q^2}\, ,
\end{equation}
corresponding to the Cauchy horizon $\{ r =r_- \}$ and the outer horizon, or horizon of the black hole, $\{ r=r_0 \}$. The spacetime is made of three blocks
\[ \bI = \{ 0<r<r_- \} \, ,~ \bII = \{ r_- < r < r_0 \} \, ,~ \bIII = \{ r_0 < r \} \, .\]
\item Extreme case: for $M=|Q|$, $r_0=r_-=M$ is the only root of $F$ and there is only one horizon. Block II is no longer there and the spacetime is made of blocks I and III.
\item Superextremal case: for $M< |Q|$, the function $F$ has no real root. There are no horizons, the space-time contains no black hole and the singularity $\{ r=0 \}$ is naked (\emph{i.e.}\ not hidden beyond a horizon).
\end{enumerate}

\subsubsection{The Regge-Wheeler coordinate}

The Regge-Wheeler coordinate is a function $r_*$ of $r$ such that
\begin{equation}
 \frac{{\mathrm d}r_*}{{\mathrm d}r} = \frac{1}{F(r)} .
\end{equation}
Expressed in coordinates $(t,r_*,\omega)$, the radial null geodesics, which are also the principal null geodesics (their tangent vectors are double roots of the Weyl tensor), are described as the straight lines
\[ t = \pm r_* + C \, ,~ C\in \R \, .\]
Moreover the metric $g$ takes the form:
\begin{equation}
 g = F(r) ({\mathrm d}t^2 - {\mathrm d}r^2_*) - r^2 {\mathrm d} \omega^2 .
\end{equation}
Let us give a more precise description of $r_*$ in the situations we mean to study.
\begin{itemize}
\item {\bf In the subextremal de Sitter-Reissner-Nordström case}, the function $F$ can be written as:
\[F(r) = - \Lambda \frac{(r-r_{n})(r-r_-)(r-r_0)(r-r_+)}{r^2} , \]
with $r_{n} < 0 < r_- < r_0 < r_+$.
The variable $r_*$ is given by
\begin{equation} \label{RstarDSRN}
 r_* = \frac{1}{ \kappa_n} \log \vert r-r_{n} \vert+ \frac{1}{ \kappa_-} \log \vert r-r_{-} \vert + \frac{1}{ \kappa_0}\log \vert r-r_{0} \vert 
 + \frac{1}{ \kappa_+} \log\vert r-r_{+}\vert + R_0 \, ,
\end{equation}
where
\begin{eqnarray*}
\kappa_n = F'(r_n) &=& -\Lambda \frac{(r_{n} - r_-)(r_{n}-r_0)(r_{n} - r_+)}{r_{n}^2 } >0\, ,\\
\kappa_- = F'(r_-) &=& -\Lambda \frac{(r_--r_n)(r_--r_0)(r_--r_+)}{r_-^2} <0 \, , \\
\kappa_0 = F'(r_0) &=& -\Lambda \frac{(r_0-r_n)(r_0-r_-)(r_0-r_+)}{r_0^2} >0 \, ,\\
\kappa_+ = F'(r_+) &=& -\Lambda \frac{(r_+-r_n)(r_+-r_-)(r_+-r_0)}{r_+^2} <0
\end{eqnarray*}
and $R_0$ is a constant of integration. We see that in block III, as $r \rightarrow r_0$, $r_* \rightarrow -\infty$ and as $r \rightarrow r_+$, $r_* \rightarrow +\infty$; $r_*$ is an analytic diffeomorphism from $]r_0,r_+[$ onto $\R$ and $\bIII$ is described in $(t,r_*,\omega)$ coordinates by
\[  \bIII = \mathbb{R}_t \times \Sigma \, ,~ \Sigma =  \mathbb{R}_{r_*} \times \mathbb{S}^2_{\omega}  \, . \]
Moreover,
\[ r-r_0 \simeq e^{\kappa_0 r_*} \mbox{ as } r_* \rightarrow -\infty \, ;~ r_+-r\simeq e^{\kappa_+r_*}  \mbox{ as } r_* \rightarrow +\infty \, . \]

\item {\bf In the subextremal Reissner-Nordström situation},
\[ F(r) = \frac{r^2-2Mr+Q^2}{r^2} = \frac{(r-r_-)(r-r_0)}{r^2} \]
and we have
\begin{equation} \label{RstarRN}
r_* = r +  \frac{1}{\kappa_-} \log \vert r-r_- \vert +\frac{1}{\kappa_0} \log \vert r-r_0 \vert +R_0 \, ,
\end{equation}
with
\[ \kappa_- = F'(r_-) =\frac{r_- -r_0}{r_-^2} <0\, , ~ \kappa_0 = F'(r_0) = \frac{r_0 -r_-}{r_0^2} >0, \]
and $R_0$ a constant of integration. In $\bIII$, as $r \rightarrow r_0$, $r_*$ tends to $-\infty$ and as $r \rightarrow +\infty$, $r_* \rightarrow +\infty$; $r_*$ is an analytic diffeomorphism from $]r_0,+\infty[$ onto $\R$ and $\bIII$ is described in $(t,r_*,\omega)$ coordinates by
\[  \bIII = \mathbb{R}_t \times \Sigma \, ,~ \Sigma =  \mathbb{R}_{r_*} \times \mathbb{S}^2_{\omega}  \, . \]
Moreover,
\[ r-r_0 \simeq e^{\kappa_0 r_*} \mbox{ as } r_* \rightarrow -\infty \, ;~ r\simeq r_* \mbox{ as } r_* \rightarrow +\infty \, . \]
\end{itemize}

\subsection{Charged Klein-Gordon equation, conserved energy current and ergoregion}

In the present article, we shall focus on the evolution of charged scalar fields, in block III of subextremal Reissner-Nordström and de Sitter-Reissner-Nordström black holes, seen by an observer whose perception of time is described by the variable $t$.
Such observers are completely natural in asymptotically flat situations since they are the analytic continuation inside the spacetime of static observers at infinity.
In an asymptotically de Sitter universe, there is no such interpretation. In the cases we consider however, they are natural for a different reason, which is equally valid in the asymptotically flat and asymptotically de Sitter cases: they are associated with the only future-oriented timelike Killing vector field in block III that is orthogonal to a family of spacelike slices ($\partial_t$ in Schwarzschild coordinates).
 
The charged Klein-Gordon equation on our backgrounds is given by
\begin{equation}
\label{CKG1}
 \Box^A_g f + m^2 f = 0 \, ,~ \Box^A_g f = (\nabla_a - i q A_a ) (\nabla^a - i q A^a ) ,
\end{equation}
where $A$ is the electromagnetic potential $1$-form of the spacetime, given by
\[ A_a {\mathrm d}x^a = \frac{Q}{r} {\mathrm d}t \, .\]
Then, \eqref{CKG1} can be explicitly written as
\begin{equation} \label{CKG2}
 \left[\frac{1}{F} \left( \frac{\partial }{ \partial t} - i \frac{q Q}{r} \right)^2 
 - \frac{1}{r^2} \frac{\partial}{\partial r} r^2 F \frac{\partial}{\partial r}
 - \frac{1}{r^2} \Delta_{S^2} + m^2 \right] f = 0 ,
\end{equation}
with $\Delta_{S^2}$ the Laplace-Beltrami operator on $\mathbb{S}^2$.
Putting $\phi = r f$ and using the Regge-Wheeler coordinate $r_*$ defined above, \eqref{CKG2} has the following expression:
\begin{equation}
\label{CKG3}
 \left( \partial^2_t - \partial^2_{r_*} \right) \phi - 2 i \frac{q Q}{r} \partial_t \phi + F(r) \left[ \frac{-\Delta_{S^2}\phi }{r^2} + m^2 \phi + \frac{F'(r)}{r} \phi \right] - \frac{q^2 Q^2}{r^2} \phi = 0 \, .
\end{equation}
As was observed in Ref.~\cite{DiNi2015}, \eqref{CKG1} admits a conserved energy current which can be obtained from a non conserved stress-energy tensor.
Indeed, \eqref{CKG1} is not covariant in the sense that it is not coupled back to the Maxwell equations. This means that there is no natural conserved stress-energy tensor for \eqref{CKG1}. But we can introduce in the stress-energy tensor for the uncharged Klein-Gordon equation a natural modification involving the electromagnetic potential. Putting
\begin{equation}
T_{ab} = \sum_{j=1}^2 \left( \partial_a f_j \partial_b f_j  - \frac{1}{2} g^{cd} \left( \partial_c f_j \partial_d f_j +  q^2 A_c A_d \left( f _j\right)^2 \right) g_{ab} + \frac{m^2}{2} (f_j )^2 g_{ab} \right)  \, ,\label{Stress-Energy}
\end{equation}
where $f_1 = \Re f$ and $f_2 = \Im f$.
We construct an energy current by contracting $T_{ab}$ with $\partial_t$
\begin{eqnarray}
J^a \partial_a := T_0^a \partial_a &=& \sum_{j=1}^2 \left( \partial_t f_j \nabla f_j + \frac12 \left[ -\langle \nabla f_j , \nabla f_j \rangle_g - q^2 \langle A , A \rangle_g f_j^2 + m^2 f_j^2 \right] \partial_t \right) \nonumber \\
&=& \sum_{j=1}^2 \left( \partial_t f_j \nabla f_j + \frac12 \left[ -\langle \nabla f_j , \nabla f_j \rangle_g + (m^2 - F^{-1} \frac{q^2Q^2}{r^2} ) f_j^2 \right] \partial_t \right) \, .\label{Poynting}
\end{eqnarray}
It was established in Ref.~\cite{DiNi2015} that, although $\nabla^a T_{ab} \neq 0$, we have $\nabla^a J_a = 0$ for any solution $f$ of Ref.~\eqref{CKG1} (the proof in Ref.~\cite{DiNi2015} was written for the Reissner-Nordström case but it extends without modification to the de Sitter-Reissner-Nordström framework).

The energy flux across a hypersurface
\[ \Sigma_t := \{ t \} \times \R_{r_*} \times \mathbb{S}^2 \, ,\]
is given by (see Ref.~\cite{DiNi2015} for details)
\begin{eqnarray}
{\cal F}_{\Sigma_t} &=& \frac12 \int_{\Sigma_t} \left( \vert \partial_t f \vert^2 + \vert\partial_{r_*} f \vert^2 + \frac{F}{r^2} \vert \nabla_{S^2} f \vert^2 \right. \nonumber \\
&& \hspace{0.5in} \left. + \left( Fm^2 - \frac{q^2Q^2}{r^2} \right) \vert f \vert^2 \right) r^2 \d r_* \d^2 \omega \nonumber  \\ 
&=& \frac12 \int_{\Sigma_t} \left( \vert \partial_t \phi \vert^2 + \vert \partial_{r_*} \phi \vert^2  + \frac{F}{r^2} \vert \nabla_{S^2} \phi \vert^2 \right. \nonumber \\
&& \hspace{0.5in} \left.+ \left( \frac{FF'}{r} +Fm^2- \frac{q^2Q^2}{r^2} \right) \vert \phi \vert^2 \right) \d r_* \d^2 \omega \, ,\label{EnSigmatphi}
\end{eqnarray}
where
\[ \vert \nabla_{S^2} f \vert^2 = \vert\partial_\theta f \vert^2 + \frac1{\sin^2 \theta} \vert \partial_\varphi f\vert^2 \mbox{ and } \d^2 \omega = \sin \theta \d \theta \d \varphi\, .\]
Across a hypersurface
\[ [t_1 , t_2 ] \times S_R \, ,~ S_R = \{ R \}_{r_*} \times \mathbb{S}^2_{\omega} \, ,\]
the outgoing energy flux has the form
\begin{eqnarray}
{\cal F}_{[t_1 , t_2 ] \times S_R} &=& \int_{[t_1 , t_2 ] \times S_R} \left( -\partial_t f_1 \partial_{r_*} f_1 - \partial_t f_2 \partial_{r_*} f_2 \right)  r^2 \d t\d^2 \omega \nonumber \\ 
&=& - \sum_{j=1}^2 \int_{[t_1 , t_2 ] \times S_R}  \partial_t \phi_j \left( \partial_{r_*} \phi_j - \frac{F}{r} \phi_j \right) \d t \d^2 \omega \, .\label{FluxRSortantphi}
\end{eqnarray}
If we consider a solution of \eqref{CKG1} with given angular momentum $l(l+1)$, $l\in \N$, \emph{i.e.}\ such that $-\Delta_{{\mathbb{S}}^2} f = l(l+1) f$, or equivalently $-\Delta_{{\mathbb{S}}^2} \phi = l(l+1) \phi$, we have
\begin{equation} \label{EnSigmatphil}
{\cal F}_{\Sigma_t} = \frac12 \int_{\Sigma_t} \left( \vert \partial_t \phi \vert^2 + \vert \partial_{r_*} \phi \vert^2  + \left( F \frac{l(l+1)}{r^2} + \frac{FF'}{r} +Fm^2- \frac{q^2Q^2}{r^2} \right) \vert \phi \vert^2 \right) \d r_* \d^2 \omega \, .
\end{equation}
Denoting
\begin{equation} \label{Potentials}
P= F \frac{l(l+1)}{r^2} + \frac{FF'}{r} +Fm^2\, ,~ V = \frac{qQ}{r}\, ,
\end{equation}
the potential $P-V^2$ appearing in \eqref{EnSigmatphil} becomes negative as $r_* \rightarrow -\infty$ in the Reissner-Nordström case and as $r_* \rightarrow \pm \infty$ in the de Sitter-Reissner-Nordström case.
\begin{definition}[Ergoregion]
The ergoregion $\cal E$ is the region where the energy density in \eqref{EnSigmatphi} is allowed to become negative. For a fixed angular momentum $l \in \N$, this is exactly the region where $P$ is negative.
\end{definition}
We see that in the Reissner-Nordström situation, $\cal E$ contains a domain of the form
\[ \R_t \times ]-\infty , R [_{r_*} \times \mathbb{S}^2 \, ,~ R \in \R \, ,\]
while in the de Sitter-Reissner-Nordström case, $\cal E$ contains a domain of the form
\[ \R_t \times \left( ]-\infty , R_1 [_{r_*} \cup ]R_2 , +\infty [_{r_*} \right) \times \mathbb{S}^2 \, ,~-\infty < R_1 < R_2 <+\infty \, .\]
\begin{remark}
The ergoregion could also be defined as the region where the energy density on $\Sigma_t$ expressed in terms of $f$ is allowed to become negative.
In this case it would be characterised for a given angular momentum $l$ as the region where
\[  F \frac{l(l+1)}{r^2} +Fm^2- \frac{q^2Q^2}{r^2} <0 \, .\]
The two definitions are not equivalent but the property immediately above is valid for both.
\end{remark}

\section{Black hole bombs}\label{BHBombs}

The three models of black hole bombs that we discuss in this article are constructed by setting spherical mirrors outside a subextremal Reissner-Nordström or de Sitter-Reissner-Nordström black hole. A mirror is simply described in the evolution system as a homogeneous Neumann (resp. Dirichlet) boundary condition, \emph{i.e.}\ the normal derivative of the field (resp. the field) is set to zero at the surface of the mirror. These two conditions are commonly used when dealing with partial differential equations set on bounded domains. As it will be seen below, they lead to different properties for the corresponding solutions.

\begin{figure}[htb]
\centering
  \begin{tabular}{@{}ccc@{}}
   \hspace{-0.1in}
   \includegraphics[width=.3\textwidth]{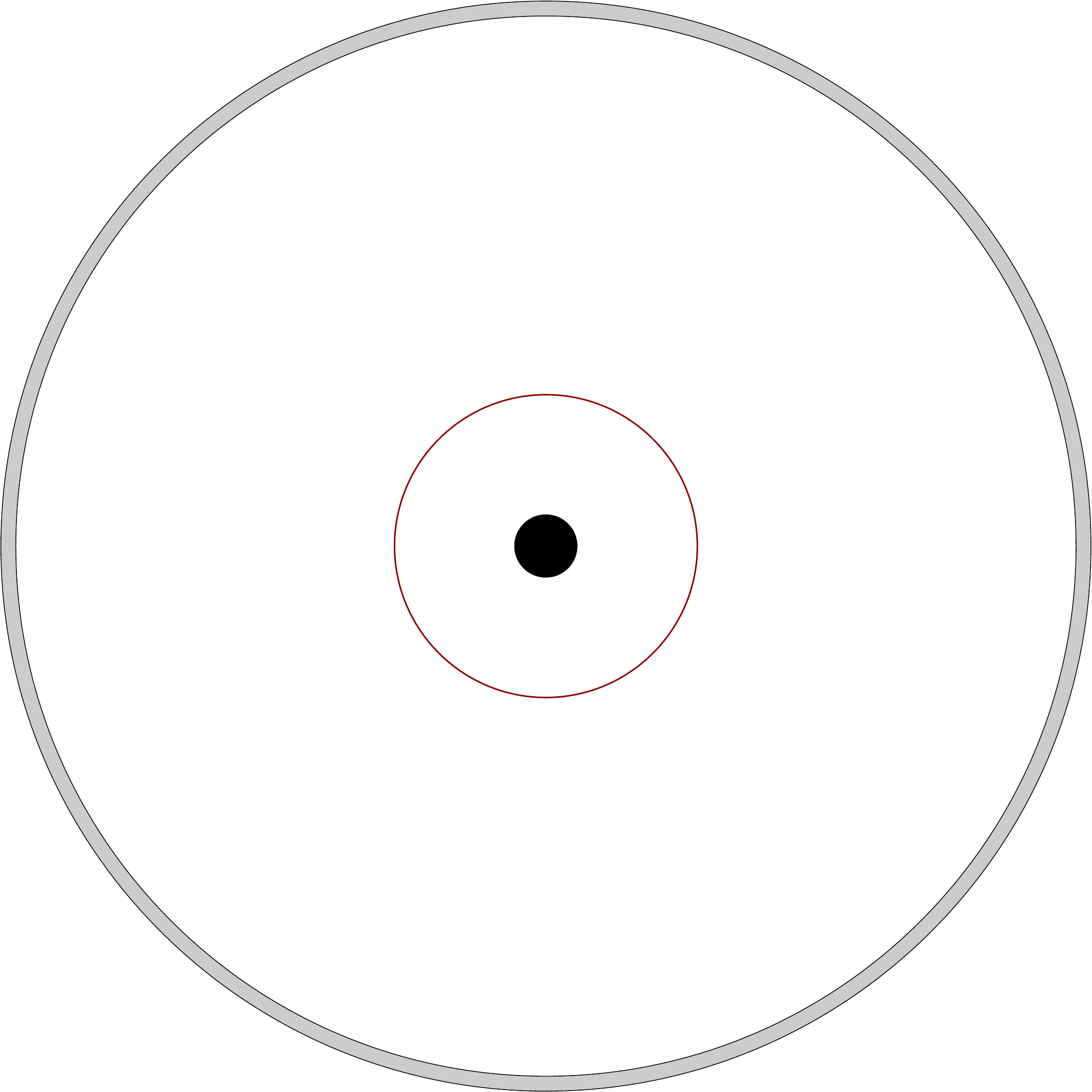} &
   \includegraphics[width=.3\textwidth]{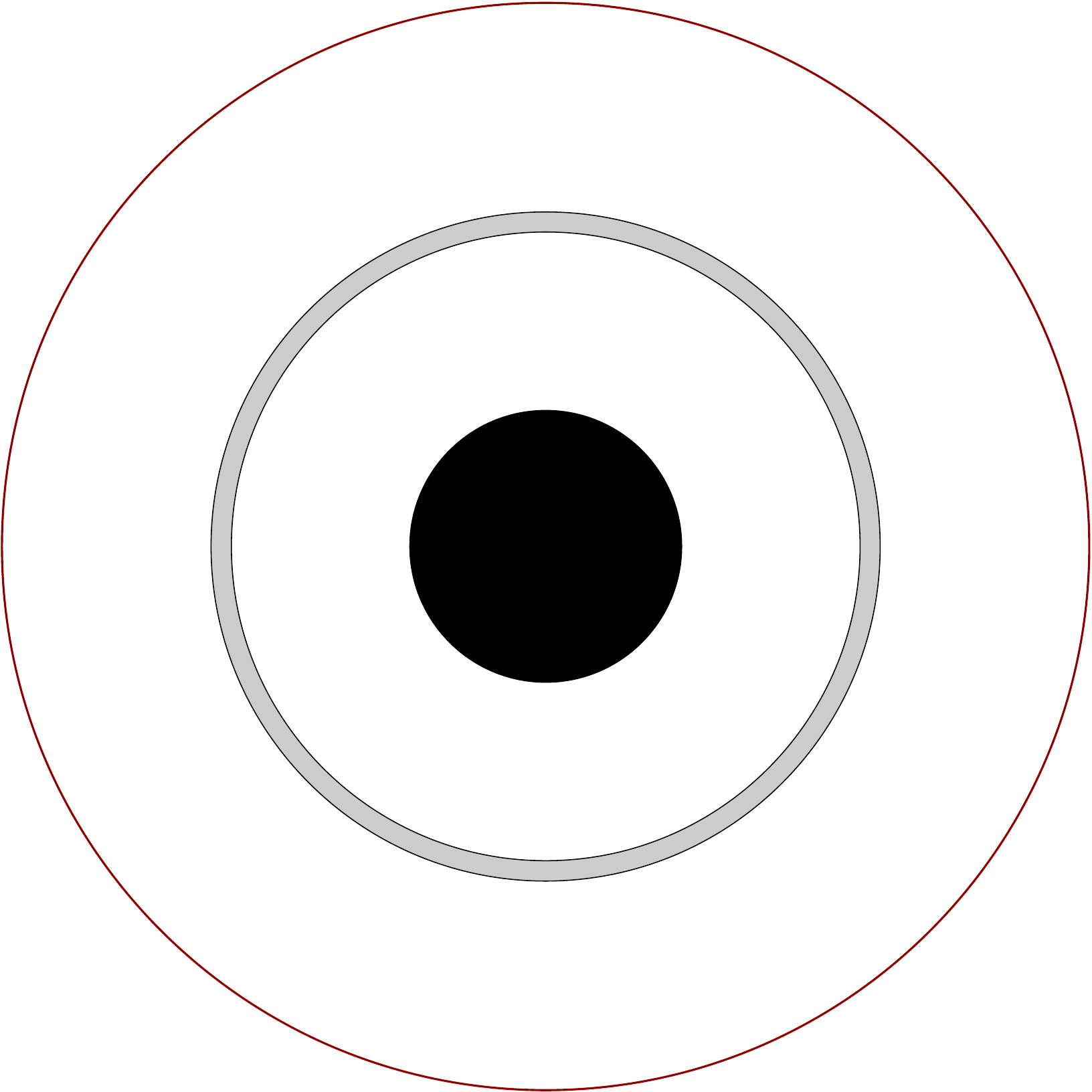} &
   \includegraphics[width=.3\textwidth]{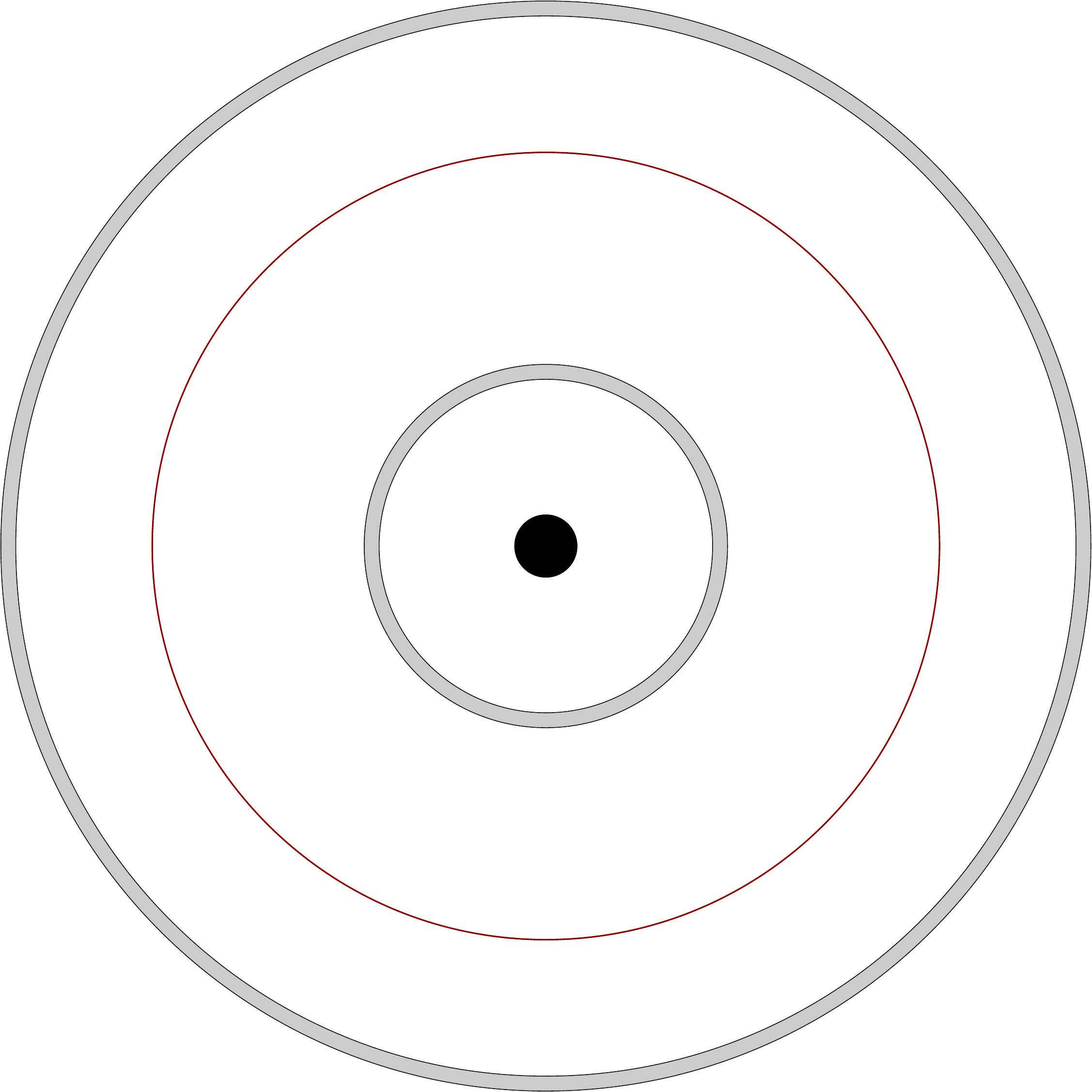}
  \end{tabular}
\caption{The three models of black hole bombs: types I, II, and III from left to right.
The black disc is the black hole horizon and what is beyond, the red circle is the outer boundary of the ergoregion and mirrors are represented as thick grey circles.}\label{fig1}
\end{figure}

We describe each type of bomb with boundary conditions involving an abstract boundary operator $B$: Dirichlet conditions correspond to $Bu=u$ and Neumann conditions to $Bu=\partial_{r_*} u$. We set at $t=0$ some flare type data\footnote{The datum for the field is zero and its time derivative is a smooth compactly supported function.
In our numerical simulations, we choose a Gaussian with some oscillation, which, due to the finite precision of the computer, can be assumed to be
compactly supported.}, entirely contained inside the ergoregion.
Let $R\in \R$ such that $\R_t \times ]-\infty , R [_{r_*} \times \mathbb{S}^2 \subset {\cal E}$.
\begin{itemize}
\item {\bf Type I bombs.} They are directly inspired by the Press-Teukolsky construction \cite{PreTe1972}.
The mirror may surround the whole ergoregion or only a part of it and the field evolves beneath the mirror outside the black hole.
See Figure \ref{fig1} on the left.
Let $R_1 \in \R$, not necessarily lower than $R$, we set the mirror on the hypersurface $\{ r_* = R_1\}$ and we study the boundary initial value problem: 
\begin{equation} \label{EvolBHB1}
\left\{ \begin{array}{l} {\phi \mathrm{~solution~of~}\eqref{CKG3} \mathrm{~on~} \R^+_t \times ]-\infty , R_1[_{r_*} \times \mathbb{S}^2_\omega \, ,} \\ {B\phi \vert_{r_*=R_1} = 0\, ,~\forall\, t\geq 0 \, ,} \\ {\phi \vert_{t=0} = 0 \, ,~ \partial_t \phi \vert_{t=0} = \phi_1 \in {\cal C}^\infty_{c} (]-\infty, \min \{ R,R_1 \}[_{r_*} \times \mathbb{S}^2_\omega) \, ,} \end{array} \right.
\end{equation}
where ${\cal C}^\infty_{c} (\Omega)$ denotes the space of smooth functions on $\Omega$ whose support is a compact subset of $\Omega$. 
\item {\bf Type II bombs.} The mirror is set within the ergoregion so that a part of $\cal E$ lies outside of it.
The field propagates outside of the mirror.
See Figure \ref{fig1} in the centre. Precisely, let $R_1 \in ] -\infty , R[$, we set the mirror on the hypersurface $\{ r_* = R_1\}$ and we study the boundary initial value problem: 
\begin{equation} \label{EvolBHB2}
\left\{ \begin{array}{l} {\phi \mathrm{~solution~of~}\eqref{CKG3} \mathrm{~on~} \R^+_t \times ]R_1,+\infty[_{r_*} \times \mathbb{S}^2_\omega \, ,} \\ {B\phi \vert_{r_*=R_1} = 0\, ,~\forall\, t\geq 0 \, ,} \\ {\phi \vert_{t=0} = 0 \, ,~ \partial_t \phi \vert_{t=0} = \phi_1 \in {\cal C}^\infty_{c} (]R_1,R[_{r_*} \times \mathbb{S}^2_\omega) \, .} \end{array} \right.
\end{equation}
\item {\bf Type III bombs.} There are two mirrors, the inner one being set inside the ergoregion, the outer one may be set inside or outside of the ergoregion.
The field propagates between the two mirrors.
See Figure \ref{fig1} on the right. Let $-\infty < R_1 < R_2 <+\infty$ with $R_1< R$. The mirrors are set on the hypersurfaces $\{ r_* = R_1\}$ and $\{ r_* = R_2\}$ and we study the boundary initial value problem:
\begin{equation} \label{EvolBHB3}
\left\{ \begin{array}{l} {\phi \mathrm{~solution~of~}\eqref{CKG3} \mathrm{~on~} \R^+_t \times ]R_1,R_2[_{r_*} \times \mathbb{S}^2_\omega \, ,} \\ {B\phi \vert_{r_*=R_1} = 0\, ,~B\phi \vert_{r_*=R_2} = 0 \, ,~ \forall\, t \geq 0 \, ,} \\ {\phi \vert_{t=0} = 0 \, ,~ \partial_t \phi \vert_{t=0} = \phi_1 \in {\cal C}^\infty_{c} (]R_1,\min \{ R,R_2\} [_{r_*} \times \mathbb{S}^2_\omega) \, .} \end{array} \right.
\end{equation}
\end{itemize}
By general theorems for hyperbolic equations, the solutions of \eqref{EvolBHB1}, \eqref{EvolBHB2}, and \eqref{EvolBHB3} exist and are unique in the class of functions such that $^\mathrm{t} (\phi \, ,~ \partial_t \phi )$ is continuous on $[0,+\infty[$ with values in the finite energy space on $\tilde{\Sigma}$, where
\[ \tilde{\Sigma} = \left\{ \begin{array}{ll} {]-\infty , R_1]_{r_*} \times \mathbb{S}^2_\omega} & {\mbox{for type I,}}\\ {[R_1,+\infty[_{r_*} \times \mathbb{S}^2_\omega} & {\mbox{for type II,}} \\ {[R_1,R_2]_{r_*} \times \mathbb{S}^2_\omega} & {\mbox{for type III.}} \end{array} \right. \]
Moreover, they are smooth on $\R^+_t \times \tilde{\Sigma}$.
Imposing homogeneous Dirichlet or Neumann boundary conditions as we do
does not alter the divergence-free property of the energy current vector within the 
closed tubes where the solutions live.
In particular, we have the following useful result:
\begin{proposition}
Let $\phi$ be a solution of \eqref{EvolBHB1}, \eqref{EvolBHB2} or \eqref{EvolBHB3}, then the energy flux through the time slice $\tilde{\Sigma}_t = \{ t \} \times \tilde{\Sigma}$:
\[ {\cal F}_{\tilde{\Sigma}_t} = \frac12 \int_{\tilde\Sigma_t} \left( \vert \partial_t \phi \vert^2 + \vert \partial_{r_*} \phi \vert^2  + \left( F \frac{l(l+1)}{r^2} + \frac{FF'}{r} +Fm^2- \frac{q^2Q^2}{r^2} \right) \vert \phi \vert^2 \right) \d r_* \d^2 \omega \, \]
is conserved.
\end{proposition}
\noindent{\bf Proof.--}~This is a simple integration by parts. The homogeneous boundary conditions 
ensure that the boundary terms of the form $\partial_{r_*}\phi \,\partial_t \bar{\phi}$ 
vanish.\qed

\section{Presentation of the numerical strategy}\label{SIM}

In order to investigate the dynamics of our three types of black hole bombs, we solve \eqref{CKG3} using a finite 
differences scheme, in the spirit of the numerical simulations that have been performed in Ref.~\cite{DiNi2015}.
Working with solutions of given angular momentum, \eqref{CKG3} reduces to a partial differential equation in $1+1$-dimensions:
\begin{equation} \label{CKGl}
\left( \partial^2_t - \partial^2_{r_*} \right) \phi - 2 i V \partial_t \phi + (P-V^2) \phi =0 \, ,
\end{equation}
where $P$ and $V$ are given by \eqref{Potentials}.
We discretise it from a new vectorial formulation of \eqref{CKGl}
using a semi-implicit method, which requires at each time step an inversion procedure that is moderate in terms of CPU times and increases the stability of the scheme. We describe the numerical method in some details before presenting the results of our simulations in the next section.
Our purpose in this article is not to present a systematic study of the three types of black hole bombs, but rather to point out that different configurations from the one usually considered do lead to linear instabilities.
The results of our simulations for the new type II bombs are presented first.
Then we describe the results for the types I and III bombs. In this section and the next, the variable $r_*$ will also be denoted by $x$. 

\subsection{Reformulation of the initial problem}
We begin by expressing (\ref{CKGl}) as the first-order linear evolution system
\begin{equation}
\partial _t \left(
\begin{array}{l}
u\\v
\end{array}
 \right)
-
\left( 
\begin{array}{cc}
iV & 1\\ \partial _x^2 -P & iV
\end{array}
\right)
 \left(
\begin{array}{l}
u\\v
\end{array}
 \right)
=
 \left(
\begin{array}{l}
0\\0
\end{array}
 \right)
\label{syst}
\end{equation}
by setting $u:=\phi$ and $v:=(\partial _t - i V)u$. This means that we deal with a
first-order system instead of a second-order equation. Consequently, we calculate
the two quantities $u$ and $v$ as independent functions even if $v$ is expressed in terms
of $v$. This is a classical strategy that is intensively used for second-order
differential equations. 

Since we intent to deal with
a numerical scheme relevant for the simulation of superradiance, it is essential
to focus on the problem of energy conservation. Indeed, a given scheme does not
necessarily mimic the conservation properties of the exact solution of the
problem.

We first reformulate the energy evaluated on the space-like slice 
$\left\{t\right\}\times \R$ in terms of $u$ and $v$, starting from the Cauchy
data $(u(0),v(0))=(u_0,v_0)$.
\begin{theorem}\label{energcons}
Let $(u,v)$ be solutions of \eqref{syst}. 
For $t\ge 0$, one has the energy conservation $E(u(t),v(t))=E(u_0,v_0)$, with 
\begin{equation}
E(u,v)=\frac{1}{2}\int_\R \Big( |v|^2 +|\partial_x u|^2 + P|u|^2\Big)\,{\mathrm d}x
+ \int_\R\Im (V\bar{u}v)\,{\mathrm d}x.
\label{venerg}
\end{equation}
\end{theorem}
\noindent
{\bf Proof.--}~We first multiply the second equation of \eqref{syst} by $\bar{v}$ and integrate in space. 
We obtain
$$
\int_\R \Big(\partial _tv \bar{v} - iV |v|^2 \Big) \,{\mathrm d}x
=\int_\R \Big(\partial_x^2u - Pu\Big)\bar{v}\,{\mathrm d}x
=\int_\R \Big((\partial_x^2u - Pu)(\partial_t \bar{u}+iV\bar{u})\Big)\,{\mathrm d}x
$$
when $v$ is expressed in terms of $u$. Taking the real part gives
\begin{eqnarray*}
\frac{1}{2}\frac{\d}{\d t}
\int_\R |v|^2 \,{\mathrm d}x &=&\Re \int_\R \Big(\partial_x^2u \partial _t\bar{u} +
iV (\partial_x^2 u) \bar{u} -Pu\partial_t \bar{u} \Big)\,{\mathrm d}x\\
&=&
\Re \int_\R \Big(\partial_x^2u \partial _t\bar{u}-Pu\partial_t \bar{u} \Big)\,{\mathrm d}x-
\int_\R \Im (V (\partial_x^2 u) \bar{u}) \,{\mathrm d}x
\end{eqnarray*}
after cancellation of all purely imaginary contributions. 
This gives three integral terms. The first two are rewritten in terms of
time derivatives, when using the well-known transformations
$$
\begin{array}{l}
\displaystyle
\Re \int_\R \partial_x^2u\partial_t \bar{u}\,{\mathrm d}x
=-\Re \int_\R \partial_x u\partial_{xt}^2 \bar{u}\,{\mathrm d}x
=-\Re \int_\R \partial_x u\partial_t (\partial_x \bar{u})\,{\mathrm d}x
=-\frac{1}{2}\frac{\d}{\d t}\int_\R |\partial_x u|^2\,{\mathrm d}x,\\
\vspace*{-0.2cm}\\
\displaystyle
\Re \int_\R \Big(Pu\partial_t \bar{u} \Big)\,{\mathrm d}x=
\frac{1}{2}\Re \int_\R \partial_t \Big(P|u|^2 \Big)\,{\mathrm d}x= \frac{1}{2}\frac{\d}{\d t}\int_\R
P|u|^2 \,{\mathrm d}x.
\end{array}
$$
We transform the last contribution using $\partial_x^2u=Pu+\partial_tv-iVv$. From this, we 
obtain 
$$
\displaystyle
\Im  \int_\R V \partial_x^2 u \bar{u}\,{\mathrm d}x=
\Im  \int_\R \Big(PV |u|^2 + \partial_t v V\bar{u}-iV vV\bar{u}\Big)\,{\mathrm d}x
=\Im  \int_\R \Big(\partial_t v V\bar{u}-iV vV\bar{u}\Big)\,{\mathrm d}x.
$$
Using $-iV\bar{u}=\partial_t \bar{u}-\bar{v}$, we obtain
$$
\Im  \int_\R -iV vV\bar{u}\,{\mathrm d}x
=\Im  \int_\R \Big(\partial_t\bar{u}Vv -V|v|^2\Big)\,{\mathrm d}x
=\Im  \int_\R \partial_t\bar{u}Vv \,{\mathrm d}x.
$$
This leads to
$$
\Im  \int_\R V \partial_x^2 u \bar{u}\,{\mathrm d}x
=\Im  \int_\R V\Big(\partial_t\bar{u}\,v + \bar{u}\partial_t v \Big)\,{\mathrm d}x
=\Im  \int_\R V\partial_t(\bar{u}v)\,{\mathrm d}x=
\frac{d}{dt}\left(  \int_\R\Im (V\bar{u}v)\,{\mathrm dx}\right).
$$
\noindent
From this, we finally conclude that 
$$
\frac{\d}{\d t}\left(\frac{1}{2}
\int_\R \Big( |v|^2+|\partial_x u|^2 +P|u|^2\Big)\,{\mathrm dx} + \int_\R\Im (V\bar{u}v)\,{\mathrm dx}
\right)=0
$$
which completes the proof. \qed

Note that this conservation can easily be seen as the consequence
of the conservation of the initial form of the energy flux across $\Sigma_t$ for a given angular momentum
$$
{\cal F}_{\Sigma_t} =2\pi \int_\R  \Big(|\partial_t \phi|^2 + |\partial_x \phi|^2 +(P-V^2)|\phi|^2\Big) \,{\mathrm dx} \, ,
$$ 
when formally replacing $\partial_t \phi=v+iV\phi=v+iVu$. Nevertheless, it is a crucial point to
obtain the energy conservation from the vectorial formulation of the problem that
will inspire the proof of conservation for the discrete energy of the numerical scheme, 
as we shall see later.  

These considerations have been made for the problem set on the whole real line. If we now
assume that we deal with a bounded domain $\Omega=\mathopen]0,L[$, we have to specify a 
boundary condition for $u$ at the boundary of $\Omega$. We then set 
$$E_L(u,v) :=
\frac{1}{2}
\int_0^L \Big( |v|^2+|\partial_x u|^2 +P|u|^2\Big)\,{\mathrm dx} + \int_0^L
\Im (V\bar{u}v)\,{\mathrm dx}
$$
that is simply the local energy computed on $\Omega$. The energy conservation depends on
the kind of boundary condition set at the frontier of $\Omega$.

Prescribing either homogeneous Dirichlet conditions 
$u(t,0)=u(t,L)=0$ for $t\ge 0$ (meaning that $\partial_t u=0$ at $x=0$ and $x=L$) or homogeneous
Neumann conditions $\partial_x u(t,0)=\partial_x u(t,L)=0$ for $t\ge 0$, we
are now faced with a mixed problem, including both Cauchy data and boundary conditions. 
We then have 

\begin{theorem}\label{energcons2}
Let $(u,v)$ be solutions of \eqref{syst} computed on $]0,L[$ with Dirichlet or Neumann
conditions on the boundary. 
For $t\ge 0$, one has the energy conservation $E_L(u(t),v(t))=E_L(u_0,v_0)$.
\end{theorem}

\noindent
{\bf Proof.--}~The computations remain the same, except in the 
integration by parts on 
$\Omega$, which becomes
$$
\Re \bigg(\int_0^{L} \partial_x^2u\partial_t \bar{u}\,{\mathrm d}x\bigg)
=\Re \bigg(-\int_0^L \partial_x u\partial_{xt}^2 \bar{u}\,{\mathrm d}x+
\Big[\partial_x u\partial_{t} \bar{u}\Big]_0^{L}\bigg)
=-\frac{1}{2}\frac{d}{dt}\int_0^L |\partial_x u|^2\,{\mathrm d}x
$$ 
since the product $\partial_x u\partial_{t} \bar{u}$ 
vanishes in both cases at the boundary. Of course,
this conservation fails when considering other kinds of boundary conditions such
as the Robin condition $\partial_n u=\alpha u$ ($\partial_n$ denoting the normal 
derivative computed at the boundary) that would give rise to a boundary
contribution.

\subsection{Description of the numerical method}

It is possible to perform a symmetric semi-implicit time discretisation,
aiming at calculating the value of the solutions $(u^n,v^n)=(u(t_n,.),v(t_n,.))$
at discrete times 
$t_n=n\,\delta t$, where $\delta t$ stands for the time step.
The key point is to write that (\ref{syst}) is satisfied at each 
$t_{n+1/2}=(t_n+t_{n+1})/2$. Using the usual approximations
$(u^{n+1}-u^n)/\delta t$ and $(v^{n+1}-v^n)/\delta t$
for time derivatives $\partial_t u$ and $\partial_t v$, 
combined with the second order approximations
$(u^n+u^{n+1})/2$ and $(v^n+v^{n+1})/2$ of
$u(t_{n+1/2},.)$ and $v(t_{n+1/2},.)$, leads to
\begin{gather*}
\left( 
\begin{array}{cc}
\displaystyle
\Big(1 - iV \frac{\delta t}{2}\Big) \mathrm{Id}  & 
\displaystyle-\frac{\delta t}{2}\mathrm{Id} \\
\vspace*{-0.2cm}\\
 -\displaystyle
\frac{\delta t}{2}(\partial _x^2-P) & 
\displaystyle
\Big(1 - iV \frac{\delta t}{2}\Big) \mathrm{Id} 
\end{array}
\right)
 \left(
\begin{array}{l}
u^{n+1}\\ \\ \\v^{n+1}
\end{array}
 \right) \hspace{2in} \\
\hspace{1in}=
\left( 
\begin{array}{cc}
\displaystyle
\Big(1 + iV \frac{\delta t}{2}\Big) \mathrm{Id} & 
\displaystyle \frac{\delta t}{2}\mathrm{Id}\\
\vspace*{-0.2cm}\\
\displaystyle
\frac{\delta t}{2}(\partial _x^2-P) & 
\displaystyle
\Big(1 + iV \frac{\delta t}{2}\Big) \mathrm{Id}
\end{array}
\right)
 \left(
\begin{array}{l}
u^{n}\\ ~\\~\\v^{n}
\end{array}
 \right) .
\end{gather*}

We now discretise the bounded spatial domain on which we perform our 
simulation, with a 
uniform space step $h$, \emph{i.e.}\ $x_j = x_0 + jh$, $ j=0,...,J+1$. At each point 
$x_j$, $1\le j\le J$, we calculate the second derivative with the classical 
$3$ point approximation: for a given function, say $f=f(x)$, we have 
\begin{equation}
\partial _x^2 f(x_j)\approx\frac{1}{h^2}(f(x_{j+1})-2f(x_j)+f(x_{j-1}))
\label{DiscDiff}
\end{equation}
up to a second order term in $h$. Finally, we obtain the following discretisation of \eqref{syst}
as
\begin{eqnarray}
\frac{1}{\delta t}(u_j^{n+1}-u_j^n) - iV_j u_j^{n+1/2}&=& v_j^{n+1/2} 
\label{sch1}\\
\frac{1}{\delta t}(v_j^{n+1}-v_j^n) - iV_j v_j^{n+1/2}&=& 
\frac{1}{h^2}(u_{j+1}^{n+1/2}-2u_j^{n+1/2}+u_{j-1}^{n+1/2})-P_ju_j^{n+1/2} 
\label{sch2}
\end{eqnarray}

For the applications we have in mind, we have to take into account the presence of one 
mirror (for types I and II bombs) or two mirrors (for type III bombs).
The mirror, or mirrors, will be set at the extremities of $\Omega$ and 
boundary conditions have to be prescribed at one or both ends of the interval. 

When using first Dirichlet 
homogeneous boundary conditions, the value of the solution is set to 
vanish at $j=0$ and $j=J+1$, corresponding to $x=0$ and $x=L$. 
The classical approximation for the second-order
space derivative expressed at $j=1$ reduces to
\begin{equation}
\partial_x^2f(x_1)\approx \frac{1}{h^2}(f(x_2)-2f(x_1)),
\label{DiffDBC}
\end{equation}
plugging $f(x_0)=0$ in the three-point formula \eqref{DiscDiff}.
Consequently, we compute at each time
increment the unknown vectors
$U^n=(u_1^n,\ldots,u_J^n)^T\in \C^J$ and $V^n=(v_1^n,\ldots,v_J^n)^T\in 
\C^J$ as the solutions of the linear system
\begin{equation}
\left( 
\begin{array}{rr}
A & -B\\
 -C & 
A
\end{array}
\right)
 \left(
\begin{array}{l}
U^{n+1} \\V^{n+1}
\end{array}
 \right)
=
\left( 
\begin{array}{rr}
D & B\\
C & 
D
\end{array}
\right)
\left(
\begin{array}{l}
U^{n}\\V^{n}
\end{array}
 \right),
\label{systdis}
\end{equation}
where $A = (1 - iV \frac{\delta t}{2}) \mathrm{Id}$, $B=\frac{\delta t}{2}\mathrm{Id}$,
$C=\frac{\delta t}{2}(\Delta_2-P)$, $D = (1 + iV \frac{\delta t}{2}) \mathrm{Id}$, and $\Delta_2$ denotes the tridiagonal
matrix obtained from (\ref{DiscDiff}).

\begin{figure}[ht!]
\begin{center}
\begin{minipage}{15.cm}
 \includegraphics[width=15cm,height=6cm]{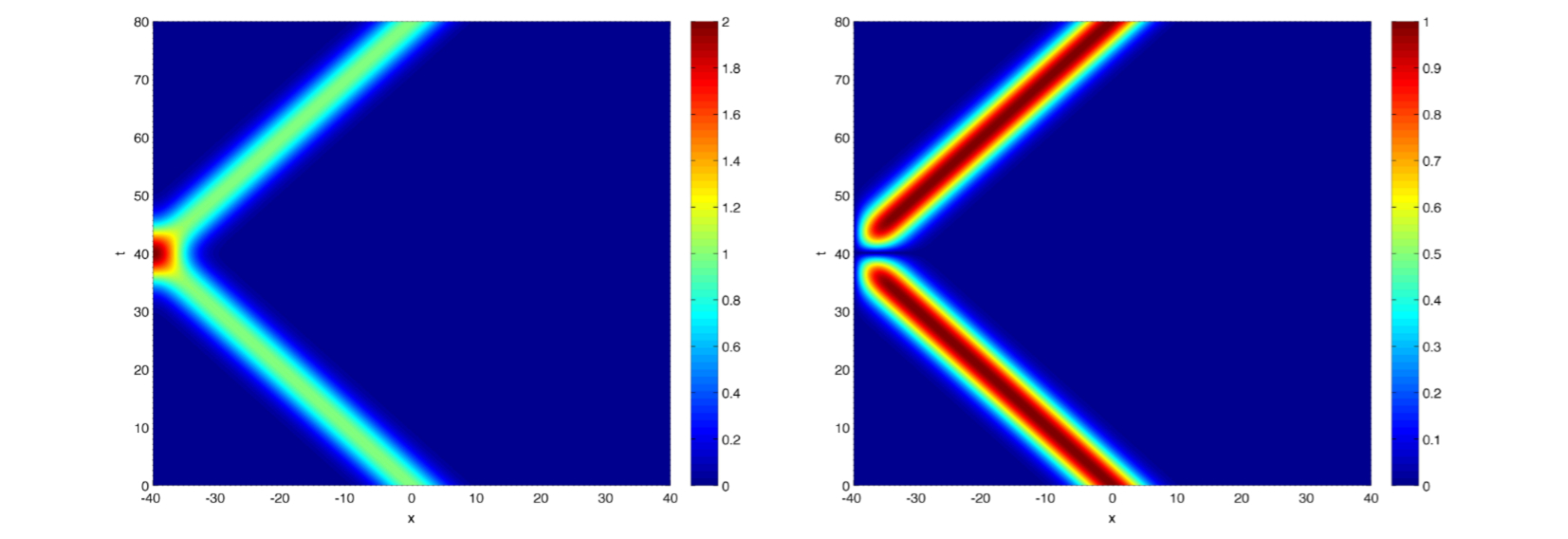}
\caption{{\small Reflection of a left-propagating wave at a mirror located at the left extremity of  the computational domain: homogeneous Neumann boundary conditions (left), homogeneous Dirichlet boundary conditions (right).}}
\label{fig2}
\end{minipage}
\end{center}
\end{figure}

If we now deal with Neumann
condition $\partial_x u=0$, it means that the value of the solution is not prescribed at
the boundary. For the same space step value as before, we have to compute the solution at
all the spatial points $x_0$, $\ldots$, $x_{J+1}$ and the vectors to be determined 
are now $U^n=(u_0^n,\ldots,u_{J+1}^n)^T\in \C^{J+2}$ and $V^n=(v_0^n,\ldots,v_{J+1}^n)^T\in 
\C^{J+2}$. For example at the left extremity, the second-order
Taylor expansion writes
$$
f(x_1)=f(x_0)+h\partial_xf(x_0)+\frac{h^2}{2}\partial_x^2f(x_0)+{\mathcal O}(h^3),
$$
which leads to the approximation
\begin{equation}
\partial_x^2f(x_0)\approx \frac{2}{h^2}(f(x_1)-f(x_0))
\label{DiffNBC}
\end{equation}
since $\partial_xf(x_0)=0$. Consequently, the new discrete system to be solved still
has the form \eqref{systdis}, but it now involves $(J+2)\times (J+2)$ matrices, where the
approximation matrix $\Delta_2$ of the second-order derivative, taking into account the
Neumann boundary condition at each extremity, now reads
$$
\Delta_2=\frac{1}{h^2}\left[\begin{array}{rrrrrrrrr}
-2 & 2 & 0 & \ddots  \\
1 & -2 & 1 & 0 &\ddots \\
0 & 1 & -2 & 1 & 0 &\ddots\\

 & \ddots & \ddots & \ddots & \ddots & \ddots \\
& \ddots & 0 & 1 & -2 & 1\\
&& \ddots & 0 & 2 & -2
\end{array} \right].
$$
Let us point out that dealing with Neumann conditions affects the computation of the
second derivative of $u$ at the boundary whereas for Dirichlet conditions, we set
$u_0^n=u_{J+1}^n=0$.
The effect of these two kinds of boundary conditions is shown in Figure \ref{fig2}.

In cases where only one mirror is considered, we 
impose outgoing Sommerfeld boundary conditions at the other end of the computational domain.
These conditions are only approximately absorbing, but we make sure that this other extremity is sufficiently far away so that before the final time $T$, the solution may reach it but the spurious reflected wave will not have time to come back to the points at which we measure the amplitude of the solution and the energy flux. 

\begin{figure}[ht!]
\begin{center}
\begin{minipage}{7.cm}
\includegraphics[width=7cm,height=6.cm]{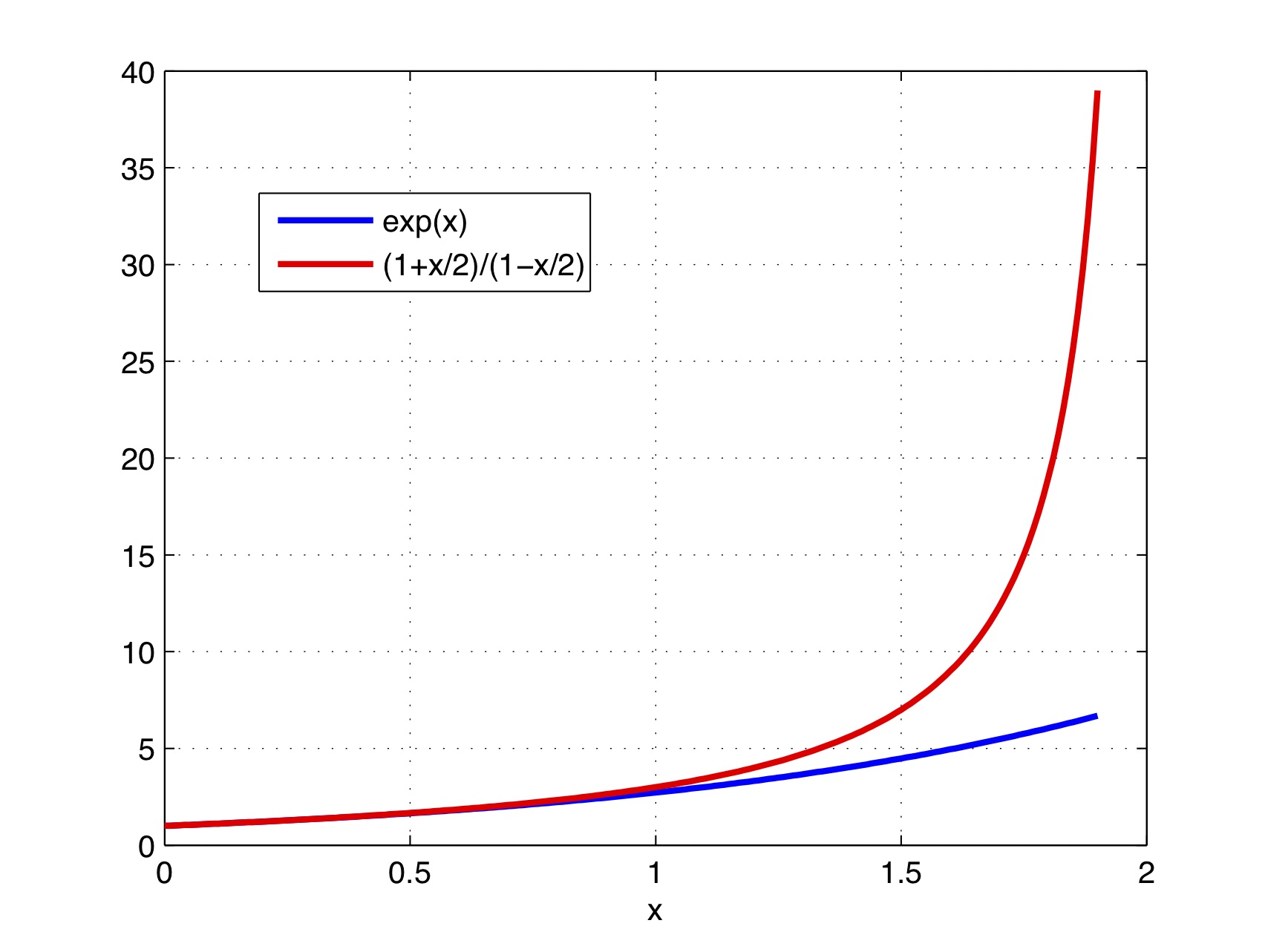}
\vspace*{-0.9cm}
\caption{Plots of the two functions $e^x$ and $\frac{1+x/2}{1-x/2}$ for $x\in [0,2[$.}
\label{fig3}
\end{minipage}
\hspace*{0.5cm}
\begin{minipage}{7.cm}
\includegraphics[width=7cm,height=6.cm]{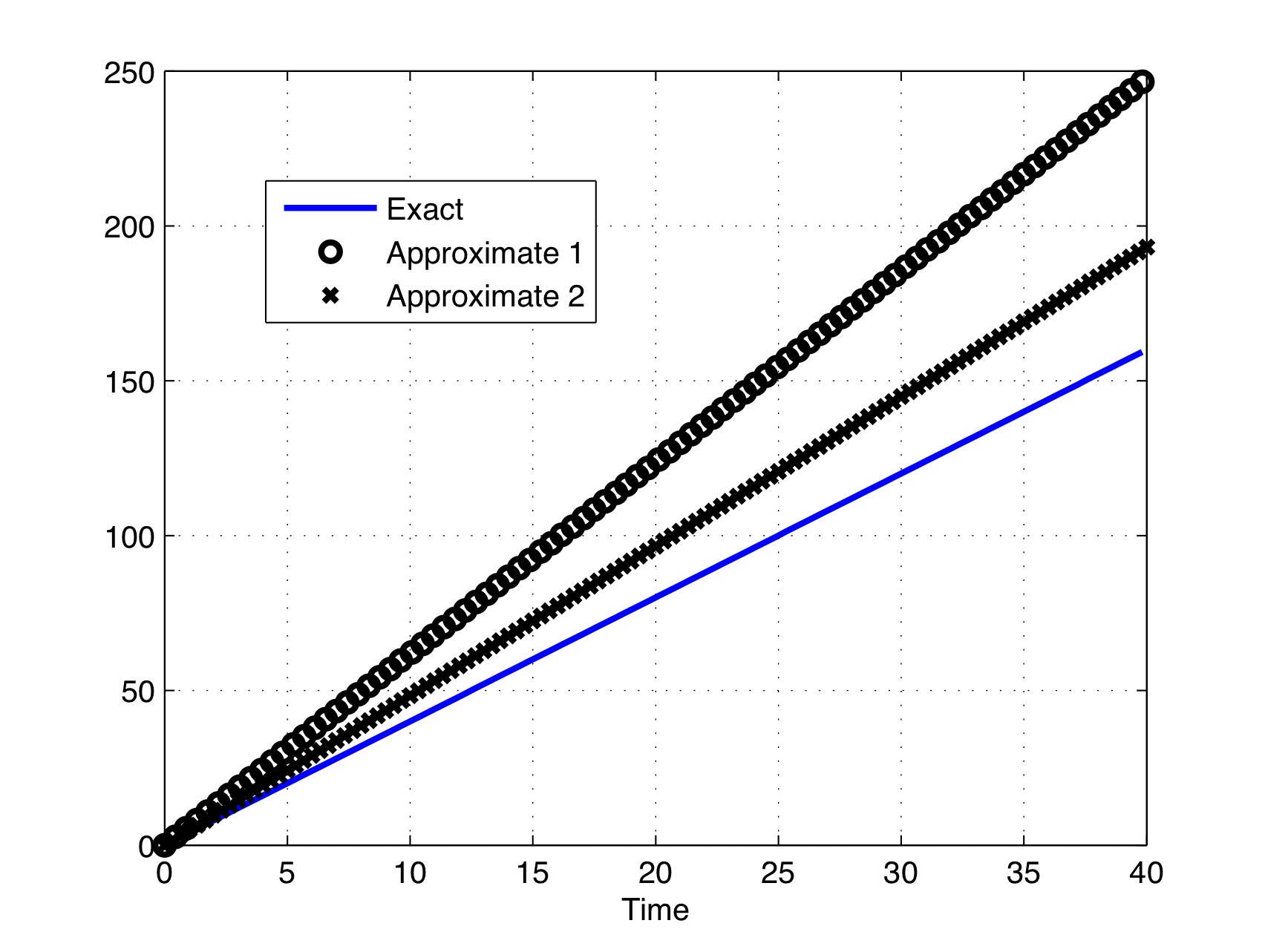}
\vspace*{-0.9cm}
\caption{log of the solution: exact profile and
profiles computed with $x=7/4$, $x=3/2$.}
\label{fig4}
\end{minipage}
\end{center}
\end{figure}

We aim at observing solutions whose amplitude increases exponentially in time.
This requires choosing a well-adapted time step. As an illustration of this, let us 
consider the very simple first-order differential equation 
$y'(t)=\alpha y(t)$ ($t\ge 0$), with initial data $y(0)=y_0$ and where
$\alpha>0$. 
In this case, the exact solution $y(t)=e^{\alpha t}y_0$ grows exponentially. Looking for an approximate solution
$y^n\approx y(t_n)$ computed by means of the time discretisation given above, we obtain the following numerical scheme
\[\Big(1-\frac{\alpha \delta t}{2}\Big) y^{n+1}= 
\Big(1+\frac{\alpha \delta t}{2}\Big) y^{n}\, .\]
The value of the approximate solution at computational time $T=t_N = N\delta t$ is explicitly given by
$$
y^N=\left(\frac{1+\frac{\alpha \delta t}{2}}
{1-\frac{\alpha \delta t}{2}}\right)^N y_0
$$
which turns out to be an approximation of $e^{\alpha T}y_0$. By Taylor
expansion with respect to $\delta t$, one finds that 
\begin{eqnarray*}
\frac{y^N -y(T)}{y(T)}&=&\exp \Bigg(N \left(\log \left( 1+\frac{\alpha \delta t}{2} \right) - \log \left( 1-\frac{\alpha \delta t}{2}\right)\right) -\alpha T \Bigg) -1 \\
&=& \exp \Big(\frac{1}{12}\alpha^3 
T\delta t^2+ {\cal O}(\delta t^3)\Big) -1 \\
&\simeq& \frac{1}{12}\alpha^3 
T\delta t^2 \mbox{ for } T \mbox{ fixed and } \delta t \rightarrow 0 \, .
\end{eqnarray*}
The numerical growth rate differs from the exact one, and the relative error at time $T$ is of magnitude $\frac{1}{12}\alpha^3 
T\delta t^2$. When dealing with a given final time $T\gg 1$ in order to 
study long time asymptotics, one has to prescribe a time step in such a
way that $\frac{1}{12}\alpha^3 
T\delta t^2\ll 1$. One can notice that the accuracy of the profile is 
governed by the value of $\alpha \delta t=:x$, which is involved in the
two factors $e^{\alpha \delta t}$ and $(1+\frac{\alpha \delta t}{2})/(
1-\frac{\alpha \delta t}{2})$. A plot of the two corresponding functions
is given with respect to $x$ in Figure \ref{fig3}, clearly showing the 
different behaviour as $x$ becomes large.
Furthermore, we plot in Figure \ref{fig4} the logarithm of the time evolution 
of the solution, comparing the exact profile with two numerical values 
considered with the two
cases $x=7/4$ and $x=3/2$. In each case, there is an amplification of the
exponential growth that appears to be larger for large $x$. As these experiments clearly show, the value of $\delta t$ has to be moderate enough in 
order to measure accurately the exponential growth of the solution.

\vspace*{0.2cm}

We shall observe the linear instability caused by the presence of the mirror, not only by measuring the amplitude of the wave, but also by computing the outgoing energy gain through a fixed sphere. In order to do so in a meaningful manner, we have to be careful with the way we evaluate
the energy in the discrete case. In what follow, we present a discrete version of
the energy that is numerically conserved and grants the validity of our numerical
strategy.

\subsection{Discrete energy conservation on the whole space}

We first deal with \eqref{systdis} that is written for all indices $j\in \mathbb Z$, meaning
that we consider the discrete values $(u^n,v^n)=(u_j^n,v_j^n)_{j\in \Z}$. Thus, we study
the energy conservation of the theoretical scheme, that differs from the one involving boundary
conditions that will be implemented for numerical simulations. 

\begin{theorem}\label{discons}
For approximate solutions of \eqref{syst} 
given by \eqref{sch1}-\eqref{sch2}, the quantity
$$
E_n=\frac{1}{2}\sum_{j\in {\mathbb Z}}\Big(
|v_j^n|^2 + \Big|\frac{u_{j+1}^n-u_j^n}{h}\Big|^2 + P_j |u_j^n|^2 \Big)
+\sum_{j\in {\mathbb Z}}\Im(V_j\bar{u}_j^nv_j^n)
$$
is conserved for each index $n$.
\end{theorem}
Before proving this result, let us first notice that this discrete version of
energy is nothing but a discretisation of the continuous energy given by \eqref{venerg}
using an upwind discretisation of the space derivative of $u$. This shows that 
this energy is consistent with the exact one.

\vspace*{0.2cm}

\noindent
{\bf Proof.--}~We follow the exact same way of deriving the energy in terms of $u$ ad $v$
in the continuous case, as shown in the proof of Theorem \ref{energcons}. We
first multiply \eqref{sch2} by $\bar{v}_j^{n+1/2}$, which gives
\begin{gather*}
\frac{1}{2\delta t}\Big(v_j^{n+1}-v_j^n\Big)\Big(\bar{v}_j^{n+1}+\bar{v}_j^n\Big)
-iV_j|v_j^{n+1/2}|^2 \hspace{1in}\\
\hspace{1in} = \left(\frac{1}{h^2}\Big( u_{j+1}^{n+1/2}-2u_j^{n+1/2}+u_{j-1}^{n+1/2}\Big)\!\!-P_j u_j^{n+1/2}
\right)\!\bar{v}_j^{n+1/2}.
\end{gather*}
Taking the real part gives us
\begin{equation}
\frac{1}{2\delta t}\Re
\Big(\big(v_j^{n+1}-v_j^n\big)\Big(\bar{v}_j^{n+1}+\bar{v}_j^n\Big)
\Big)=\alpha_j^n-\beta_j^n ,
\label{first}
\end{equation}
with
$$
\begin{array}{lll}
\alpha_j^n &= &
\displaystyle \frac{1}{h^2}\Re\Big(
\big( u_{j+1}^{n+1/2}-2u_j^{n+1/2}+u_{j-1}^{n+1/2}\big)\bar{v}_j^{n+1/2}\Big)\\
&=& \displaystyle\frac{1}{h^2}\Re\Big(
\big( u_{j+1}^{n+1/2}-2u_j^{n+1/2}+u_{j-1}^{n+1/2}\big)
\big(\frac{1}{\delta t}\big( \bar{u}_j^{n+1}-\bar{u}_j^n\big)+iV_j \bar{u}_j^{n+1/2}\big)
\Big),
\end{array}
$$
with $\bar{v}_j^{n+1/2}$ being expressed in terms of $\bar{u}_j^{n+1}$, $\bar{u}_j^n$, 
$\bar{u}_j^{n+1/2}$ and similarly
$$
\beta_j^n=\Re\Big(P_j
\Big(\frac{1}{\delta t}
\big( \bar{u}_j^{n+1}-\bar{u}_j^n\big)u_j^{n+1/2}+iV_j |u_j^{n+1/2}|^2\Big)\Big)
=\Re\Big(P_j
\Big(\frac{1}{\delta t}
\big( \bar{u}_j^{n+1}-\bar{u}_j^n\big)u_j^{n+1/2}\Big)\Big).
$$
We now give two technical lemmas that will be useful to simplify the
computations. The first one can be seen as the discrete version of the relation
$\Re(\partial_t v \bar{v})=\frac{1}{2}\partial_t |v|^2$:
\begin{lemma}
One has
\begin{equation}
\displaystyle
\frac{1}{2}\Re\Big((v_j^{n+1}-v_j^n)(\bar{v}_j^{n+1}+\bar{v}_j^n)\Big)=
\frac{1}{2}\Big(|v_j^{n+1}|^2-|v_j^n|^2\Big).
\label{transf}
\end{equation}
\label{lem1}
\end{lemma}
\noindent
{\bf Proof.--}~Developing the proposed expression simply leads us to
$$
\Re\Big((v_j^{n+1}-v_j^n)(\bar{v}_j^{n+1}+\bar{v}_j^n)\Big)
=\Re(v_j^{n+1}\bar{v}_j^{n+1}-v_j^n\bar{v}_j^{n+1}+v_j^{n+1}\bar{v}_j^n-
v_j^n\bar{v}_j^n)= |v_j^{n+1}|^2 -|v_j^{n}|^2
$$
since $v_j^n\bar{v}_j^{n+1}$ and $v_j^{n+1}\bar{v}_j^n$ are complex-conjugated. The lemma is proved. \qed

\vspace*{0.2cm} 

The second lemma can be interpreted as the discrete version of
an integration by parts:
\begin{lemma}
One has
\begin{equation}
\Re \Big(\sum_{j\in {\mathbb Z}}(u_{j+1}^{n+1/2}-2u_j^{n+1/2}+u_{j-1}^{n+1/2})
(\bar{u}_j^{n+1}-\bar{u}_j^n)
\Big)=-\frac{1}{2}\sum_{j\in {\mathbb Z}}\left(\big|u_{j+1}^{n+1}-u_j^{n+1}\big|^2 
-\big|u_{j+1}^{n}-u_j^{n}\big|^2\right).
\label{ipp}
\end{equation}
\label{lem2}
\end{lemma}
\noindent
{\bf Proof.--}~We have for each index $j$
$$
\begin{array}{lll}
 &\Re\Big((u_{j+1}^{n+1/2}-2u_j^{n+1/2}+u_{j-1}^{n+1/2})
(\bar{u}_j^{n+1}-\bar{u}_j^n)\Big)\\
= &
\frac{1}{2}
\Re\Big(
u_{j+1}^{n+1}\bar{u}_j^{n+1}-2|u_j^{n+1}|^2+u_{j-1}^{n+1}\bar{u}_j^{n+1}
-u_{j+1}^{n+1}\bar{u}_j^{n}+2u_j^{n+1}\bar{u}_j^n-u_{j-1}^{n+1}\bar{u}_j^{n}\\

&~~+ u_{j+1}^{n}\bar{u}_j^{n+1}-2u_j^{n}\bar{u}_j^{n+1}+u_{j-1}^{n}\bar{u}_j^{n+1}
-u_{j+1}^{n}\bar{u}_j^{n}+2|u_j^{n}|^2-u_{j-1}^{n}\bar{u}_j^{n}\Big)\\
= &
\frac{1}{2}
\Re\Big(
u_{j+1}^{n+1}\bar{u}_j^{n+1}-2|u_j^{n+1}|^2+u_{j-1}^{n+1}\bar{u}_j^{n+1}
-u_{j+1}^{n+1}\bar{u}_j^{n}-u_{j-1}^{n+1}\bar{u}_j^{n}\\
&~~+ u_{j+1}^{n}\bar{u}_j^{n+1}+u_{j-1}^{n}\bar{u}_j^{n+1}
-u_{j+1}^{n}\bar{u}_j^{n}+2|u_j^{n}|^2-u_{j-1}^{n}\bar{u}_j^{n}\Big).
\end{array}
$$
Taking the sum over ${\mathbb Z}$ leads to
$$
\begin{array}{lll}
 &
\displaystyle
\Re \Big(\sum_{j\in {\mathbb Z}}(u_{j+1}^{n+1/2}-2u_j^{n+1/2}+u_{j-1}^{n+1/2})
(\bar{u}_j^{n+1}-\bar{u}_j^n)
\Big)\\ 
=& \displaystyle
\frac{1}{2}\Re \Big(\sum_{j\in {\mathbb Z}}
\Big(
u_{j+1}^{n+1}\bar{u}_j^{n+1}-2|u_j^{n+1}|^2+u_{j}^{n+1}\bar{u}_{j+1}^{n+1}
-u_{j+1}^{n+1}\bar{u}_j^{n}-u_{j}^{n+1}\bar{u}_{j+1}^{n+1}\\

&~~~~~~~~~~~+ u_{j+1}^{n}\bar{u}_j^{n+1}+u_{j}^{n}\bar{u}_{j+1}^{n+1}
-u_{j+1}^{n}\bar{u}_j^{n}+2|u_j^{n}|^2-u_{j}^{n}\bar{u}_{j+1}^{n}\Big)\Big),
\end{array}
$$
using an index translation as often as required 
to get rid of $j-1$ contributions. If we simplify this
expression taking into account all the complex conjugations, it leads to

$$
\begin{array}{lll}
 &
\displaystyle
\Re \Big(\sum_{j\in {\mathbb Z}}(u_{j+1}^{n+1/2}-2u_j^{n+1/2}+u_{j-1}^{n+1/2})
(\bar{u}_j^{n+1}-\bar{u}_j^n)
\Big)\\ 
=& -\displaystyle\frac{1}{2}
\Re \Big(\sum_{j\in {\mathbb Z}}
\Big(
\big(2|u_j^{n+1}|^2 -2u_{j}^{n+1}\bar{u}_{j+1}^{n+1}\big) -
\big(2|u_j^{n}|^2 -2u_{j}^{n}\bar{u}_{j+1}^{n}\big)\Big)\Big)\\
=& \displaystyle -\frac{1}{2}
\sum_{j\in {\mathbb Z}}\Big(\big(|u_{j+1}^{n+1}|^2 +|u_{j}^{n+1}|^2-
2\Re(u_{j}^{n+1}\bar{u}_{j+1}^{n+1})\big)
-\big(|u_{j+1}^{n}|^2 +|u_{j}^{n}|^2-
2\Re(u_{j}^{n}\bar{u}_{j+1}^{n})\big)\Big)\\
=& \displaystyle -\frac{1}{2}
\sum_{j\in {\mathbb Z}}\Big(
|u_{j+1}^{n+1}-u_j^{n+1}|^2 - |u_{j+1}^{n}-u_j^{n}|^2
\Big),
\end{array}
$$
which proves the lemma. \qed

\vspace*{0.2cm}

\noindent
Using Lemma \ref{lem2} gives
{\small $$
\begin{array}{lll}
\displaystyle 
\sum_{j\in {\mathbb Z}}\alpha_j^n\!\!\!\!\!&=&\!\!\!\!\!\!\displaystyle
-\sum_{j\in {\mathbb Z}}\left(
\frac{1}{2}\frac{1}{\delta t}\frac{\left|u_{j+1}^{n+1}-u_j^{n+1}\right|^2 
-\left|u_{j+1}^{n}-u_j^{n}\right|^2}{h^2}-
\Re\left(\frac{i}{h^2}\Big(u_{j+1}^{n+1/2}-2u_j^{n+1/2}+u_{j-1}^{n+1/2}\Big)
V_j\bar{u}_j^{n+1/2}\right)\right)\\
~\!\!\!\!\!&=&\!\!\!\!\!\!-\displaystyle
\sum_{j\in {\mathbb Z}}\left(\frac{1}{2}\frac{1}{\delta t}
\frac{\left|u_{j+1}^{n+1}-u_j^{n+1}\right|^2 
-\left|u_{j+1}^{n}-u_j^{n}\right|^2}{h^2}
°+\Im\left(\frac{1}{h^2}\Big(u_{j+1}^{n+1/2}-2u_j^{n+1/2}+u_{j-1}^{n+1/2}\Big)
V_j\bar{u}_j^{n+1/2}\right)\right).
\end{array}
$$}
We then express the second-order finite difference term with respect to $v$. It gives
$$
\begin{array}{lll}
~& &\displaystyle 
\Im\left(\frac{1}{h^2}\Big(u_{j+1}^{n+1/2}-2u_j^{n+1/2}+u_{j-1}^{n+1/2}\Big)

V_j\bar{u}_j^{n+1/2}\right)\\
&= &
\displaystyle 
\Im \left(\Big(\frac{1}{\delta t}\Big(v_j^{n+1}-v_j^n\Big) - iV_j v_j^{n+1/2}+P_j u_j^{n+1/2}
 \Big)V_j\bar{u}_j^{n+1/2}\right)\\
& = &\displaystyle 
\Im \left(\frac{1}{\delta t}\Big(v_j^{n+1}-v_j^n\Big) V_j\bar{u}_j^{n+1/2}- iV_j v_j^{n+1/2}
V_j\bar{u}_j^{n+1/2}\right) .\\
\end{array}
$$
Since $-iV_j\bar{u}_j^{n+1/2}=\frac{1}{\delta t}(\bar{u}_j^{n+1}-\bar{u}_j^n)-\bar{v}_j^{n+1/2}$,
this implies
$$
\begin{array}{lll}
~& &
\displaystyle 
\Im\left(\frac{1}{h^2}\Big(u_{j+1}^{n+1/2}-2u_j^{n+1/2}+u_{j-1}^{n+1/2}\Big)
V_j\bar{u}_j^{n+1/2}\right)\\

& = &\displaystyle 
\Im \left(\frac{1}{\delta t}\Big(v_j^{n+1}-v_j^n\Big) V_j\bar{u}_j^{n+1/2}
+\frac{1}{\delta t}\Big(\bar{u}_j^{n+1}-\bar{u}_j^n\Big) V_j v_j^{n+1/2}
+V_j \big|v_j^{n+1/2}\big|^2\right)\\
& = &
\displaystyle 
\Im \left(V_j\Big(\frac{1}{\delta t}\Big(v_j^{n+1}-v_j^n\Big) \bar{u}_j^{n+1/2}
+\frac{1}{\delta t}\Big(\bar{u}_j^{n+1}-\bar{u}_j^n\Big)v_j^{n+1/2}\Big)\right).
\displaystyle 
\end{array}
$$
Rewriting now
$$
\begin{array}{lll}
& &\Im\Big(V_j\Big(v_j^{n+1}-v_j^n\Big) \bar{u}_j^{n+1/2}+\Big(\bar{u}_j^{n+1}-\bar{u}_j^n\Big)v_j^{n+1/2}\Big)\\
\vspace*{-0.2cm}\\
&= &\displaystyle\frac{1}{2}\Im\Big(
V_j( v_j^{n+1}\bar{u}_j^{n+1}+v_j^{n+1}\bar{u}_j^{n}-v_j^{n}\bar{u}_j^{n+1}-v_j^{n}\bar{u}_j^{n}
+\bar{u}_j^{n+1}v_j^{n+1}+\bar{u}_j^{n+1}v_j^{n}-\bar{u}_j^{n}v_j^{n+1}-\bar{u}_j^{n}v_j^{n})\Big)\\
\vspace*{-0.2cm}\\
&= &\displaystyle\frac{1}{2}\Im\Big(
V_j( v_j^{n+1}\bar{u}_j^{n+1}-v_j^{n}\bar{u}_j^{n}
+\bar{u}_j^{n+1}v_j^{n+1}-\bar{u}_j^{n}v_j^{n})\Big)
=\Im\Big(
V_j(\bar{u}_j^{n+1}v_j^{n+1}-\bar{u}_j^{n}v_j^{n})\Big),
\end{array}
$$
it follows that
$$\sum_{j\in {\mathbb Z}}\alpha_j^n=
-\frac{1}{\delta t}\sum_{j\in {\mathbb Z}}\Bigg(\frac{1}{2}
\frac{\left|u_{j+1}^{n+1}-u_j^{n+1}\right|^2 
-\left|u_{j+1}^{n}-u_j^{n}\right|^2}{h^2}
+\Im\Big(
V_j( \bar{u}_j^{n+1}v_j^{n+1}-\bar{u}_j^{n}v_j^{n})\Big)\Bigg).$$

\noindent
Using Lemma \ref{lem1}, we also express the sum of all $\beta_j^n$ as
$$
\sum_{j\in {\mathbb Z}}\beta_j^n=
\frac{1}{2}\frac{1}{\delta t}\sum_{j\in {\mathbb Z}}P_j
\Big(|u_j^{n+1}|^2-|u_j^n|^2\Big).
$$
\noindent
Finally, summing \eqref{first} over all indices $j$ and 
separating $n+1$ contributions in the left-hand side and
$n$ contributions in the right-hand side, we find that
$$
\begin{array}{lll}
~& &\displaystyle 
\frac{1}{\delta t}\sum_{j\in {\mathbb Z}}
\Big(\frac{1}{2}\Big( |v_j^{n+1}|^2 + \Big|\frac{u_{j+1}^{n+1}-u_j^{n+1}}{h}
\Big|^2 + P_j |u_j^{n+1}|^2 \Big)
+\Im(V_j\bar{u}_j^{n+1}v_j^{n+1})\Big)\\
&= &\displaystyle
\frac{1}{\delta t}\sum_{j\in {\mathbb Z}}\Big(\frac{1}{2}
\Big(|v_j^n|^2 + \Big|\frac{u_{j+1}^n-u_j^n}{h}\Big|^2 +
 P_j |u_j^n|^2 \Big)+\Im(V_j\bar{u}_j^nv_j^n)\Big),
\end{array}
$$
which completes the proof of Theorem \ref{discons}. \qed

Note that the successive steps in this proof are similar
to the ones that have been used for the conservation of energy in the continuous case. 

\vspace*{0.2cm}

Let us point out that this discrete conservation holds for the
{\em exact} solution of the semi-implicit numerical scheme. However, one needs to use 
a linear 
algebra routine in order to calculate the solution $(u^{n+1},v^{n+1})$, 
solving a linear system. This means that
the obtained numerical solution solves \eqref{sch1}-\eqref{sch2} up to
some numerical error. It follows that the discrete energy calculated with 
the computed $(u^{n+1},v^{n+1})$ is not exactly conserved. It can be also noticed 
that the discrete version of the space derivative is a good approximation of the
exact one only if the space step is small enough. It turns out that there may be a 
discrepancy between the exact energy and the approximate one (exactly conserved
by the exact solution $(u^n,v^n)$ given by \eqref{sch1} and \eqref{sch2}) when dealing
with highly oscillating functions.

\subsection{Influence of boundary conditions on energy conservation}

In numerical experiments, the finite difference scheme \eqref{sch1}-\eqref{sch2} is 
always used on a bounded domain for obvious reasons. We 
now analyse how this affects the energy conservation. For the sake of simplicity, we first 
deal here with the half-line $[0,+\infty[$, that requires
to calculate the values $u_j^n$ and $v_j^n$ for nonnegative values of indices $j$. Taking
bounded domain $\Omega$ will lead us to similar considerations 
on the opposite extremity. 

\vspace*{0.2cm}

We first assume
that the Dirichlet boundary condition $u(t,0)=0$ is prescribed at each time.
The following proposition shows
that the discrete version of the classical integration by parts 
$$
\int_0^{+\infty}\partial_x^2 u\,\bar{u}\,{\mathrm d}x= -\int_0^{+\infty}|\partial_x u|^2
\,{\mathrm d}x
$$
(when $u(t,0)=0$ and $u(t,x)=0$ if $x\gg 1$) still holds. 
From now on, the quantity $u$ is assumed to depend only on the space index $j$.

\begin{proposition}\label{IPP}
Let $w_j$ denote the approximation of the second derivative of $u$ at index $j$ computed in terms
of $(u_k)_{k\ge 1}$. We then have
$$
\sum_{j=1}^{+\infty} w_j \bar{u}_j=-\frac{1}{h^2}|u_1|^2-\sum_{j=2}^{+\infty}
\Big|\frac{u_j-u_{j-1}}{h}\Big|^2.
$$ 
\end{proposition}
\noindent
{\bf Proof.--}~Setting $w_1=(u_2-2u_1)/h^2$ (by consistency with \eqref{DiffDBC})
and $w_j=(u_{j+1}-2u_j+u_{j-1})/h^2$ for $j\ge 2$ (using \eqref{DiscDiff}),
we have
$$
\begin{array}{lll}
\displaystyle \sum_{j=1}^{+\infty} w_j \bar{u}_j& =& \displaystyle \frac{1}{h^2}\Big((u_2-2u_1)
\bar{u}_1+
\sum_{j=2}^{+\infty} (u_{j+1}-2u_j+u_{j-1})\bar{u}_j\Big)\\
& = & \displaystyle \frac{1}{h^2}\Big(
u_2\bar{u}_1 - 2|u_1|^2  
-2\sum_{j=2}^{+\infty} |u_j|^2 +
 \sum_{j=2}^{+\infty} u_{j+1}\bar{u}_j+\sum_{j=2}^{+\infty} u_{j-1}\bar{u}_j \Big)\\
& = & \displaystyle
\frac{1}{h^2}\Big(-|u_1|^2-\sum_{j=2}^{+\infty} |u_j|^2 -\sum_{j=2}^{+\infty} |u_{j-1}|^2+ 
2\Re\Big(\sum_{j=2}^{+\infty} u_{j-1}\bar{u}_j\Big)
\Big)\\
& = & \displaystyle
\frac{1}{h^2}\Big(-|u_1|^2-\sum_{j=2}^{+\infty}|u_j-u_{j-1}|^2 \Big).
\end{array} 
$$
Hence, the proposition is proved. \qed

\vspace*{0.2cm}

Note that when dealing with Dirichlet boundary conditions, it can be assumed that the
value at $x=0$ is always $u_0=0$. Consequently, the identity of Proposition \ref{IPP} 
can be rewritten as
$$
\sum_{j=0}^{+\infty} w_j \bar{u}_j=-\sum_{j=1}^{+\infty}
\Big|\frac{u_j-u_{j-1}}{h}\Big|^2.
$$ 
It turns out that all the computations performed in the proof of Theorem \ref{discons}
remain valid and the discrete conservation of the invariant computed on the positive 
half-line still holds. The same considerations made on the right extremity of the domain
lead us to the conservation of the energy computed on $\Omega$, as states the
\begin{theorem}\label{disconsDir}
For approximate solutions of \eqref{syst} 
given by \eqref{sch1}-\eqref{sch2} computed on spatial points $x_j=a+jh$ 
($j\in \left\{0,\ldots,J+1\right\}$) with Dirichlet conditions for $u$
at the boundary (that is $u_0^n=u_{J+1}^n=0$ for each $n\ge 0$), 
the quantity
$$
E_n=\frac{1}{2}
\sum_{j=1}^{J}\Big(
|v_j^n|^2 + \Big|\frac{u_{j+1}^n-u_j^n}{h}\Big|^2 + P_j |u_j^n|^2 \Big)
+\sum_{j=1}^J\Im(V_j\bar{u}_j^nv_j^n)
$$
is conserved for each index $n$.
\end{theorem}
It can be noticed that only boundary conditions on $u$ are used in the integration by parts.
But since $\partial_tu-iVu=v$, if Dirichlet conditions are prescribed for $u$, it can be
seen that $v$ also satisfies Dirichlet conditions at the boundary.

\vspace*{0.2cm}

We then consider the Neumann boundary condition $\partial_x u(t,0)=0$. We show here
a slight modification in the conservation of energy that differs from what has been
obtained in the case of Dirichlet conditions.

\begin{proposition}\label{IPP2}
Let $w_j$ denote the approximation of the second derivative of $u$ at index $j$ computed in terms
of $(u_k)_{k\ge 0}$. We then have
$$
\frac{1}{2}w_0\bar{u}_0+\sum_{j=1}^{+\infty} w_j \bar{u}_j=
-\sum_{j=1}^{+\infty}
\Big|\frac{u_j-u_{j-1}}{h}\Big|^2.
$$ 
\end{proposition}
\noindent
{\bf Proof.--}~We now have $w_0=2(u_1-u_0)/h^2$ (by consistency with \eqref{DiffNBC})
and $w_j=(u_{j+1}-2u_j+u_{j-1})/h^2$ for $j\ge 1$ (using \eqref{DiscDiff}).
From this, the discrete sum now expresses as
$$
\begin{array}{lll}
\displaystyle \frac{1}{2}\frac{2(u_1-u_0)}{h^2}\bar{u}_0+\sum_{j=1}^{+\infty} w_j \bar{u}_j& =& \displaystyle \frac{1}{h^2}\Big(u_1\bar{u}_0-|u_0|^2
+\sum_{j=1}^{+\infty} (u_{j+1}-2u_j+u_{j-1})\bar{u}_j\Big)\\
& = & \displaystyle \frac{1}{h^2}\Big(
u_1\bar{u}_0 - |u_0|^2  
-2\sum_{j=1}^{+\infty} |u_j|^2 +
 \sum_{j=1}^{+\infty} u_{j+1}\bar{u}_j+\sum_{j=1}^{+\infty} u_{j-1}\bar{u}_j \Big)\\
& = &\displaystyle
\frac{1}{h^2}\Big(-\sum_{j=1}^{+\infty} |u_j|^2 -
\sum_{j=1}^{+\infty} |u_{j-1}|^2+ 
2\Re\Big(\sum_{j=1}^{+\infty} u_{j-1}\bar{u}_j\Big)
\Big)\\
& = & \displaystyle
-\frac{1}{h^2}\sum_{j=1}^{+\infty}|u_j-u_{j-1}|^2.
\end{array} 
$$
Hence, the proposition is proved. \qed

\vspace*{0.2cm}

Note that in this case, for a bounded domain with spatial points
$x_0=0$, $\ldots$, $x_j=a+jh$, $\ldots$, $x_{J+1}=L$, we deal with the approximation
\begin{equation}
\int_{0}^L f(x)\,{\mathrm d}x\approx h\Big(\frac{1}{2} f(0)
+\sum_{j=1}^{J}f(x_j) + \frac{1}{2} f(L)\Big)
\label{intapprox}
\end{equation}
which involves the second-order trapezoidal rule. Taking into account Proposition \ref{IPP2}
and using similar arguments at the right boundary $x=L$,
we obtain the following 
\begin{theorem}\label{disconsNeu}
For approximate solutions of \eqref{syst} 
given by \eqref{sch1}-\eqref{sch2} computed on spatial points $x_j=a+jh$ 
($j\in \left\{0,\ldots,J+1\right\}$) with Neumann conditions at the boundary, 
the quantity
$$
\begin{array}{lll}
E_n&=&\displaystyle
\frac{1}{4}\Big(|v_0^n|^2 +|v_{J+1}^n|^2 + P_0 |u_0^n|^2 + P_{J+1}|u_{J+1}^n|^2\Big)
+\frac{1}{2}\Big( \Im(V_0\bar{u}_0^nv_0^n)+\Im(V_{J+1}\bar{u}_{J+1}^nv_{J+1}^n)\Big)\\
& + & \displaystyle \frac{1}{2}
\sum_{j=1}^{J}\Big(
|v_j^n|^2 + \Big|\frac{u_{j+1}^n-u_j^n}{h}\Big|^2 + P_j |u_j^n|^2 \Big)
+\sum_{j=1}^J\Im(V_j\bar{u}_j^nv_j^n)
\end{array}
$$
is conserved for each index $n$.
\end{theorem}
Once again, the proof relies on the computations that have already been performed in the
whole space, with a $\frac{1}{2}$ coefficient for quantities evaluated at indices
$J=0$ and $j=J+1$. It can be noticed that the conservation of energy for Dirichlet
conditions is a particular case of the use of formula \eqref{intapprox} when 
boundary terms vanish, leading us to the sum between $1$ and $J$ in Theorem~\ref{disconsDir}.

\section{Numerical experiments} \label{NumExp}

We now present and discuss numerical simulations in all cases
(type I, II, and III) using the numerical scheme that has been previously described.
The solutions are evolved numerically on the bounded domain that now writes 
$[R_- , R_+ ]_{r_*}$. We choose $R_a, R_f 
\in \mathopen]R_- , R_+ [$ and we measure two types of quantities:
\begin{enumerate}
\item the amplitude of the solution at $r_* = R_a$ at all times;
\item the outgoing energy flux across the $r_* =R_f$ hypersurface at all times; in fact we measure the energy gain at all times, which is given by the outgoing energy ${\cal F}_{[0,t]\times S_{R_0}}$ divided by the energy of the initial data, \emph{i.e.}\
\begin{equation} \label{EnergyGaint}
{\cal G}_{R_0} (t) = \frac{{\cal F}_{[0,t]\times S_{R_0}}}{{\cal F}_{\tilde{\Sigma}_0}} \, .
\end{equation}
%
\end{enumerate}
For type I black hole bombs, $R_+=R_1$ and the mirror is at $r_*=R_+$; for type II bombs, $R_-= R_1$ and the mirror is located at $r_*=R_-$; in the case of type III bombs, we have one mirror at $R_-=R_1$ and another at $R_+=R_2$ (see \eqref{EvolBHB1}, \eqref{EvolBHB2}, and \eqref{EvolBHB3} for the description of the boundary initial value problems corresponding to each type of bomb).

\subsection{Black hole bombs of type II} \label{BHB2}

We investigate Type II bombs on a Reissned-Norström and a de Sitter-Reissner-Nordström background.
\begin{itemize}
\item For the Reissner-Nordström case, the values of the parameters are $M=2.001$ and $Q=2$ for the black hole and $q=1$, $m=0.1$, $l=0$ for the field. The solution has zero angular momentum and is therefore spherically symmetric.
The black hole is close to being extreme, this is however not fundamental. We have observed a behaviour similar to the one displayed below for other choices of parameters. The presentation of this particular example was motivated by the fact that the results are more striking in this case. This is probably due to the fact that superradiance becomes stronger as one approaches the extreme case.
For these values, the ergoregion spreads from $x=-\infty$ until a value between $30$ and $40$, as can be seen from the sign of the potential $P-V^2$ in Figure \ref{fig5}.
For our ``flare-like'' Cauchy data, we have set
\begin{equation} \label{FlareData}
\phi(0,.)\simeq 0 \mbox{ and } \frac{\partial \phi}{\partial t} (0,x)= \phi_1 (x) = e^{i\omega x/\alpha} e^{-((x-x_0)/\alpha)^2}\, ,
\end{equation}
with frequency $\omega =0$, scale factor $\alpha=5$, and translation coefficient $x_0 = -20$, which ensures that the data are supported (for all practical purposes) within the ergoregion. We consider both Neumann and Dirichlet boundary conditions at the mirror. Computations are first performed until final time $T=100$ on the domain $[-40,120]$
discretised with $2,000$ spatial points. This allows to observe in details the beginning of the evolution (see Figure \ref{niso1}): the wave is reflected against the mirror and sends bursts out to infinity. It seems that the wave has a tendency to stay close to the mirror.
\begin{figure}[ht!]
\begin{center}
\begin{minipage}{7.5cm}
\includegraphics[width=7.5cm,height=6.cm]{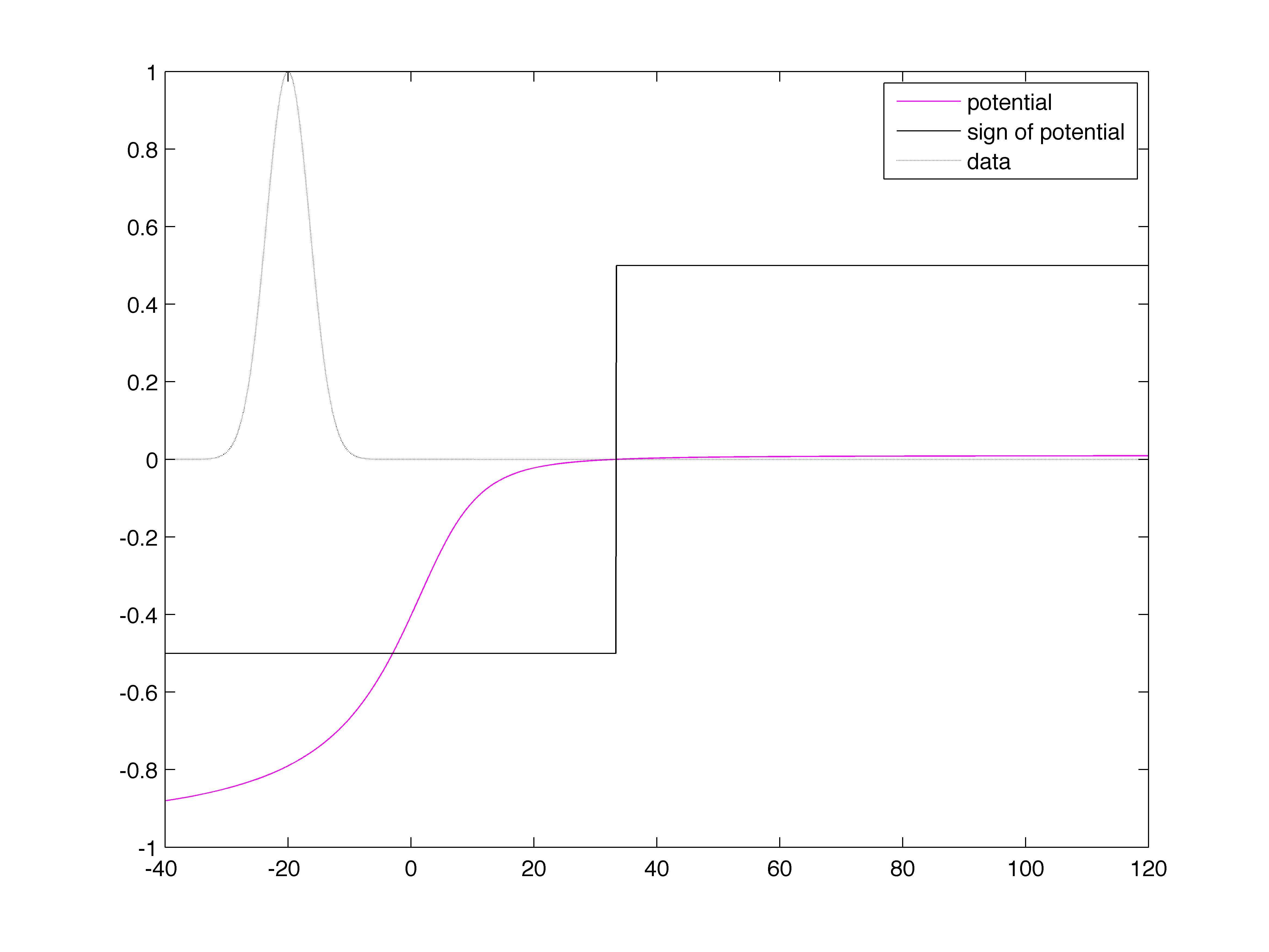}
\vspace*{-0.9cm}
\caption{Profile of data and potential.}
\label{fig5}
\end{minipage}
\end{center}
\end{figure}
\begin{figure}[ht!]
\begin{center}
\begin{minipage}{15cm}
\includegraphics[width=15cm,height=6.cm]{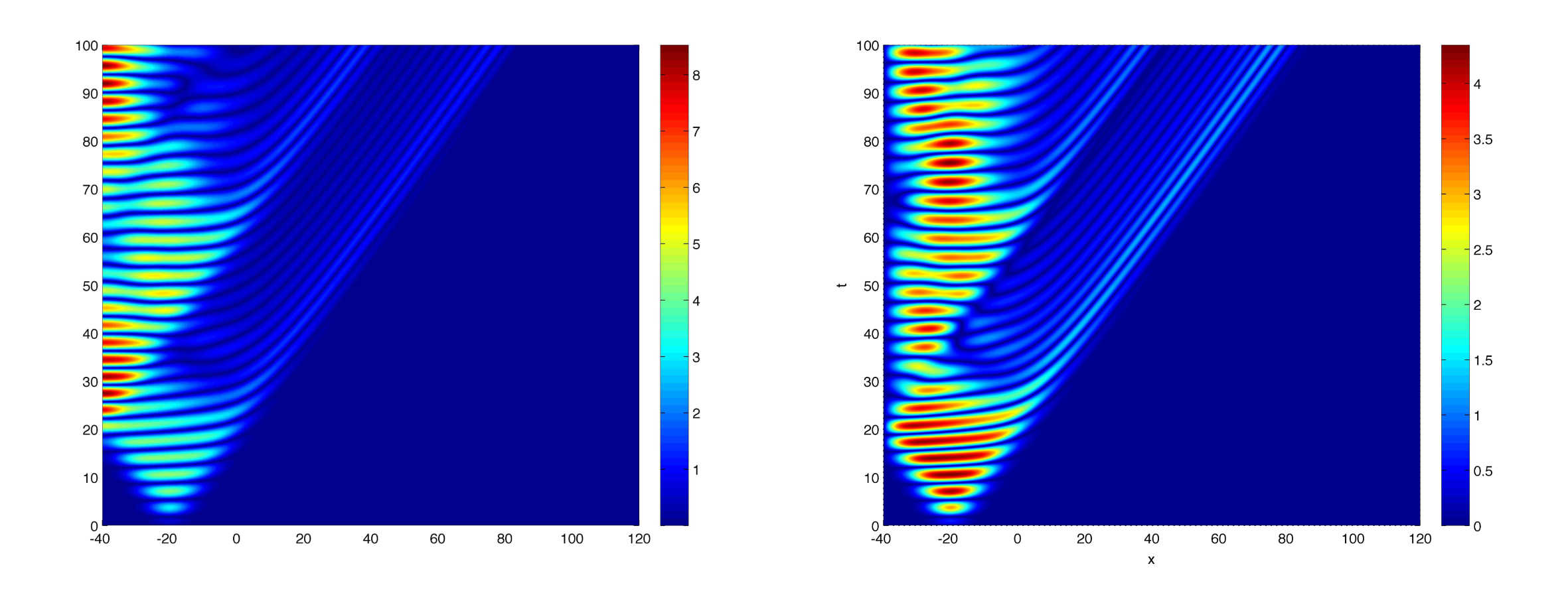}
\vspace*{-0.9cm}
\caption{$(x,t)$ isovalues of the solution: Neumann conditions (left), Dirichlet conditions (right).}
\label{niso1}
\end{minipage}
\end{center}
\end{figure}
In order to observe the phenomenon for a longer time, we enlarge the computational domain to $[-40, 920]$ with $4000$ points.
A simulation up to $T=300$ with $\delta t =h$, displayed in Figure \ref{fig7}, confirms the early observations and indicates an amplification of the field with time.
We then increase the number of spatial points to 40000 and perform the computation up to time $T=1500$.
The amplitude of the field is measured at $x=-16$ and the outgoing energy flux at $x = 56$.
The evolution with time of the logarithm of the amplitude (precisely of the absolute value of the real part of the solution) and of the energy gain are displayed on Figures \ref{fig8} and \ref{fig9}, establishing that both quantities increase exponentially in time.
\begin{figure}[ht!]
\begin{center}
\begin{minipage}{15cm}
\includegraphics[width=15cm,height=6.cm]{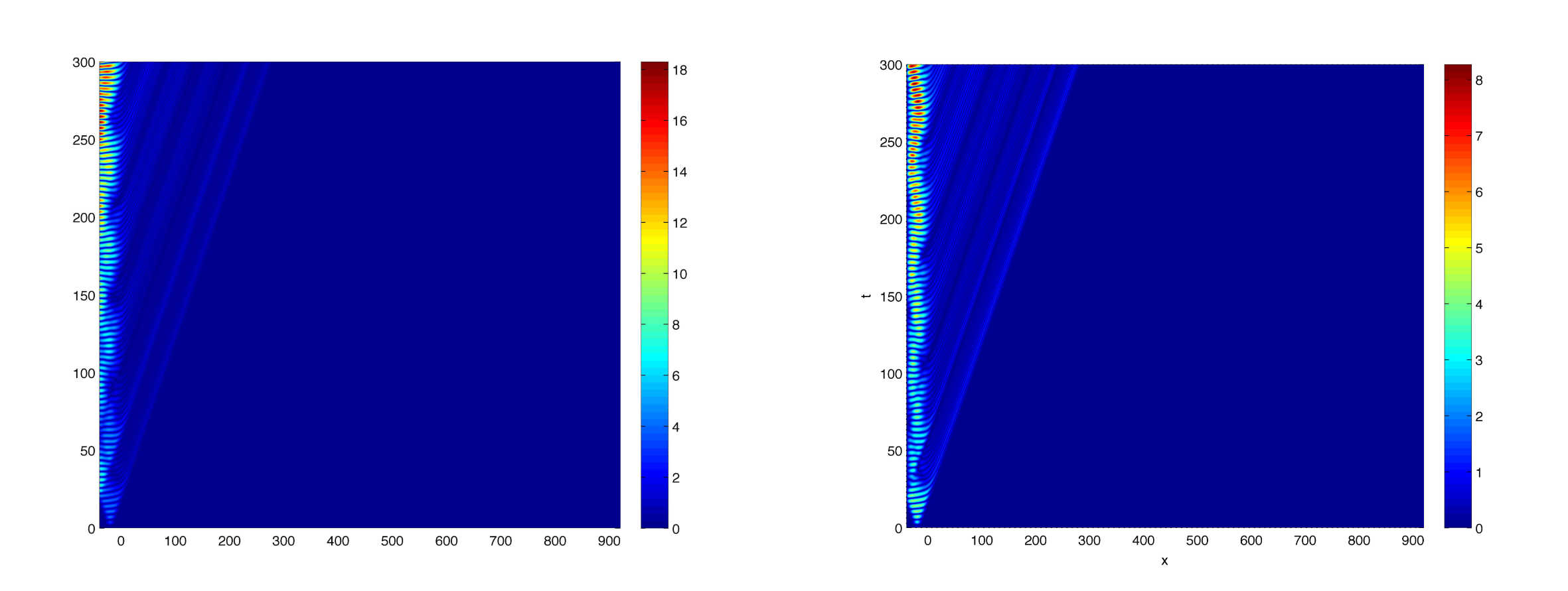}
\vspace*{-0.9cm}
\caption{$(x,t)$ isovalues of the solution: Neumann conditions (left), Dirichlet conditions (right).}
\label{fig7}
\end{minipage}
\end{center}
\end{figure}
\begin{figure}[ht!]
\begin{center}
\begin{minipage}{15cm}
\includegraphics[width=15cm,height=6.cm]{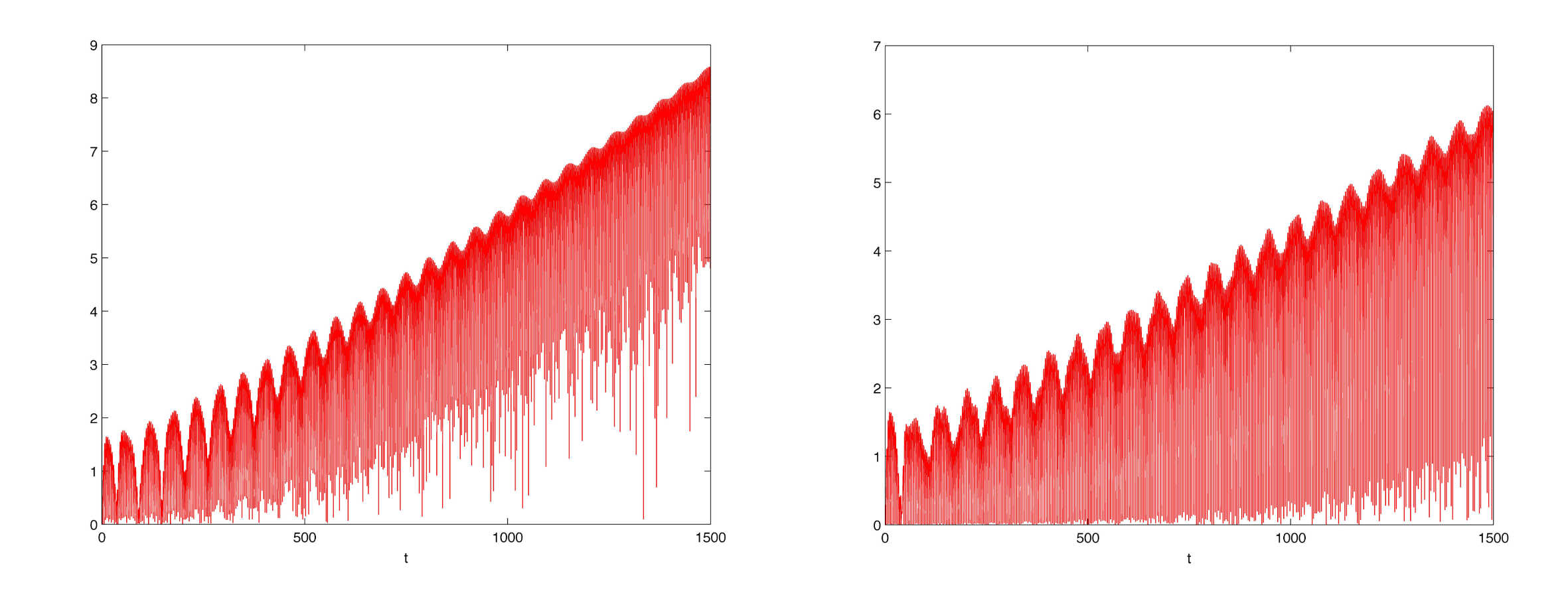}
\vspace*{-0.9cm}
\caption{Profile of the logarithm of the amplitude: Neumann conditions (left), Dirichlet 
conditions (right).}
\label{fig8}
\end{minipage}
\end{center}
\end{figure}

\begin{figure}
\begin{center}
\hspace*{0.5cm}
\begin{minipage}{15cm}
\includegraphics[width=15cm,height=6.cm]{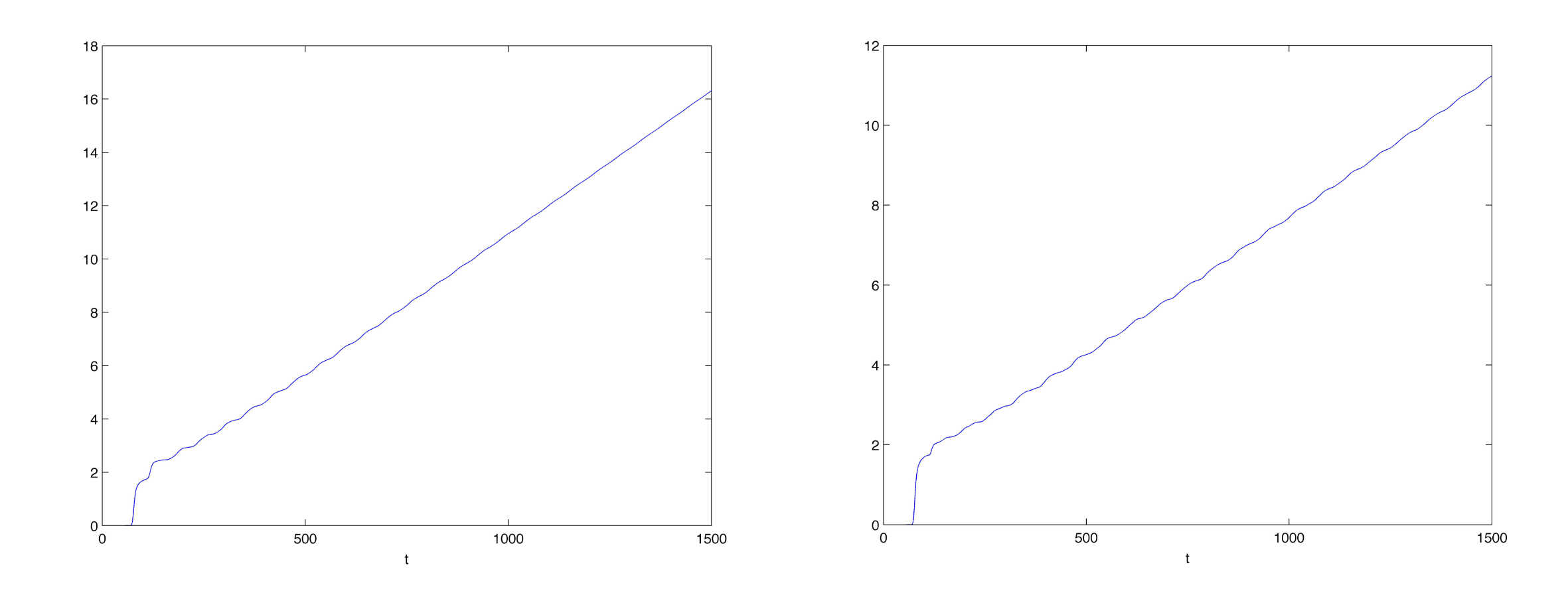}
\vspace*{-0.9cm}
\caption{Profile of logarithm of energy gain: Neumann conditions (left), 
Dirichlet conditions (right).}
\label{fig9}
\end{minipage}
\end{center}
\end{figure}
The plots in Figure \ref{fig9} exhibits two different regimes with two distinct exponential 
rates; the larger one briefly occurs at a transient stage and is followed by a smaller
growth that seems to govern the asymptotics of the energy gain.
\item We now investigate the case of a subextremal de Sitter Reissner-Nordström black hole, with parameters
$Q=2$, $M=3$ and $\Lambda = 1/(6 M)^2$. The zeros of the function $F$ (calculated by dichotomy) are given by
\begin{gather*}
r_n = -20.5361916161634,~ r_- = 0.763697274361058, \\
r_0 = 5.99999999996640, ,~ r_+ = 13.7724943418359.
\end{gather*}
\begin{figure}[ht!]
\begin{center}
\begin{minipage}{7.5cm}
\includegraphics[width=7.5cm,height=6.cm]{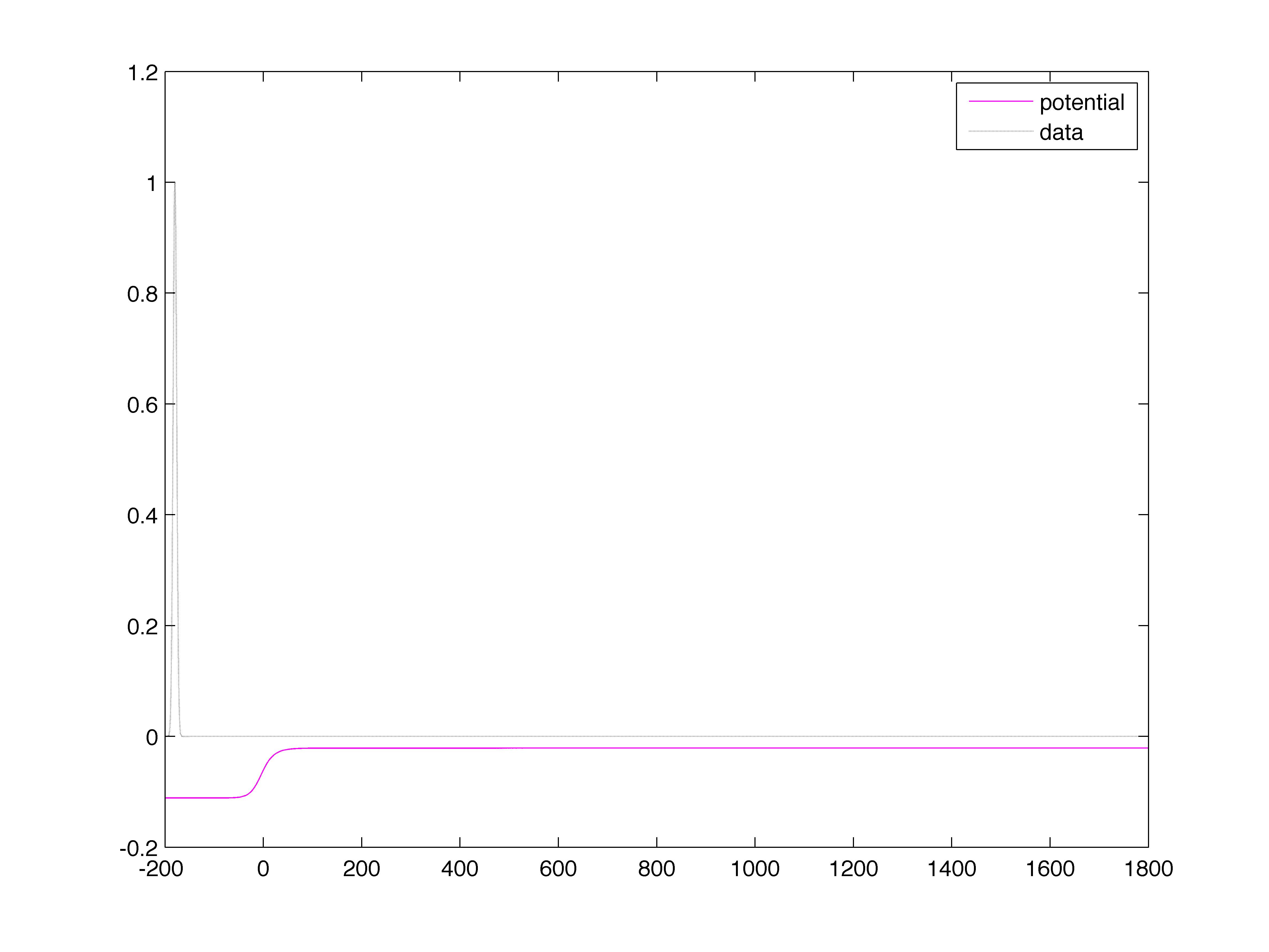}
\vspace*{-0.9cm}
\caption{Data and potential.}
\label{fig10}
\end{minipage}
\hspace*{0.5cm}
\begin{minipage}{7.5cm}
\includegraphics[width=7.5cm,height=6.cm]{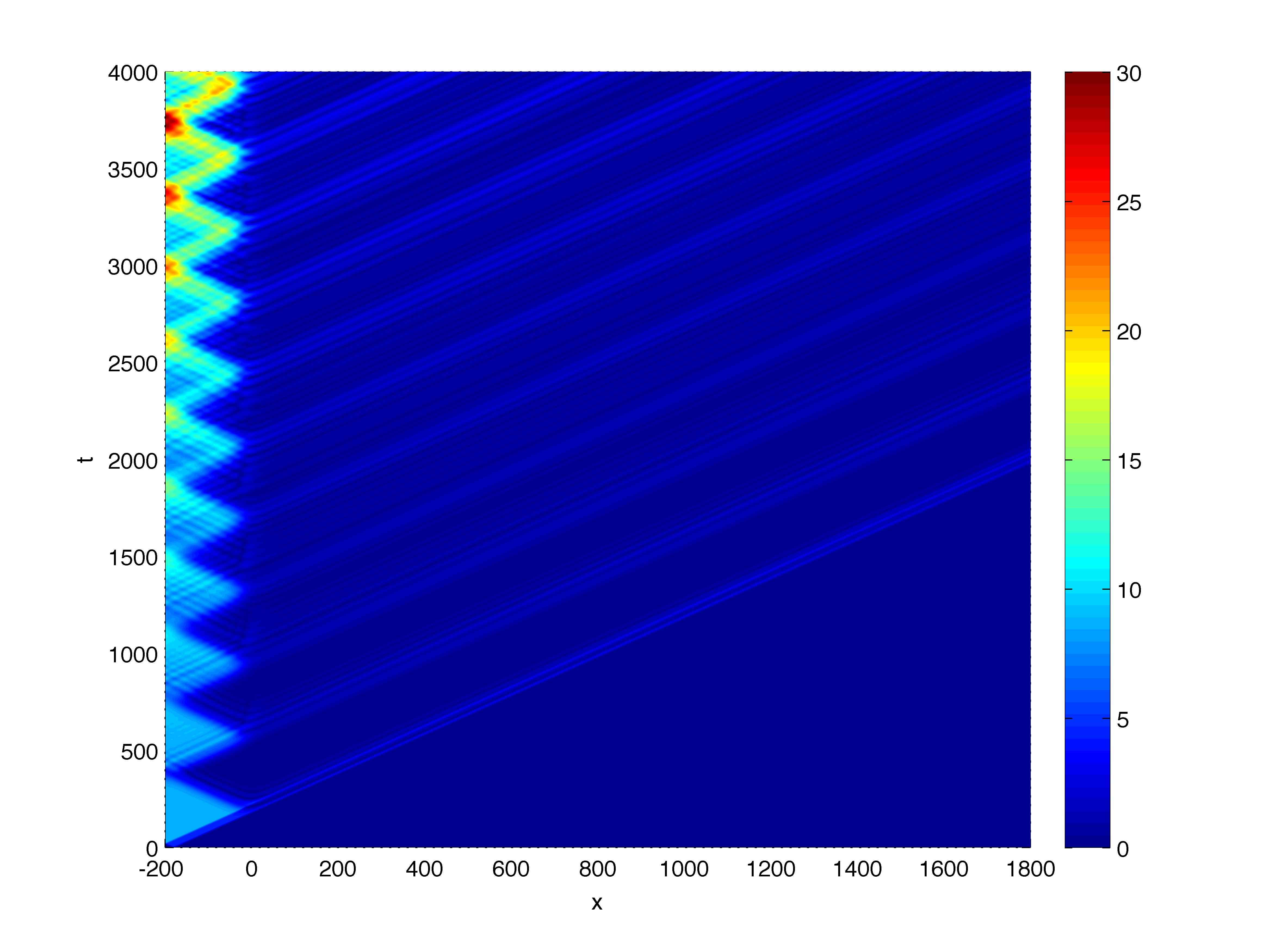}
\vspace*{-0.9cm}
\caption{$(x,t)$ Isovalues of the solution.}
\label{fig11}
\end{minipage}
\end{center}
\end{figure}
\begin{figure}[ht!]
\begin{center}
\begin{minipage}{7.5cm}
\includegraphics[width=7.5cm,height=6.cm]{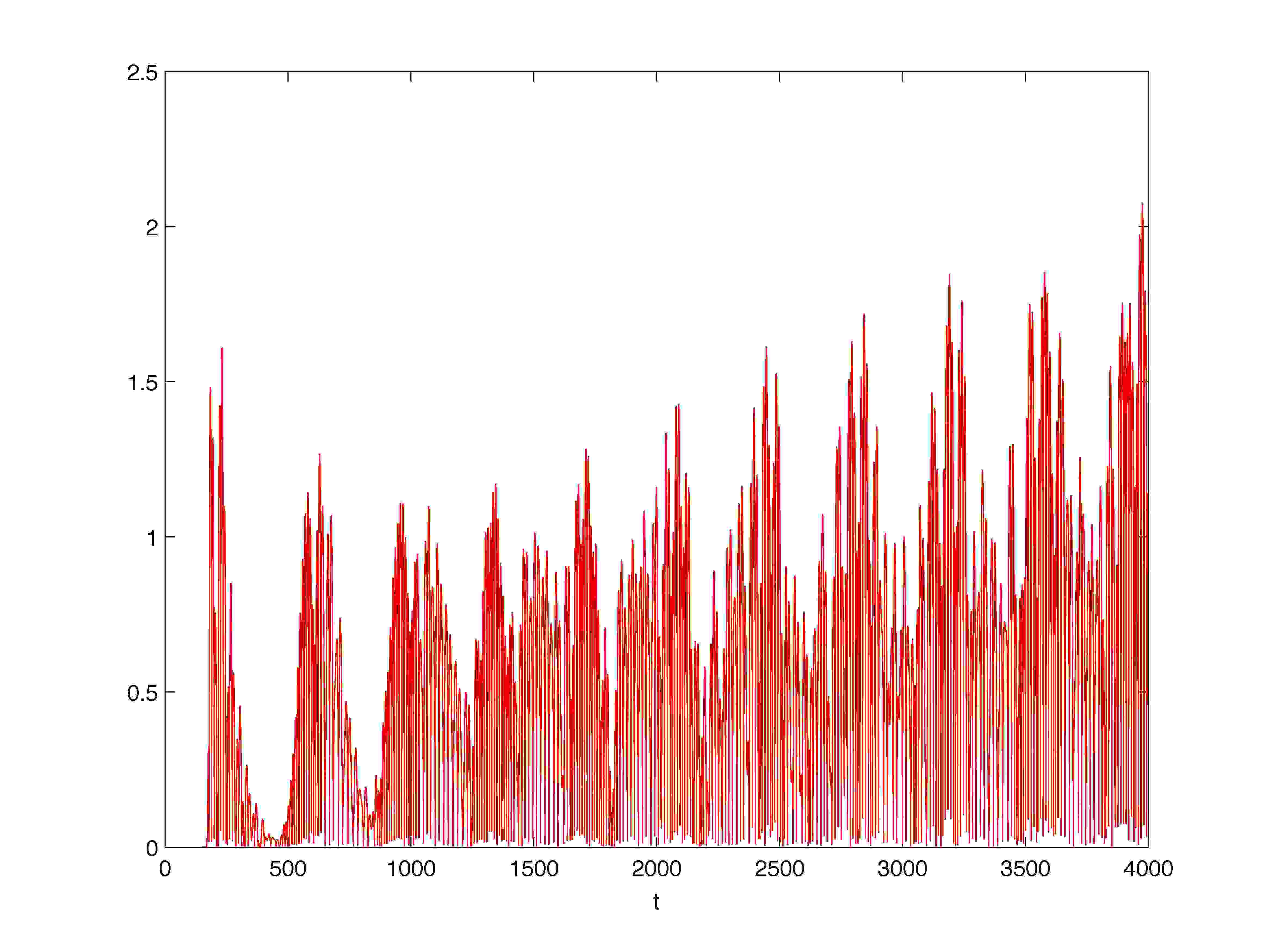}
\vspace*{-0.9cm}
\caption{Profile of the logarithm of the amplitude, $m=0.1$.}
\label{fig12}
\end{minipage}
\hspace*{0.5cm}
\begin{minipage}{7.5cm}
\includegraphics[width=7.5cm,height=6.cm]{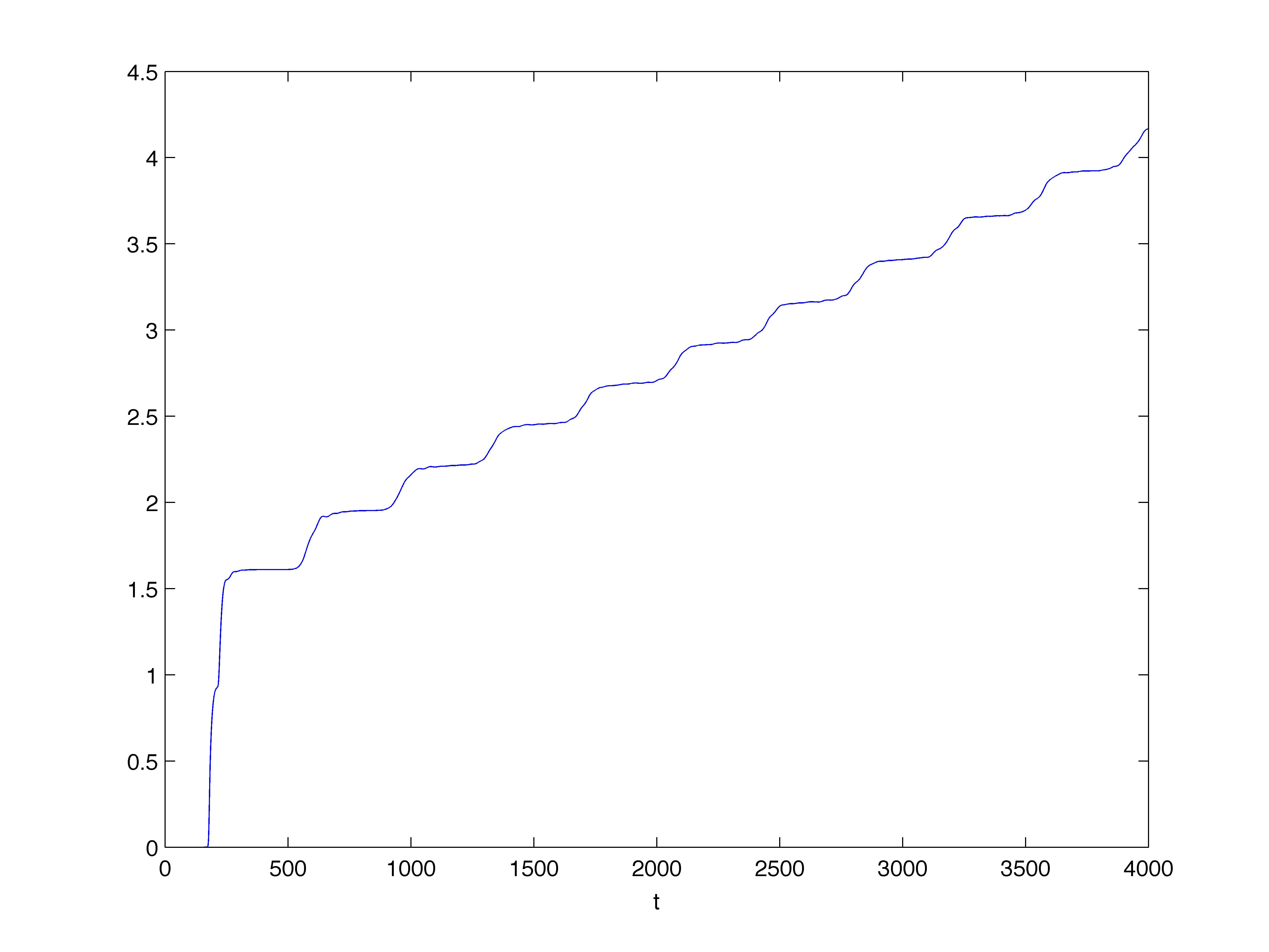}
\vspace*{-0.9cm}
\caption{Profile of the logarithm of the energy gain, $m=0.1$.}
\label{fig13}
\end{minipage}
\end{center}
\end{figure}
\begin{figure}[ht!]
\begin{center}
\begin{minipage}{7.5cm}
\includegraphics[width=7.5cm,height=6.cm]{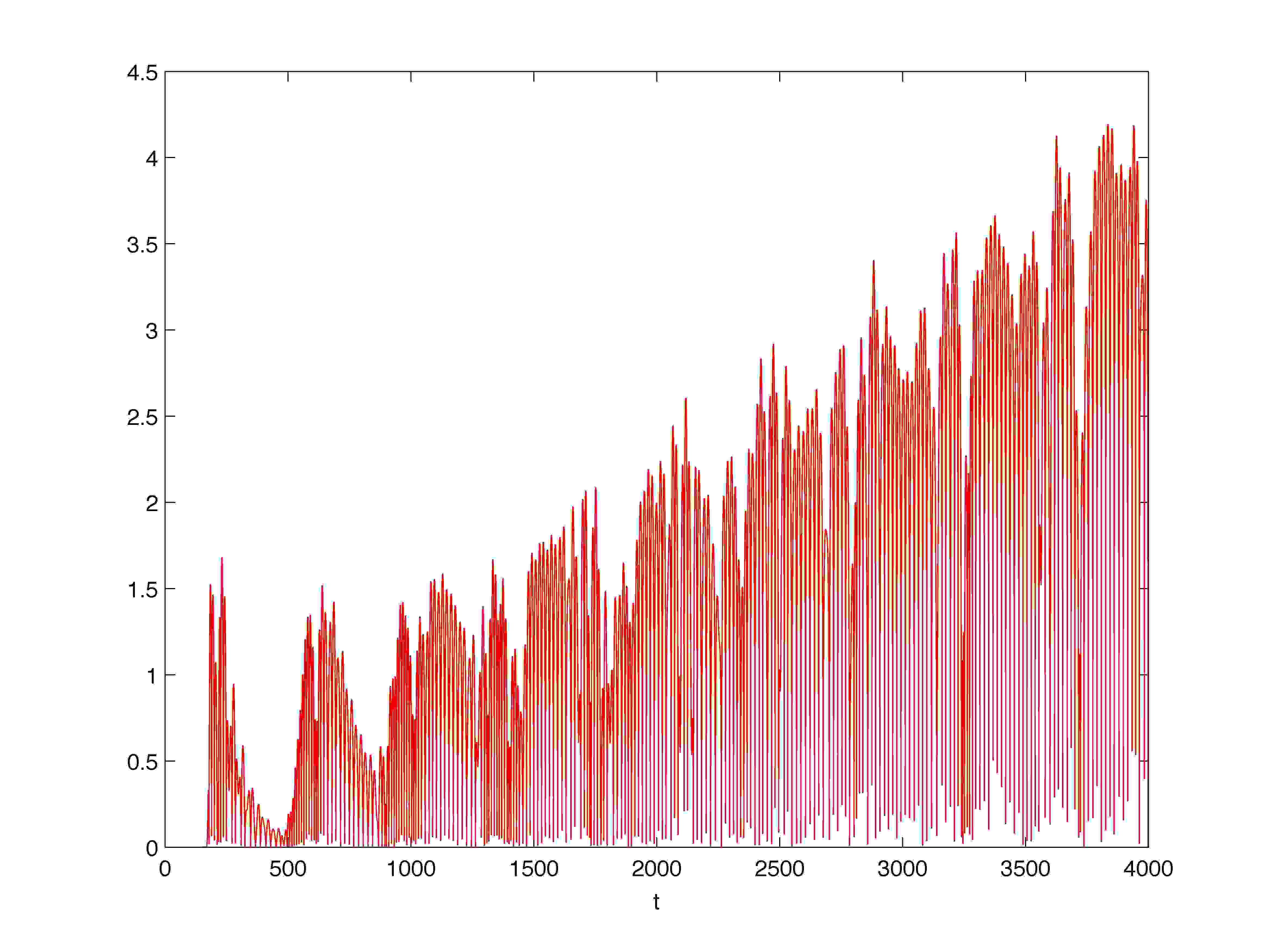}
\vspace*{-0.9cm}
\caption{Profile of the logarithm of the amplitude, $m=0$.}
\label{fig14}
\end{minipage}
\hspace*{0.5cm}
\begin{minipage}{7.5cm}
\includegraphics[width=7.5cm,height=6.cm]{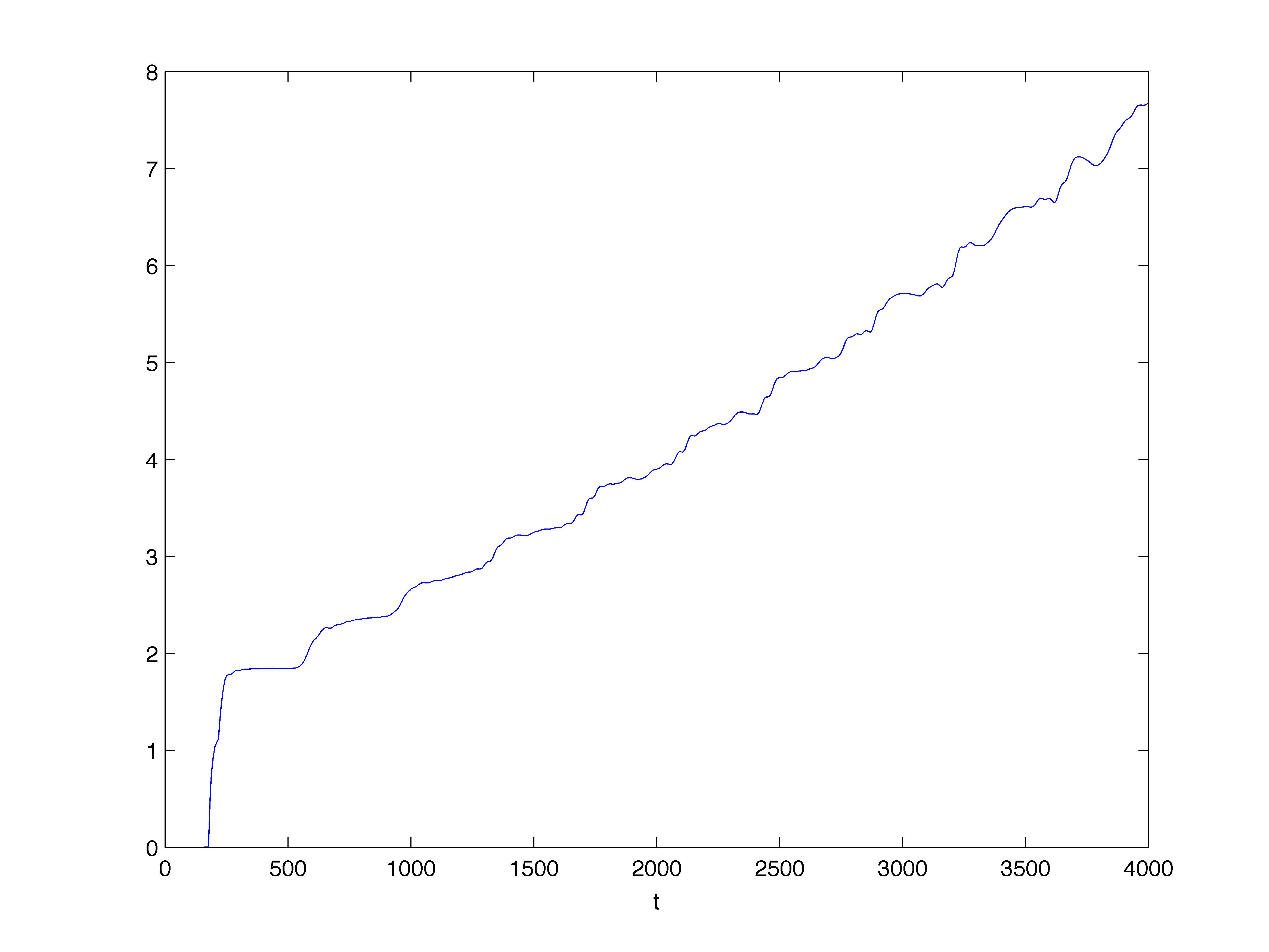}
\vspace*{-0.9cm}
\caption{Profile of the logarithm of the energy gain, $m=0$.}
\label{fig15}
\end{minipage}
\end{center}
\end{figure}
\end{itemize}
The parameters of the field are $m=0.1$, $q=1$, and $l=0$ (the solution is spherically symmetric). We work on the spatial domain $[-200,1800]$ which is entirely contained inside the ergoregion.
We discretise it with $10000$ points and perform simulations with Neumann homogeneous boundary conditions up to $T=4000$ with $\delta t = h$. The data are given by \eqref{FlareData} with $\alpha = 5$, $x_0 = -180$, and $\omega =0$.
The amplitude of the field and the outgoing energy gain are measured at $x=0$. The data and the potential are displayed in Figure \ref{fig10}.
We see that the potential undergoes a sharp transition between two approximately constant regimes, but the size of the jump is fairly small.
The evolution in Figure \ref{fig11} shows that the field bounces between the mirror and the potential barrier, each reflection of the barrier being accompanied by a burst being sent to infinity. 
The amplitude of the field appears to grow exponentially (see Figure \ref{fig12}) but with a small coefficient, which is likely due to the small size of the barrier.
The outgoing energy gain displays a faster exponential growth (Figure \ref{fig13}). The growth is slower than in the asymptotically flat case. Considering a massless field with all the other parameters unchanged, we observe in Figures \ref{fig14} and \ref{fig15} a steeper exponential growth for both the amplitude of the solution and the energy gain, the behaviour of the field being otherwise very similar. A simulation with Dirichlet boundary conditions shows a similar behaviour, but milder; we do not display it here.

\subsection{Black hole bombs of type I} \label{BHB1}

This is the analogue of the Press-Teukolski construction in the spherically symmetric setting, so one feels that one knows what to expect. This is true but we show that depending on the parameters of the field and the black hole, the behaviour can be quantitatively and qualitatively quite different.
\begin{itemize}
\item We start with a massless field on a Reissner-Nordström background. The parameters of the black hole are $M=2.001$ and $Q=2$ and for the field $q=1$, $m=0$, $l=0$. The potential and data are shown in Figure \ref{fig16}. We see that the potential remains negative on the whole computational domain $[-1700,100]$, which is consequently entirely contained inside the ergoregion.
In fact in this situation, the ergoregion covers the whole block III. The final time is $T=3000$, the spatial domain is discretised with $20000$ points and $\delta t = h$. Our initial data is as in \eqref{FlareData} with $\omega =0$, $\alpha = 5$, and $x_0 = 20$. The measure of the outgoing energy flux and of the amplitude of the solution are done at $x=55$.
Figures \ref{fig17} and \ref{loggain1} show the evolution of the logarithm of the amplitude and the energy gain for both Neumann and Dirichlet boundary conditions. The phenomenon is a mirror image of what we observed for type II bombs in the Reissner-Nordström case: the field creeps along
  the mirror with its amplitude growing exponentially in time, bursts of negative energy are sent towards the horizon and positive energy accumulates close to the mirror, growing exponentially as well.
\begin{figure}[ht!]
\begin{center}
\begin{minipage}{7.5cm}
\includegraphics[width=7.5cm,height=6.cm]{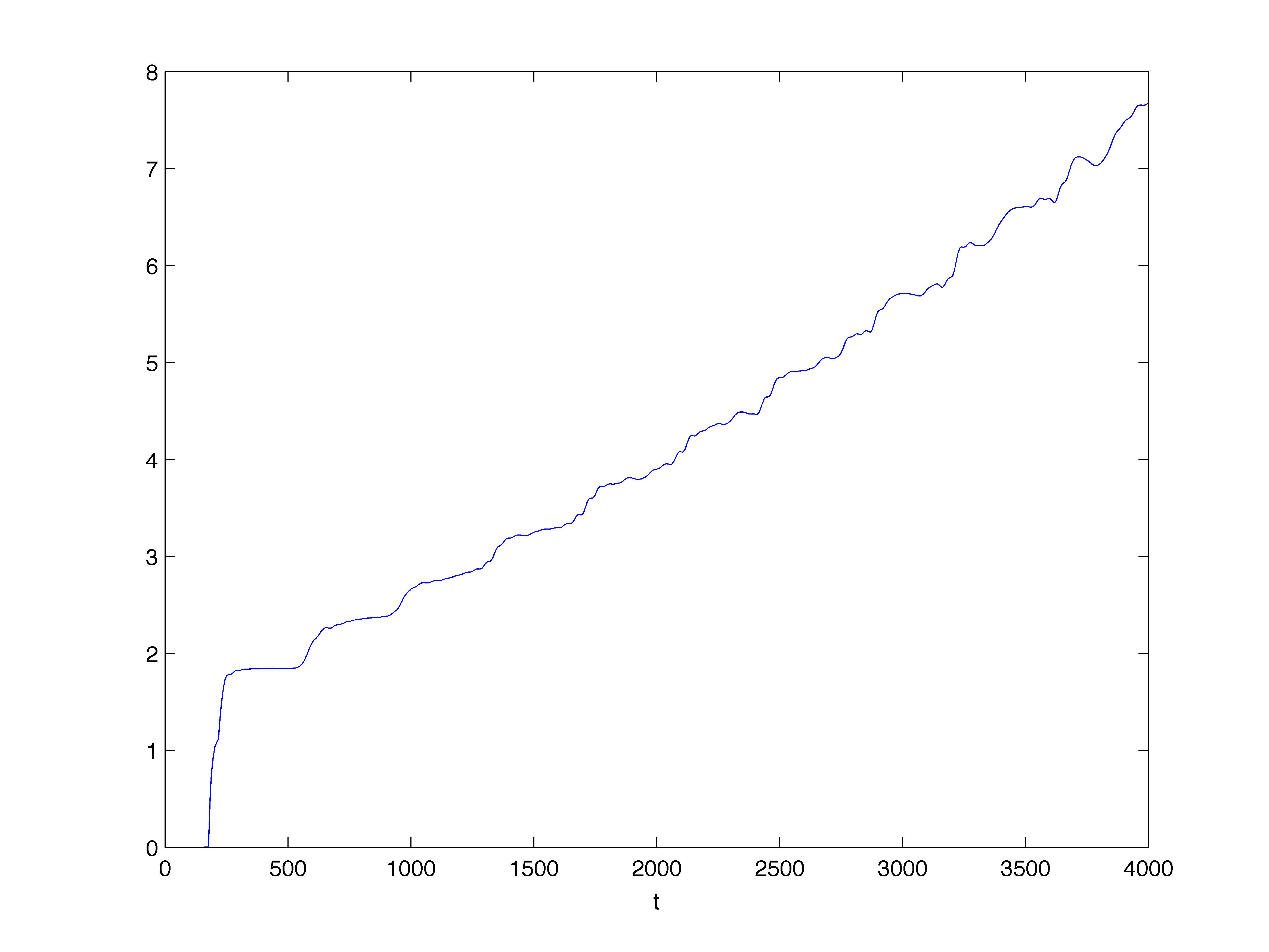}
\vspace*{-0.9cm}
\caption{Potential and initial data for $m=0$.}
\label{fig16}
\end{minipage}
\end{center}
\end{figure}
\begin{figure}
\begin{center}
\hspace*{0.5cm}
\begin{minipage}{15cm}
\includegraphics[width=15cm,height=6.cm]{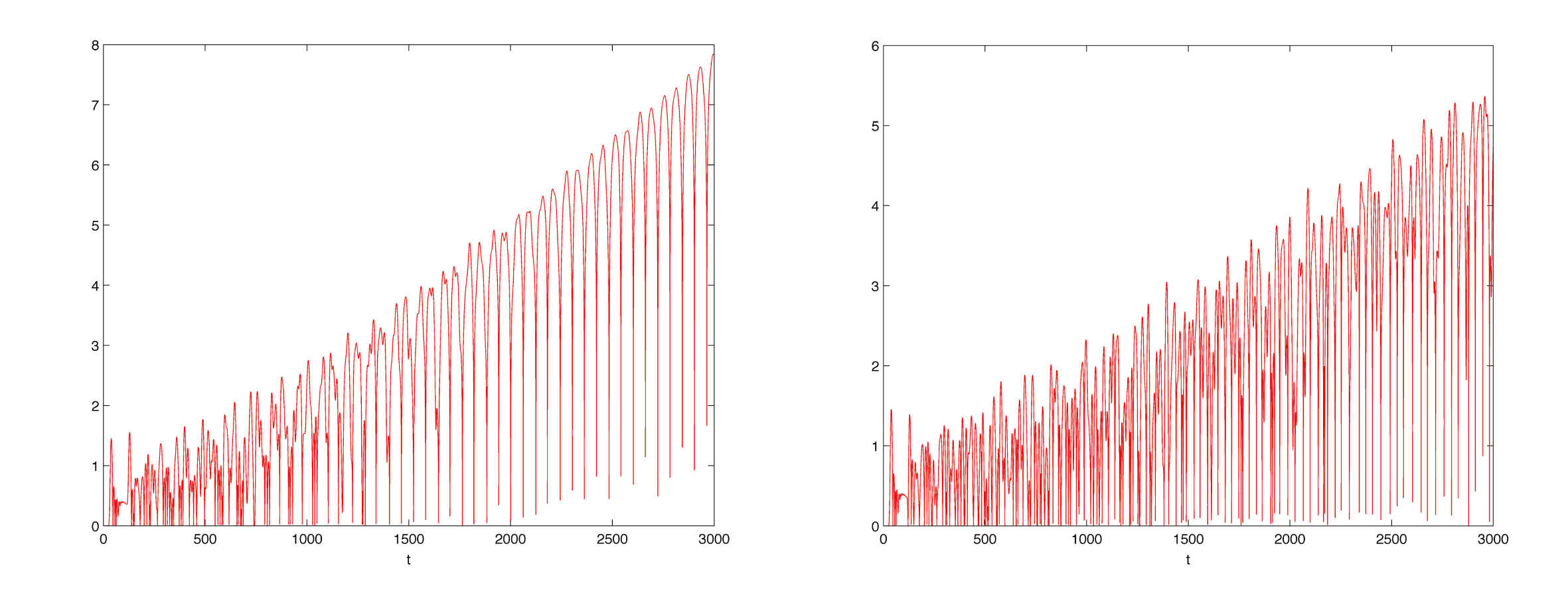}
\vspace*{-0.9cm}
\caption{Profile of logarithm of amplitude: Neumann conditions (left), Dirichlet conditions 
(right).}
\label{fig17}
\end{minipage}
\end{center}
\end{figure}
\begin{figure}
\begin{center}
\hspace*{0.5cm}
\begin{minipage}{15cm}
\includegraphics[width=15cm,height=6.cm]{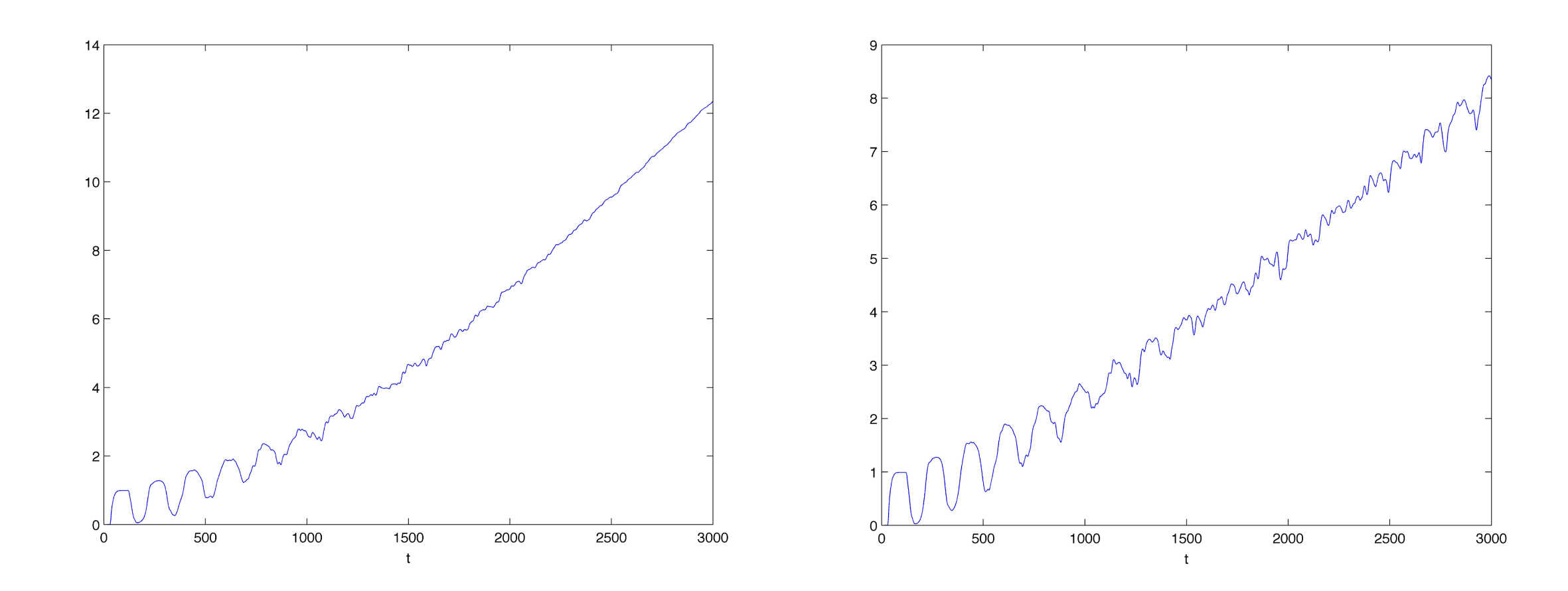}
\vspace*{-0.9cm}
\caption{Profile of logarithm of energy gain: Neumann conditions (left), Dirichlet conditions (right).}
\label{loggain1}
\end{minipage}
\end{center}
\end{figure}
\item On the same background, we study the evolution of a massive field with small mass $m=0.1$ so as to ensure that there is still superradiance. All the other parameters are the same. Due to the mass of the field, the ergoregion has shrunk and the mirror is now outside of it (see Figure \ref{fig19}). The simulations with Neumann and Dirichlet boundary conditions are almost indistinguishable, we simply display the Neumann case here.

The isovalues of the solution and the evolution of the logarithm of the amplitude and the energy gain are displayed in Figures \ref{fig20}, \ref{fig21}, and \ref{fig22}.
The behaviour of the solution is qualitatively different. A part of the field stays confined between the mirror and the boundary of the ergoregion and bounces back and forth between the two.
The reflection does not actually happen at the frontier of the ergoregion, but a little inside and it mimicks a Penrose process: the field enters the ergoregion, splits spontaneously into a part that has negative energy and falls into the black hole, and a part that has positive energy and tries to escape to infinity, but is prevented to do so by the mirror.
This is a similar phenomenon to what was observed in Ref.~\cite{DiNi2015} but here it occurs repeatedly due to the presence of the mirror.
We see from Figure \ref{fig22} that the oscillations in the logarithm of the energy gain grow linearly, which suggests that at each occurrence of the Penrose process, the same fraction of the energy of the incoming field is extracted from the ergoregion, causing an exponential accumulation of energy between the ergoregion and the mirror. The amplitude of the field also appears to grow exponentially, but this is a much milder growth than in the massless case.
\begin{figure}[ht!]
\begin{center}
\begin{minipage}{7.5cm}
\includegraphics[width=7.5cm,height=6.cm]{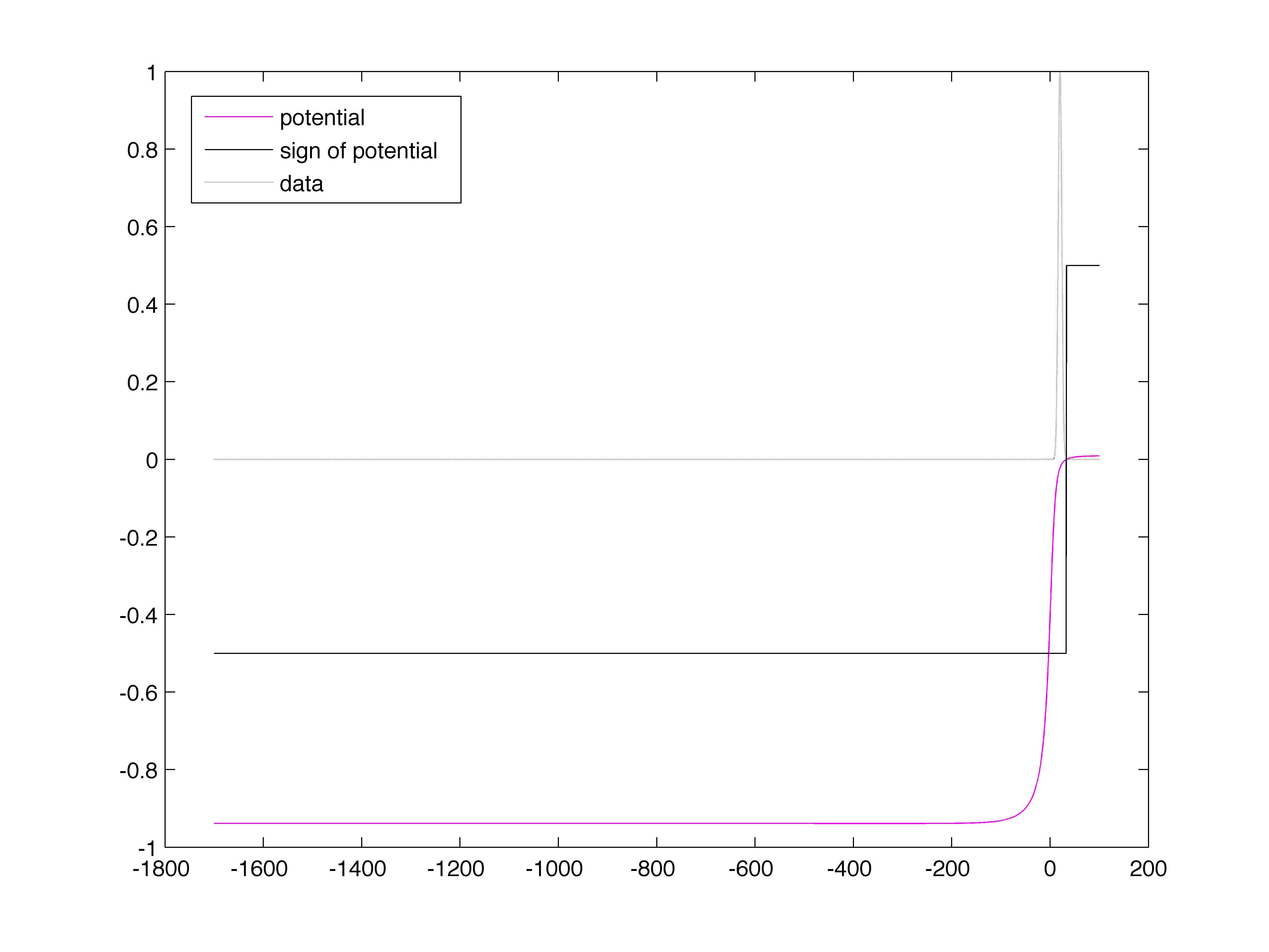}
\vspace*{-0.9cm}
\caption{Potential and initial data for $m=0.1$.}
\label{fig19}
\end{minipage}
\hspace*{0.5cm}
\begin{minipage}{7.5cm}
\includegraphics[width=7.5cm,height=6.cm]{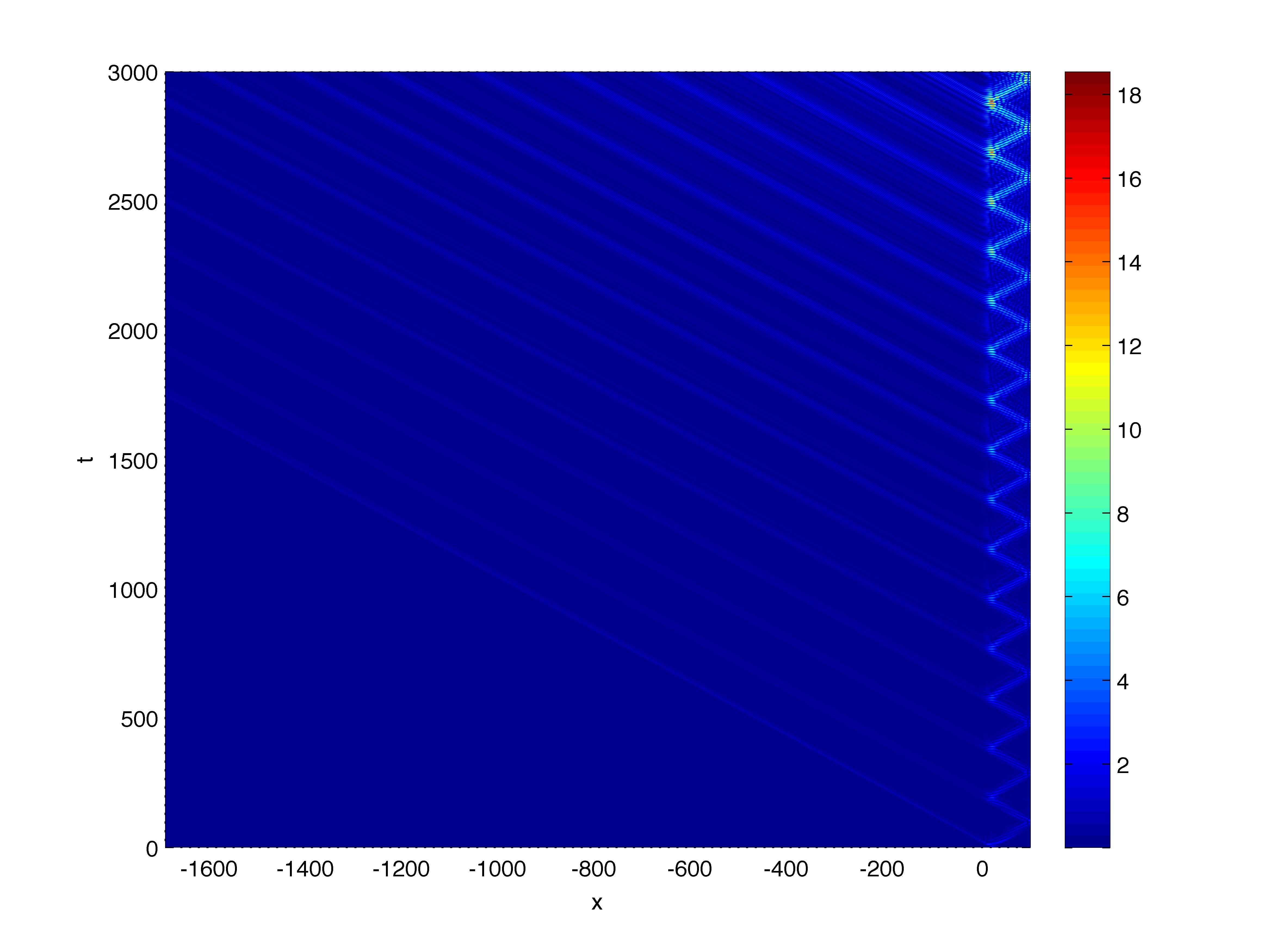}
\vspace*{-0.9cm}
\caption{$(x,t)$ isovalues of the solution, $m=0.1$.}
\label{fig20}
\end{minipage}
\end{center}
\end{figure}

\begin{figure}[ht!]
\begin{center}
\begin{minipage}{7.5cm}
\includegraphics[width=7.5cm,height=6.cm]{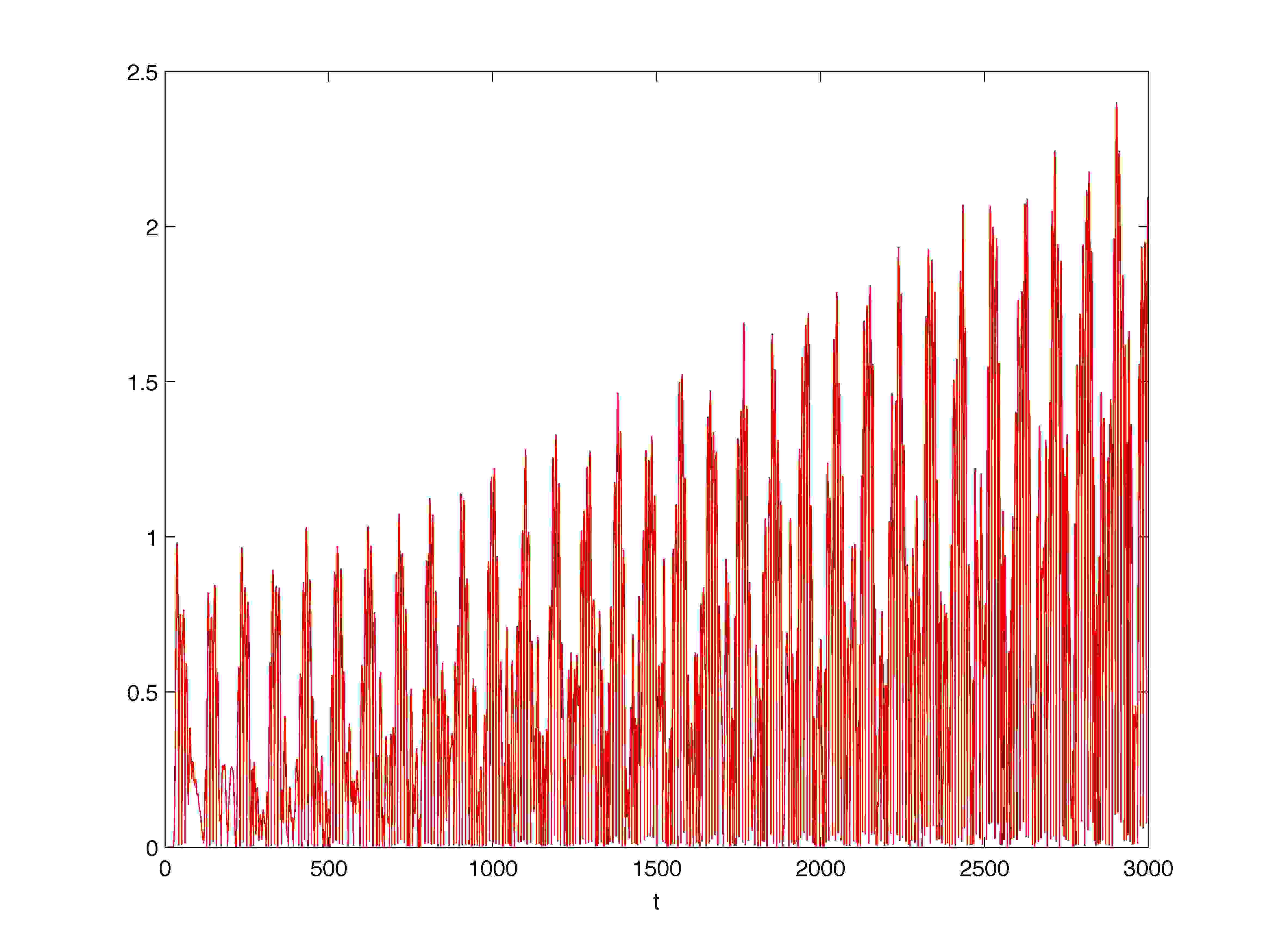}
\vspace*{-0.9cm}
\caption{Profile of the logarithm of the amplitude in the massive case.}
\label{fig21}
\end{minipage}
\hspace*{0.5cm}
\begin{minipage}{7.5cm}
\includegraphics[width=7.5cm,height=6.cm]{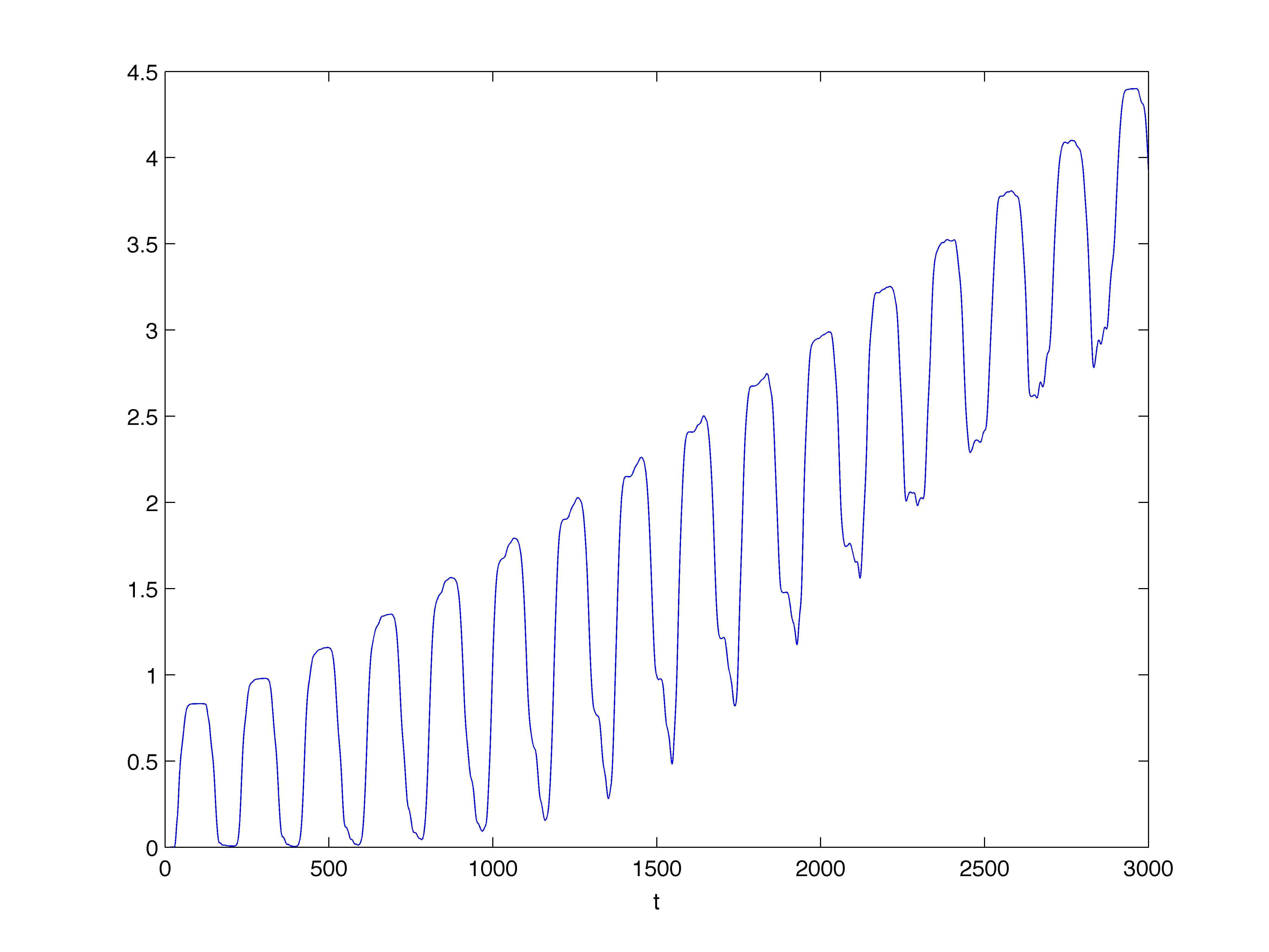}
\vspace*{-0.9cm}
\caption{Profile of the logarithm of the energy gain in the massive case.}
\label{fig22}
\end{minipage}
\end{center}
\end{figure}
\item We conclude this section on type I bombs with a study of the behaviour of a massless field outside a de Sitter Reissner-Nordström black hole with the same parameters as before: $Q=2$, $M=3$, and $\lambda = 1/(6 M)^2$. The final simulation time has been taken
equal to $T=2000$ on the domain $[-900,100]$ discretised with $10000$ spatial points, with $\delta t = h$. The flare-type Cauchy data is centred at $x_0=-100$ with $\alpha=5$ and $\omega =0$. The amplitude of the solution and the outgoing energy flux are measured at $x=0$. Figure \ref{fig23} displays a potential with a sharp jump but of small magnitude. The field being massless, we nevertheless observe in Figures \ref{fig24} and \ref{fig25} a clearcut exponential growth of the amplitude and the outgoing energy gain. In the case of Dirichlet boundary conditions, the growth is milder than for Neumann conditions.
\begin{figure}[ht!]
\begin{center}
\begin{minipage}{7.5cm}
\includegraphics[width=7.5cm,height=6.cm]{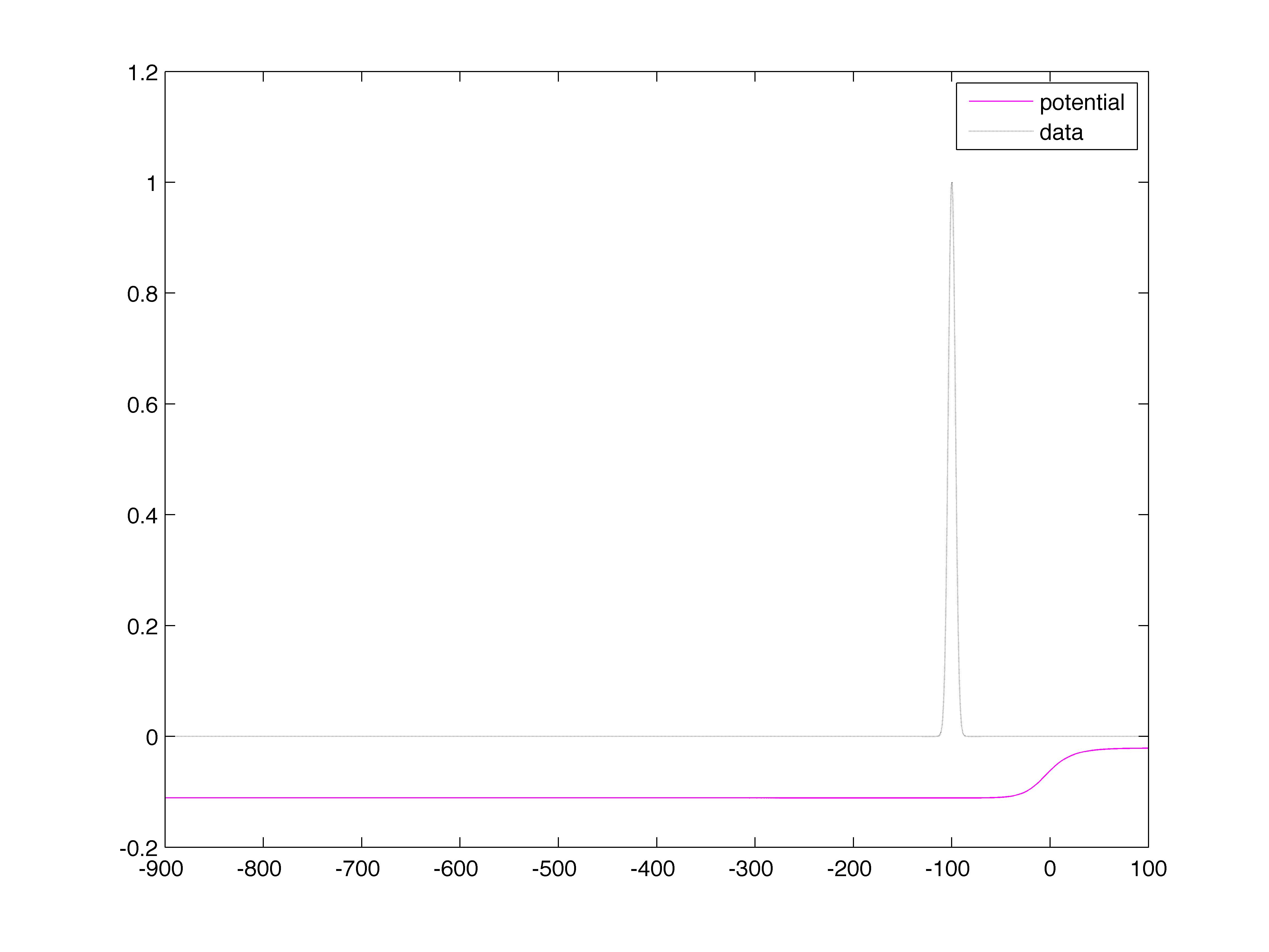}
\vspace*{-0.9cm}
\caption{Data and potential.}
\label{fig23}
\end{minipage}
\end{center}
\end{figure}
\begin{figure}[ht!]
\begin{center}
\hspace*{0.5cm}
\begin{minipage}{15cm}
\includegraphics[width=15cm,height=6.cm]{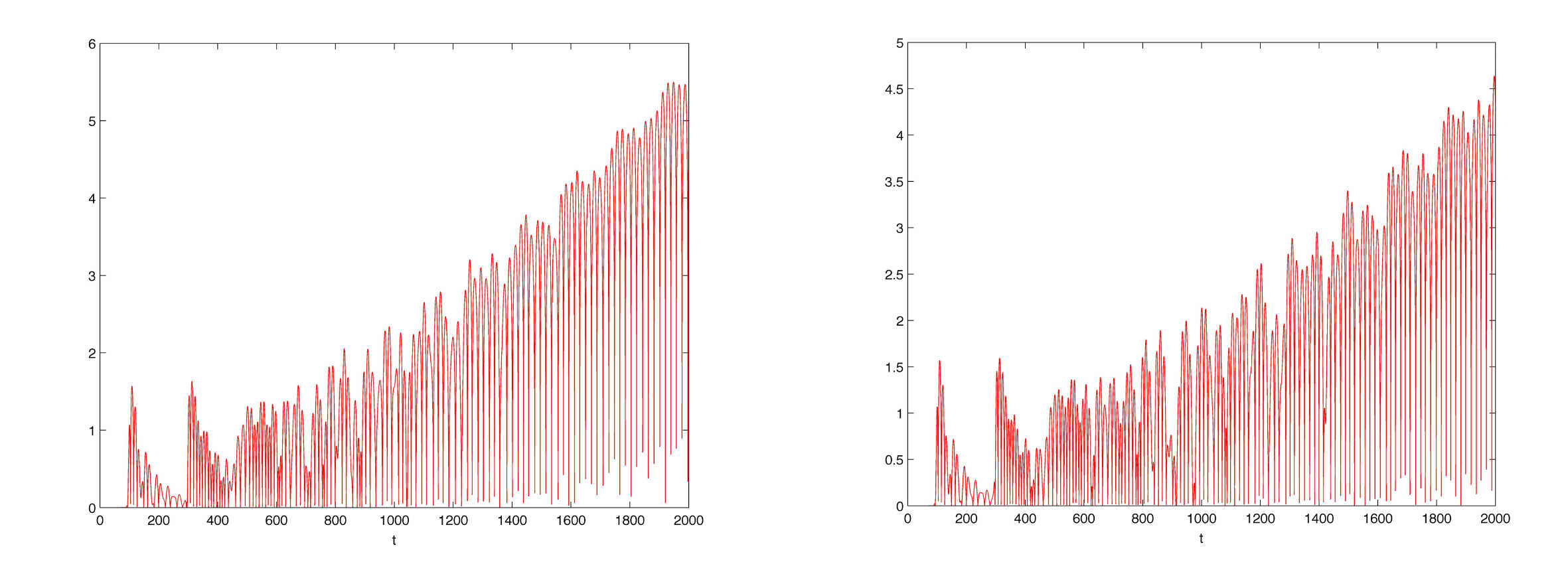}
\vspace*{-0.9cm}
\caption{Profile of logarithm of amplitude: Neumann conditions (left), Dirichlet conditions (right).}
\label{fig24}
\end{minipage}
\end{center}
\end{figure}
\begin{figure}[ht!]
\begin{center}
\hspace*{0.5cm}
\begin{minipage}{15cm}
\includegraphics[width=15cm,height=6.cm]{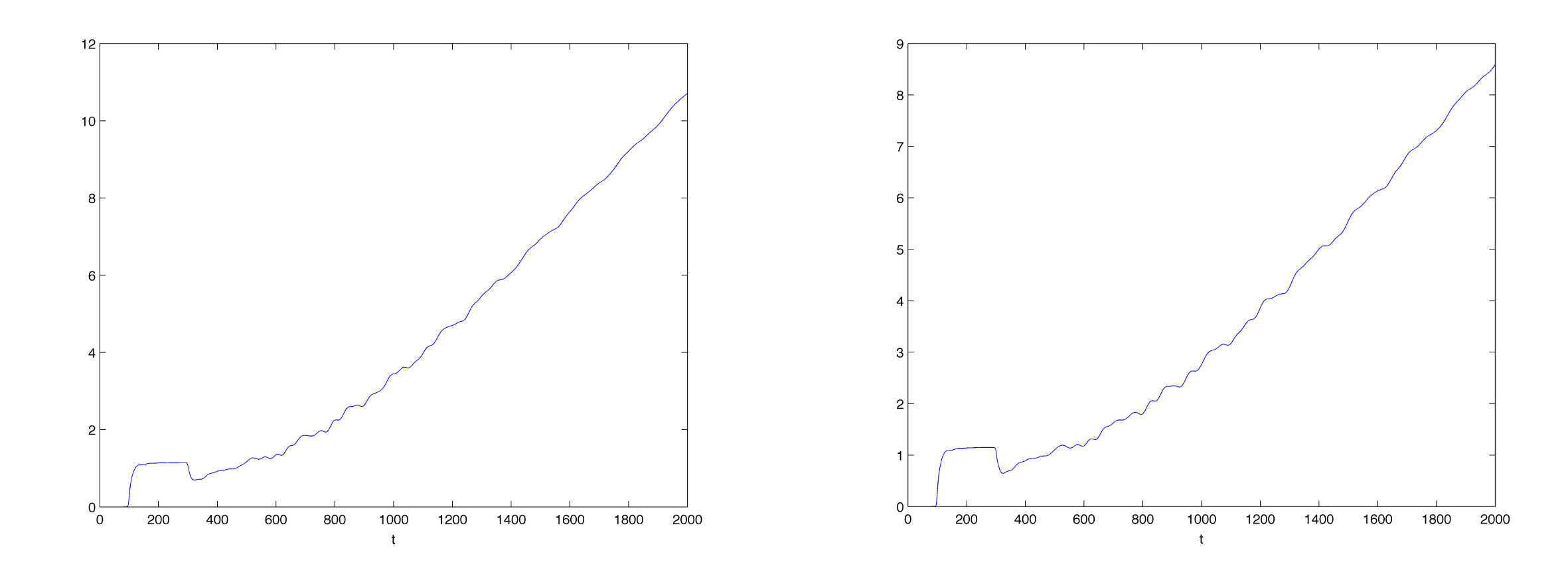}
\vspace*{-0.9cm}
\caption{Profile of logarithm of energy gain: Neumann conditions (left), Dirichlet conditions (right).}
\label{fig25}
\end{minipage}
\end{center}
\end{figure}
\end{itemize}

\subsection{Black hole bombs of type III} \label{BHB3}

We study these sandwich bombs on the two types of backgrounds studied previously.
\begin{itemize}
\item The first test is performed outside a Reissner-Nordström black hole of mass $M=2.5$ and charge $Q=2$, for a field of mass $m=0.1$, angular momentum $l=0$, and charge $q=1$. The computational domain is $[-40,40]$ discretised with $6000$ points. The initial data are of the form \eqref{FlareData} with scaling factor $\alpha = 5$, $x_0 = -20$, $\omega =0$. The amplitude and outgoing energy flux are measured at $x = 0$. For Neumann boundary conditions, the calculations are performed up to $T=1500$. For Dirichlet conditions, the growth is slower and we push the calculation to $T=2000$ with $8000$ points in order to observe something conclusive.
\begin{figure}[ht!]
\begin{center}
\begin{minipage}{7.5cm}
\includegraphics[width=7.5cm,height=6.cm]{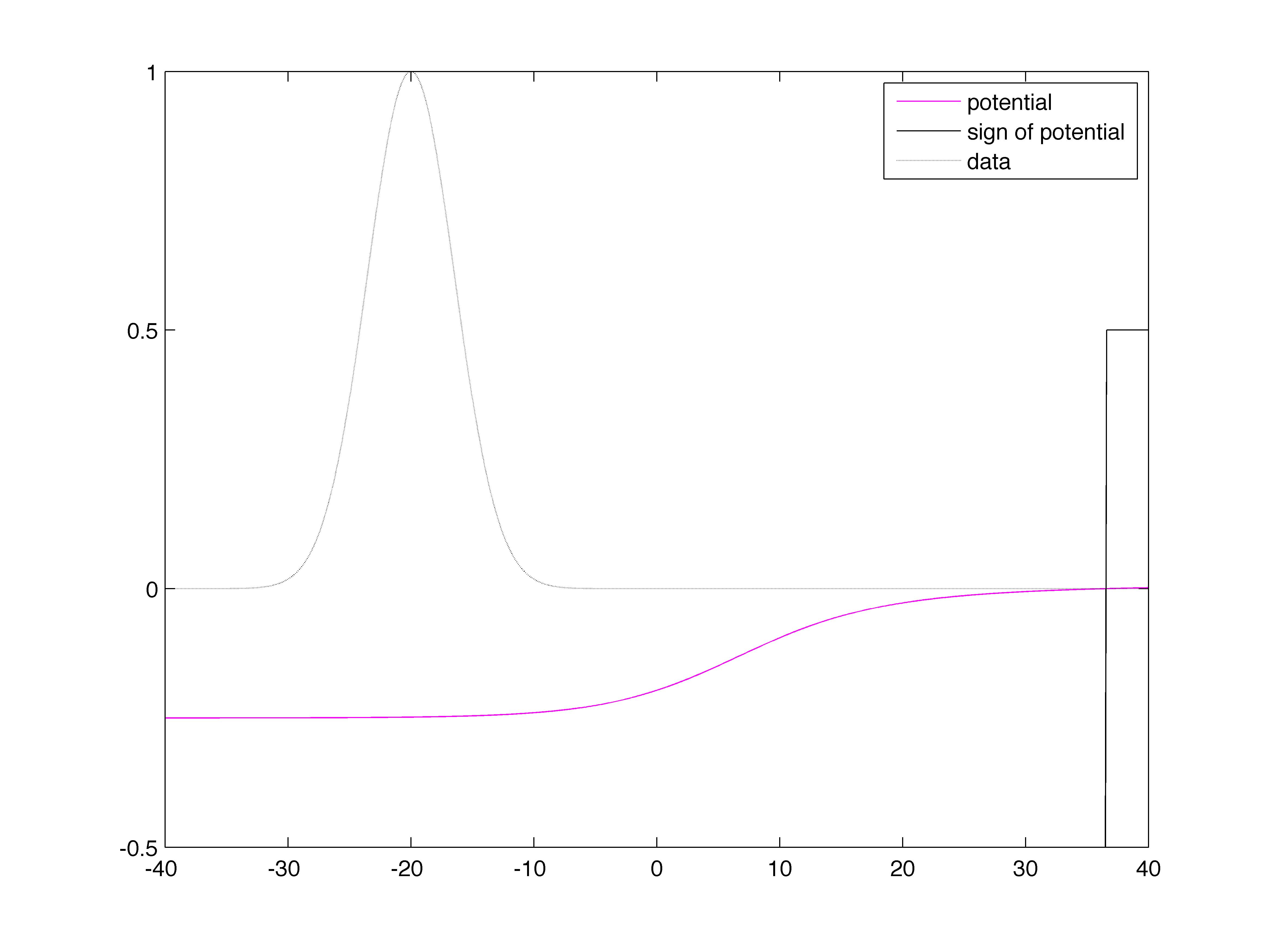}
\vspace*{-0.9cm}
\caption{Data and potential.}
\label{fig26}
\end{minipage}
\end{center}
\end{figure}
\begin{figure}[ht!]
\begin{center}
\hspace*{0.5cm}
\begin{minipage}{15cm}
\includegraphics[width=15cm,height=6.cm]{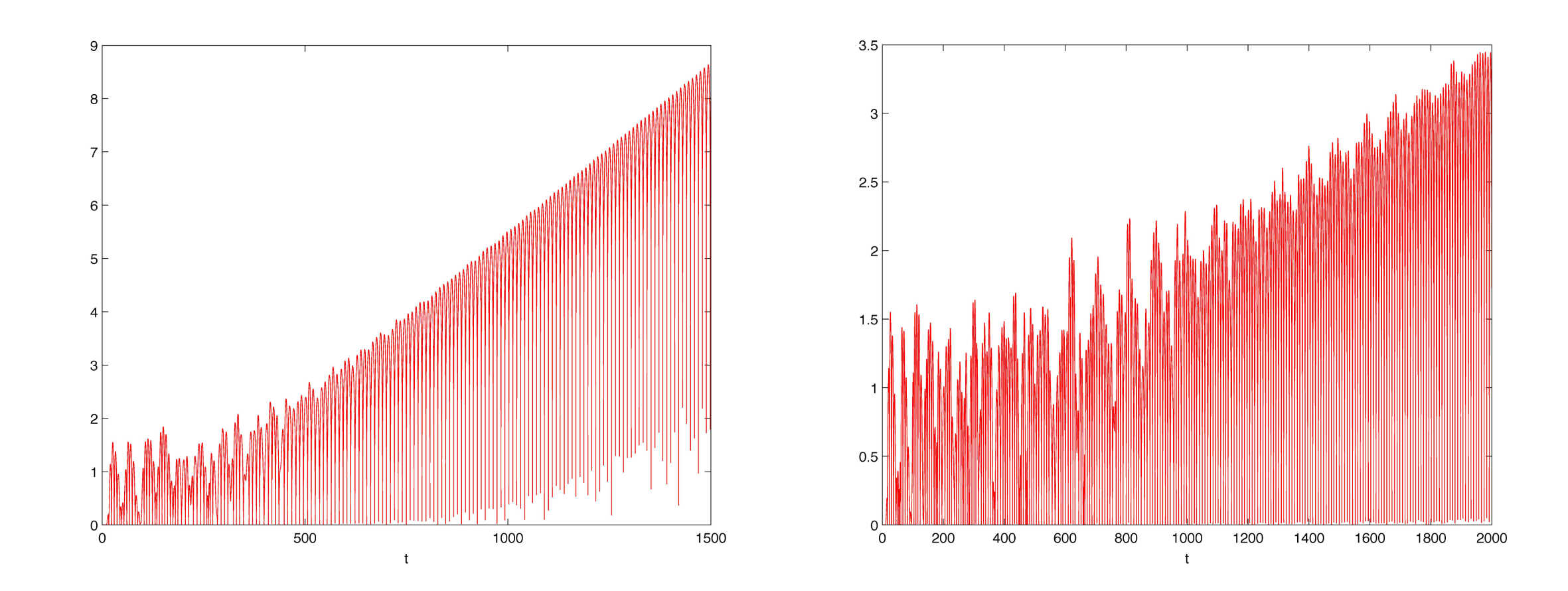}
\vspace*{-0.9cm}
\caption{Profile of logarithm of amplitude: Neumann conditions (left), Dirichlet conditions (right).}
\label{fig27}
\end{minipage}
\end{center}
\end{figure}
\begin{figure}[ht!]
\begin{center}
\hspace*{0.5cm}
\begin{minipage}{15cm}
\includegraphics[width=15cm,height=6.cm]{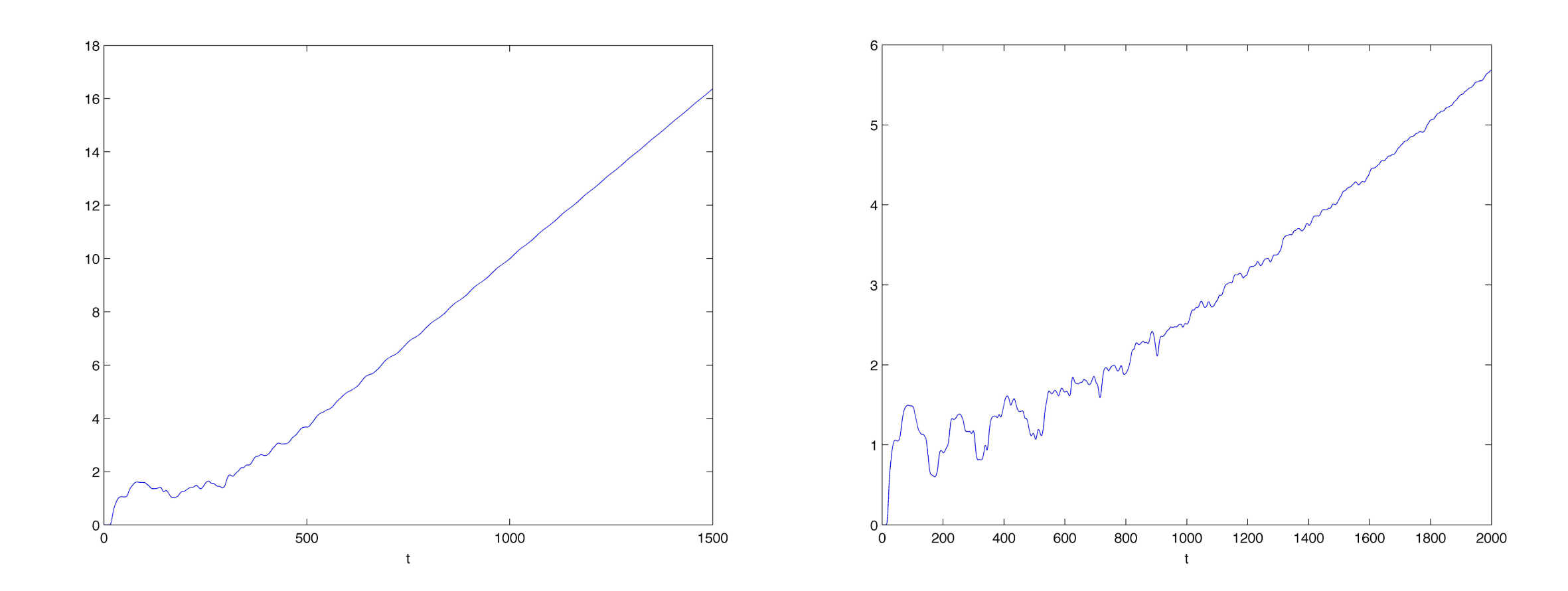}
\vspace*{-0.9cm}
\caption{Profile of logarithm of energy gain: Neumann conditions (left), Dirichlet conditions (right).}
\label{fig28}
\end{minipage}
\end{center}
\end{figure}
We note (Figure \ref{fig26}) that the potential is negative on the left-hand side of the domain and changes sign near the outer mirror. Its transition is rather mild; with incoming wave packets, this would make it difficult to observe superradiance, but with flares, superradiant behaviour is more easily captured. The amplitude of the field as well as the energy gain increase exponentially (Figures \ref{fig27} and \ref{fig28}), the latter meaning that negative energy accumulates near the inner mirror and positive energy near the outer mirror, at an exponential rate.
\item We also consider the de Sitter-Reissner-Nordström black hole that we have considered before, with $Q=2$, $M=3$, and $\lambda = 1/(6 M)^2$. The spatial domain is $[-40,40]$ with $4000$ discretisation points and we perform our simulation up to time $T=1000$ with $\delta t =h$.
\begin{figure}[ht!]
\begin{center}
\begin{minipage}{7cm}
\includegraphics[width=7cm,height=6.cm]{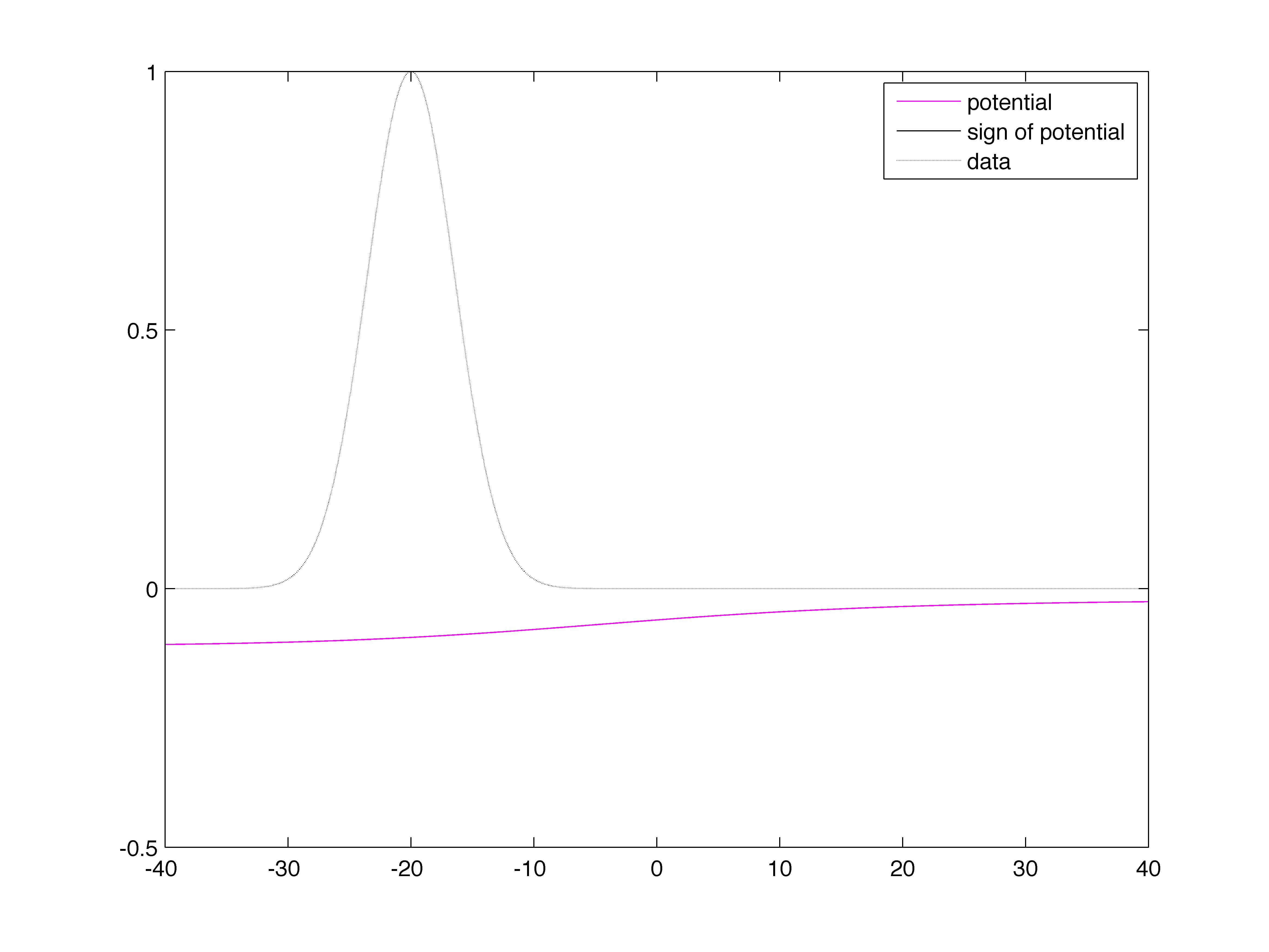}
\vspace*{-0.9cm}
\caption{Data and potential, $q=1$.}
\label{fig29}
\end{minipage}\begin{minipage}{7cm}
\includegraphics[width=7cm,height=6.cm]{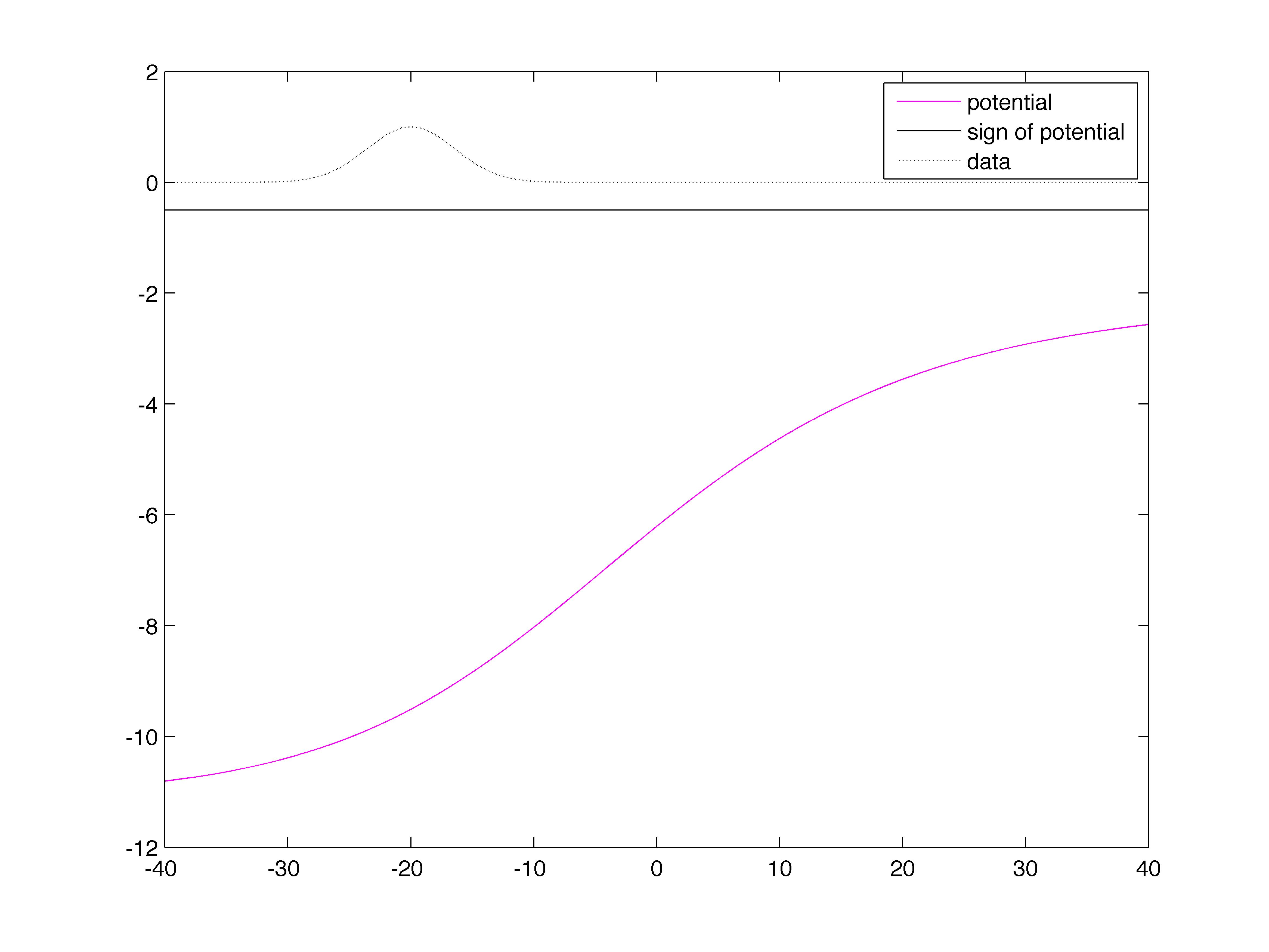}
\vspace*{-0.9cm}
\caption{Data and potential, $q=10$.}
\label{fig30}
\end{minipage}
\end{center}
\end{figure}
\begin{figure}[ht!]
\begin{center}
\hspace*{0.5cm}
\begin{minipage}{15cm}
\includegraphics[width=15cm,height=6.cm]{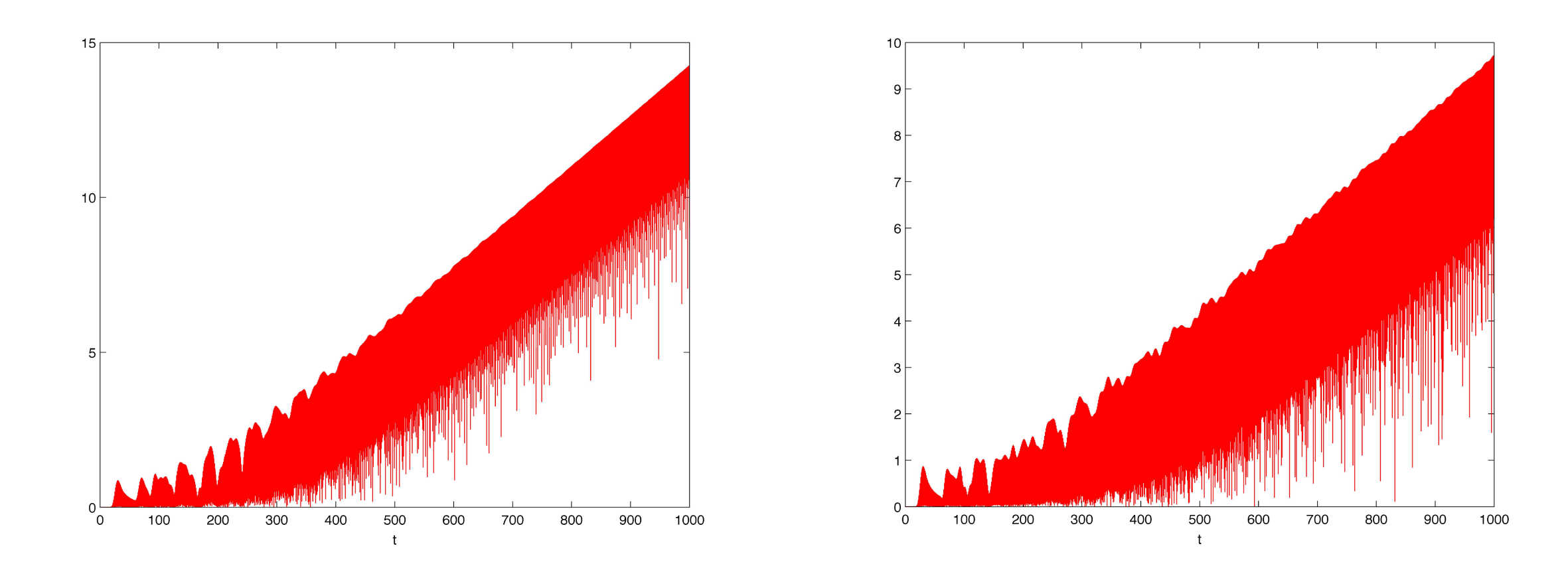}
\vspace*{-0.9cm}
\caption{Profile of logarithm of amplitude: Neumann conditions (left), Dirichlet conditions (right).}
\label{fig31}
\end{minipage}
\end{center}
\end{figure}
\begin{figure}[ht!]
\begin{center}
\hspace*{0.5cm}
\begin{minipage}{15cm}
\includegraphics[width=15cm,height=6.cm]{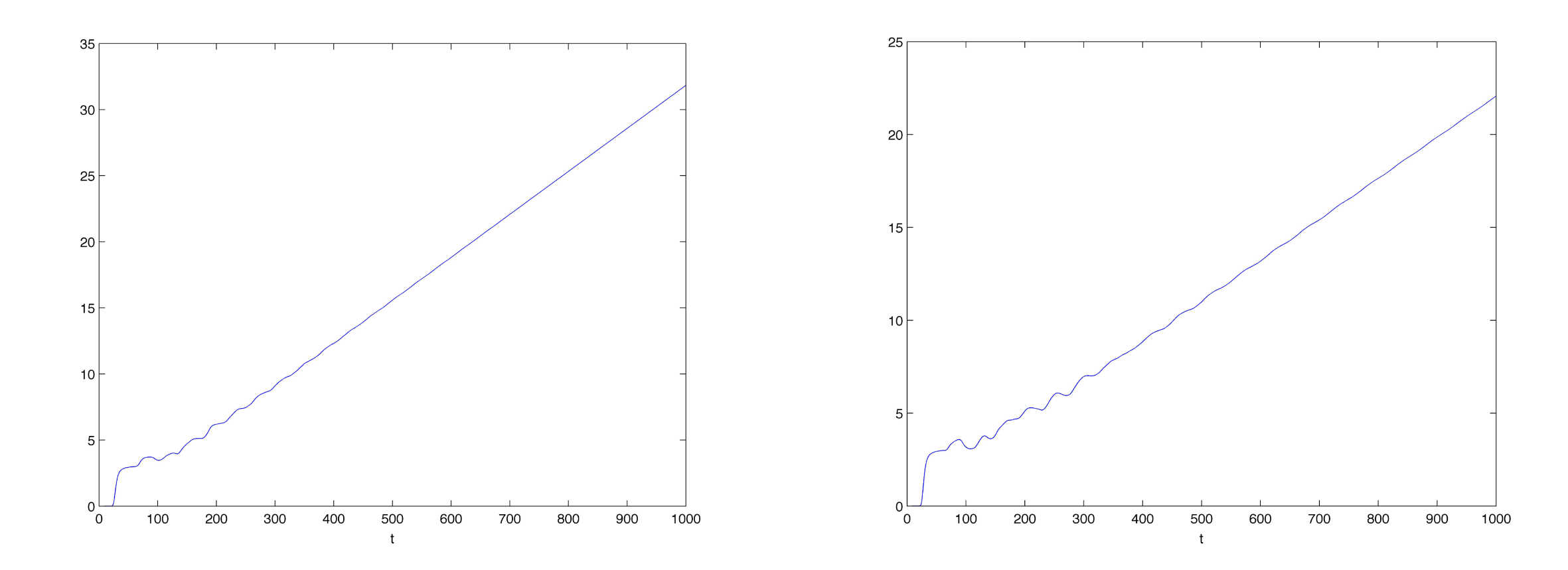}
\vspace*{-0.9cm}
\caption{Profile of logarithm of energy gain: Neumann conditions (left), Dirichlet conditions (right).}
\label{fig32}
\end{minipage}
\end{center}
\end{figure}
First, we try a field with mass $m=0.1$, $l=0$, $q=1$ with data given by \eqref{FlareData} with $\alpha =5$, $x_0 =-20$ and we measure the amplitude of the solution and the energy gain at $x=0$. The transition of the potential is extremely mild in this case (Figure \ref{fig29}) and even for large times we do not observe any linear instability. If we increase brutally the charge of the field to $q=10$, we observe a dramatic change and a severe linear instability appears (Figures \ref{fig31} and \ref{fig32}); the data and potential are displayed in Figure \ref{fig30}.
\item Finally, we show an example of high frequency behaviour of our scheme.
\begin{figure}[ht!]
\begin{center}
\begin{minipage}{7.5cm}
\includegraphics[width=7.5cm,height=6.cm]{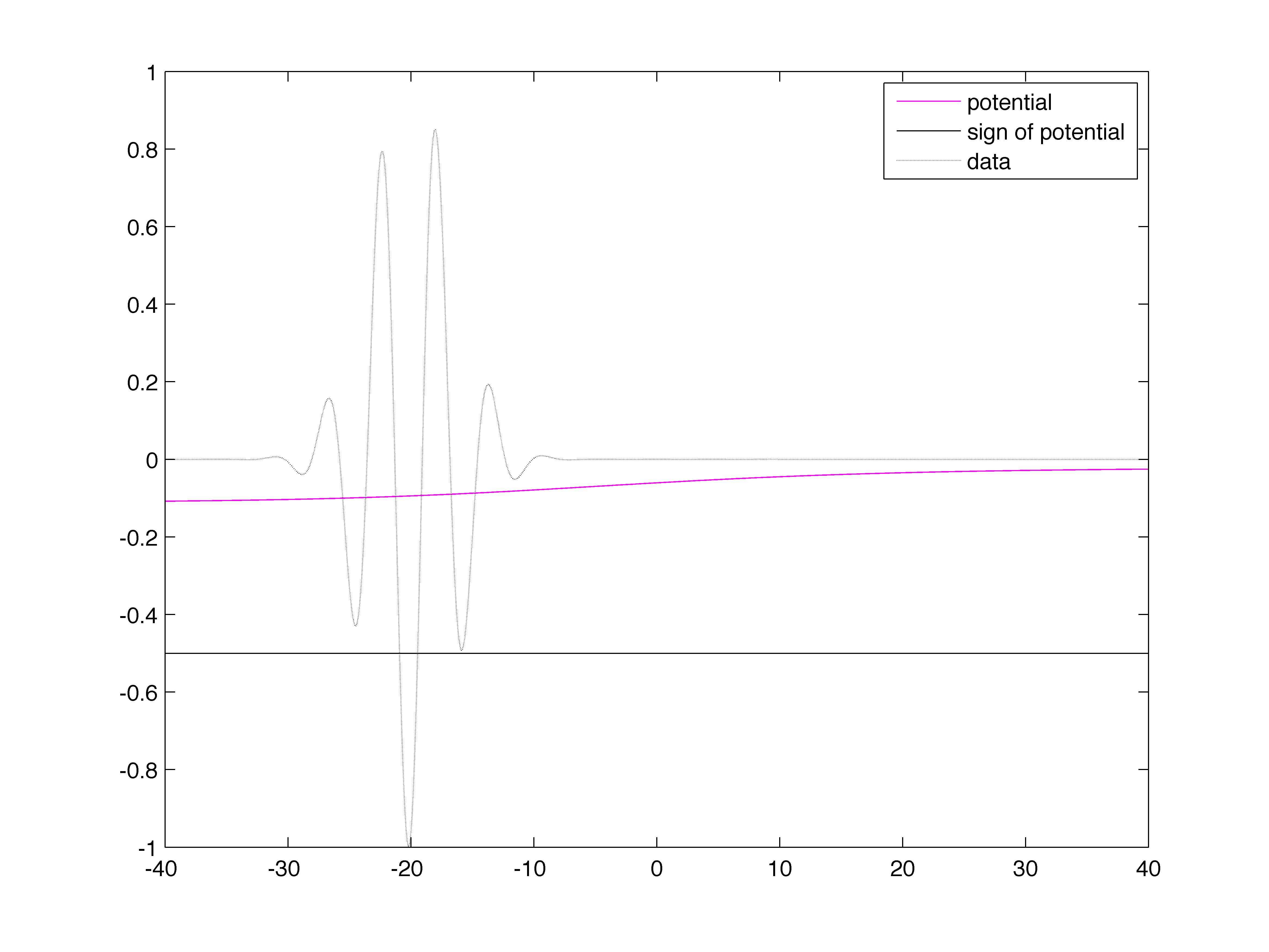}
\vspace*{-0.9cm}
\caption{Data and potential, $q=1$.}
\label{fig33}
\end{minipage}
\end{center}
\end{figure}
\begin{figure}
\begin{center}
\hspace*{0.5cm}
\begin{minipage}{15cm}
\includegraphics[width=15cm,height=6.cm]{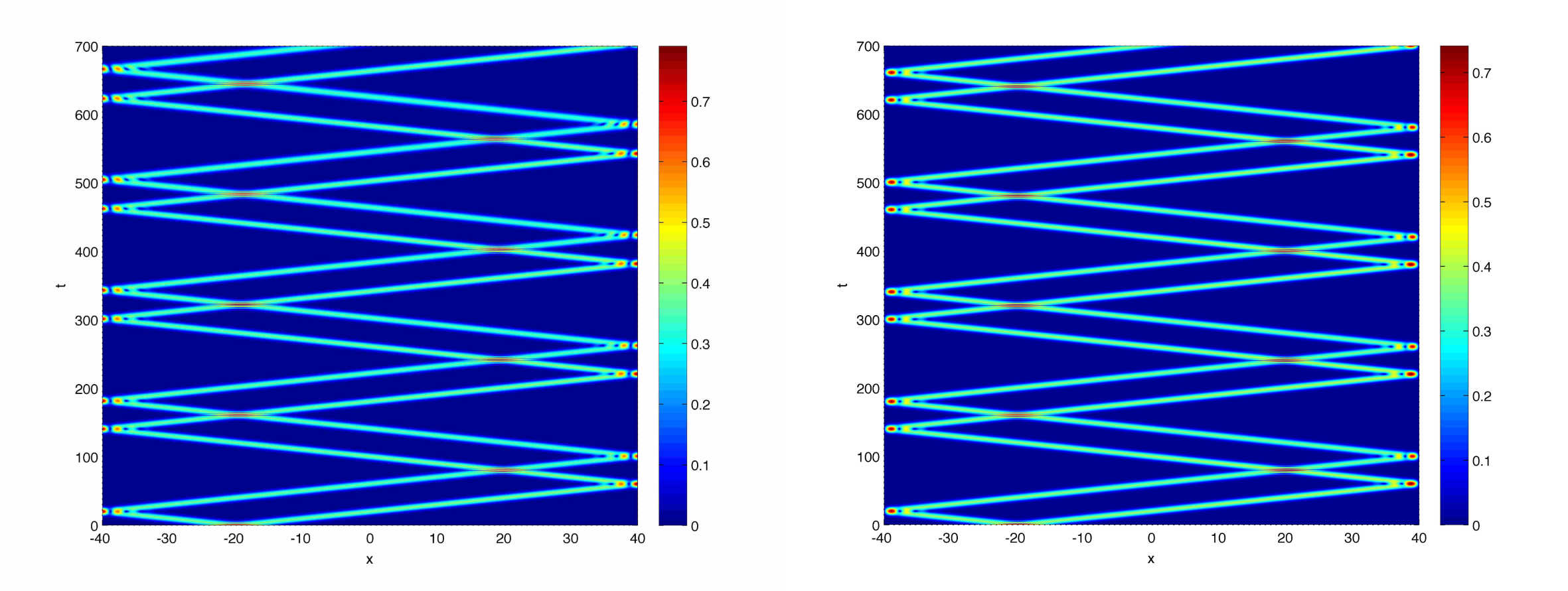}
\vspace*{-0.9cm}
\caption{$(x,t)$ isovalues, high frequency: Neumann conditions (left), Dirichlet conditions (right).}
\label{fig34}
\end{minipage}
\end{center}
\end{figure}
\begin{figure}[ht!]
\begin{center}
\begin{minipage}{7.5cm}
\includegraphics[width=7.5cm,height=6.cm]{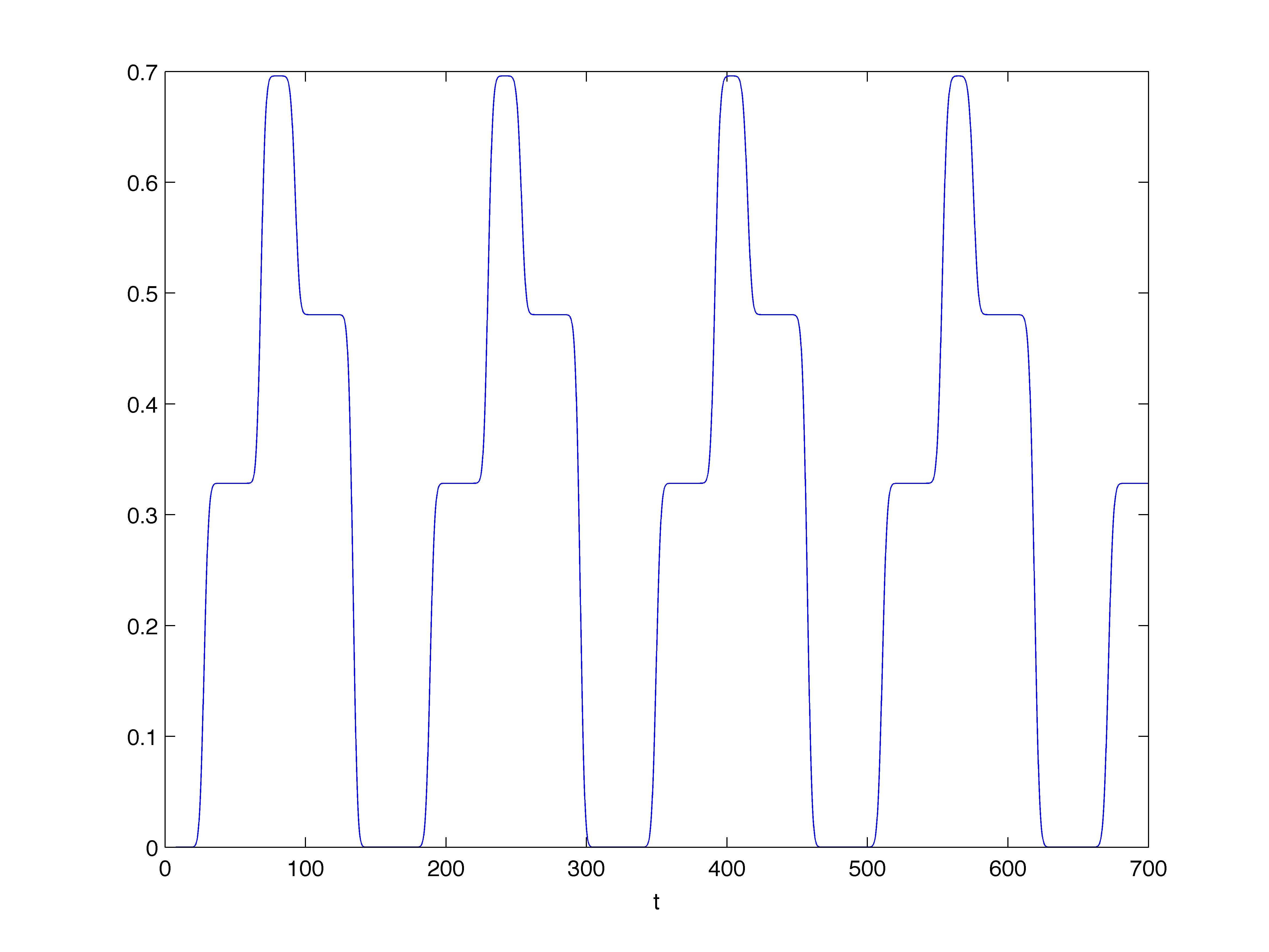}
\vspace*{-0.9cm}
\caption{Profile of log of energy gain, $q=10$.}
\label{fig35}
\end{minipage}
\end{center}
\end{figure}
We keep the same parameters as in the last simulation with $q=1$, taking now $\omega =7$. The data, the isovalues of the solution\footnote{Here it is the modulus of the field that is displayed, contrary to all the other cases where the absolute value of the real part of the solution was represented.}, and the profile of the logarithm of the energy gain are displayed in Figures \ref{fig33}, \ref{fig34}, and \ref{fig35}. For the isovalues, we show the simulations with Neumann and Dirichlet conditions; the corresponding energy gain profiles are almost identical and we only display the output for Neumann conditions.
The solution follows radial null geodesics between its reflections against the mirrors, no superradiance occurs and the field remains bounded.
\end{itemize}

\subsection{Numerical energy conservation}

As was remarked earlier on, the new form of the discrete energy that we have obtained, for both Dirichlet and Neumann boundary conditions, is conserved when considering the exact solution of the numerical scheme. However, the computed solution is only an approximate solution of the scheme and the discrete energy may not be exactly conserved. Moreover, since we are observing a regime in which the solutions display linear instabilities, the small errors made when computing the approximate solution will build up and we have to expect an exponential divergence between the exact discrete solution and the computed one. This must affect the discrete energy and cause exponential divergence. We study the evolution of the discrete energy for the type II black hole bomb on a Reissner-Nordström background that we investigated above, with Dirichlet boundary conditions. The simulation is pushed up to $T=1500$ with $16000$ points, then $32000$ and finally $64000$.
\begin{figure}[ht!]
\begin{center}
\begin{minipage}{15cm}
\includegraphics[width=15.5cm,height=4.5cm]{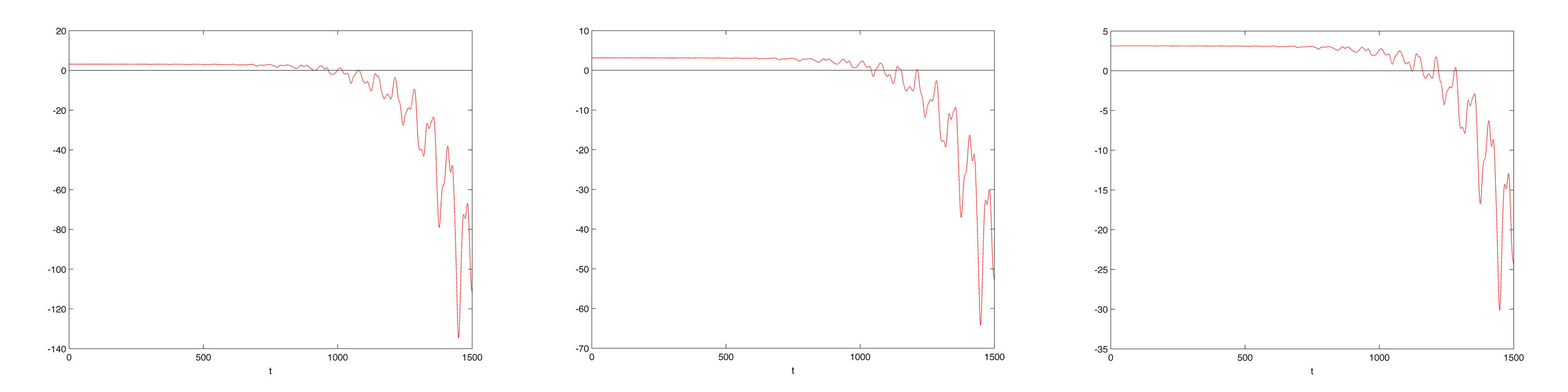}
\vspace*{-0.9cm}
\caption{Evolution of the discrete energy, $16000$ points, then $32000$ and $64000$.}
\label{fig36}
\end{minipage}
\end{center}
\end{figure}
\begin{figure}[ht!]
\begin{center}
\begin{minipage}{15cm}
\includegraphics[width=15.5cm,height=4.5cm]{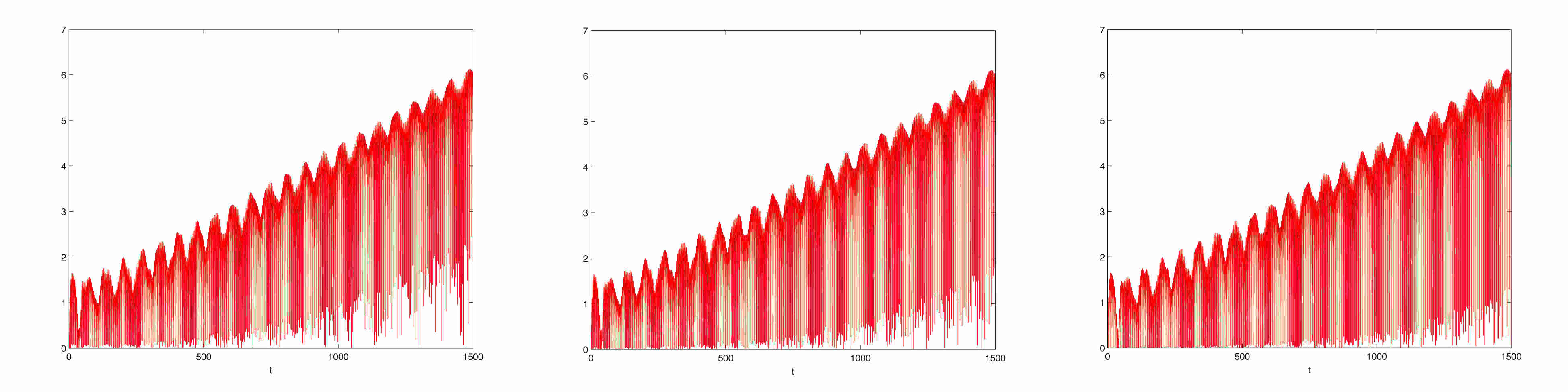}
\vspace*{-0.9cm}
\caption{Evolution of the field amplitude, $16000$ points, then $32000$ and $64000$.}
\label{fig37}
\end{minipage}
\end{center}
\end{figure}
\begin{figure}[ht!]
\begin{center}
\begin{minipage}{15cm}
\includegraphics[width=15.5cm,height=4.5cm]{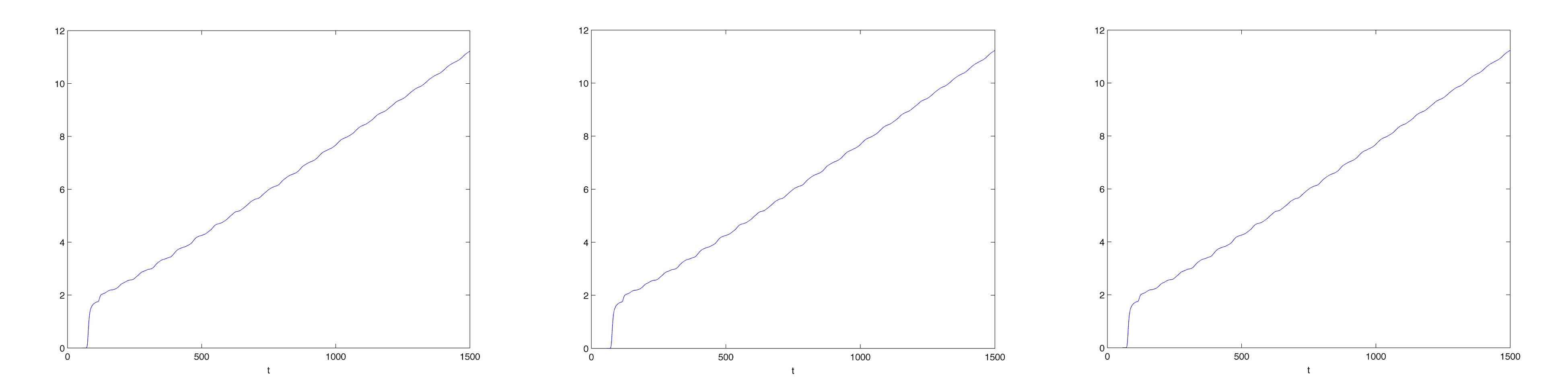}
\vspace*{-0.9cm}
\caption{Evolution of the energy gain, $16000$ points, then $32000$ and $64000$.}
\label{fig38}
\end{minipage}
\end{center}
\end{figure}
We see from Figure \ref{fig36} that the discrete energy indeed appears to diverge exponentially, but the divergence decreases like $h$ when we refine the discretisation. It seems to require a rather extreme precision to obtain a satisfactorily conserved quantity. However, the amplitude of the field and the energy gain (which relies on the calculation of the total energy only at $t=0$ and on the outgoing energy flux) do stabilise with very good accuracy when $h$ is refined (see Figures \ref{fig37} and \ref{fig38}), which indicates that our observations in this article are reliable.

\section{Conclusion}

Usually, black hole bombs refer to setting a mirror around the ergoregion.
Fields inside the mirror are thus forbidden to escape to infinity, creating linear instabilities thanks to repeated superradiant scattering in the ergoregion.
This is the construction introduced by W. H. Press and S. A. Teukolsky for rotating black holes.
In this article we have introduced a new type of black hole bomb.
Namely, the mirror is inside the ergoregion but now the fields are located outside.
Due to the presence of the mirror, the fields cannot fall into the black hole and can propagate to infinity.
Because the field can still reach the ergoregion, this system also leads to linear instability.
To our knowledge, this is the first time that this behaviour is exhibited.
In addition, we have also introduced the so-called sandwich bomb: two mirrors form a cavity where the field is trapped and which contains part of the ergoregion.
For these three types of black hole bombs, we have provided explicit numerical examples of linear instabilities, using a robust numerical method that ensures the conservation of a discrete energy
consistent with the continuous one.
The numerical study focuses on the cases of Reissner-Nordström and de Sitter-Reissner-Nordström backgrounds. The mirrors are either described by Neumann boundary conditions or by Dirichlet boundary conditions. In the case of Dirichlet conditions, apart from a single test where both boundary conditions give almost indistinguishable outputs, the instability seems to be less strong than when using Neumann boundary conditions. However, the qualitative behaviour is the same.

\section*{Acknowledgements}

This research was partially sponsored by the ANR program Horizons (ANR-16-CE40-0012-01).
MP is supported by the European Research Council Consolidator Grant NNLOforLHC2.


\end{document}